\documentclass{article}

\pdfoutput=1

\usepackage{arxiv}

\usepackage[utf8]{inputenc} % allow utf-8 input
\usepackage[T1]{fontenc}    % use 8-bit T1 fonts
\usepackage{hyperref}       % hyperlinks
\usepackage{url}            % simple URL typesetting
\usepackage{booktabs}       % professional-quality tables
\usepackage{amsfonts}       % blackboard math symbols
\usepackage{nicefrac}       % compact symbols for 1/2, etc.
\usepackage{microtype}      % microtypography
\usepackage{graphicx}
\usepackage{natbib} %this should be enough of a package to import the references
\usepackage{doi}
\usepackage{amsmath}
\usepackage{float}
\usepackage{rotating}
\usepackage{lineno}
\usepackage{multirow}
\usepackage{makecell}

\title{On the robustness of the emergent spatiotemporal dynamics in biophysically realistic and phenomenological whole-brain models at multiple network resolutions}

\author{ \hspace{1mm}Cristiana Dimulescu\,$^{1,2\dagger}$, Ronja Str\"{o}msd\"{o}rfer\,$^{1,3*\dagger}$, Agnes Fl\"{o}el\,$^{4,5}$, and Klaus Obermayer\,$^{1,2,3}$ \\
    $^{1}$Neural Information Processing Group, Fakult\"{a}t IV, Technische Universit\"{a}t Berlin, Berlin , Germany \\
    $^{2}$Bernstein Center for Computational Neuroscience, Berlin, Germany  \\
    $^{3}$Einstein Center for Neuroscience, Berlin, Germany \\
    $^{4}$Department of Neurology, University Medicine, Greifswald, Germany \\
    $^{5}$German Center for Neurodegenerative Diseases, Greifswald, Germany \\
    $^{\dagger}$shared first authorship
}

\renewcommand{\shorttitle}{Robustness of spatiotemporal whole-brain model dynamics}

\hypersetup{
pdftitle={A template for the arxiv style},
pdfsubject={q-bio.NC, q-bio.QM},
pdfauthor={David S.~Hippocampus, Elias D.~Striatum},
pdfkeywords={First keyword, Second keyword, More},
}

\begin{document}
\maketitle

\begin{abstract}
	The human brain is a complex dynamical system which displays a wide range of macroscopic and mesoscopic patterns of neural activity, whose mechanistic origin remains poorly understood. Whole-brain modelling allows us to explore candidate mechanisms causing the observed patterns. However, it is not fully established how the choice of model type and the networks’ spatial resolution influence the simulation results hence it remains unclear, to which extent conclusions drawn from these results are limited by modelling artefacts. Here, we compare the dynamics of a biophysically realistic, linear-nonlinear cascade model of whole-brain activity with a phenomenological Wilson-Cowan model using three structural connectomes based on the Schaefer parcellation scheme with 100, 200, and 500 nodes. Both neural mass models implement the same mechanistic hypotheses, which specifically address the interaction between excitation, inhibition, and a slow adaptation current which affects the excitatory populations.  We quantify the emerging dynamical states in detail and investigate how consistent results are across the different model variants. Then we apply both model types to the specific phenomenon of slow oscillations, which are a prevalent brain rhythm during deep sleep. We investigate the consistency of model predictions when exploring specific mechanistic hypotheses about the effects of both short- and long-range connections and of the antero-posterior structural connectivity gradient on key properties of these oscillations. Overall, our results demonstrate that the coarse-grained dynamics is robust to changes in both model type and network resolution. In some cases, however, model predictions do not generalize. Thus, some care must be taken when interpreting model results.
\end{abstract}

% keywords can be removed
\keywords{whole-brain modeling \and network resolution \and neural mass modeling \and spatiotemporal dynamics \and slow oscillations \and network physiology}

\section{Introduction} \label{section:introduction}
The human brain is a complex dynamical system. It exhibits a rich variety of spatiotemporally organized activity, where different patterns correspond to different functionalities and mechanisms in human cognitive processes. \cite{rasch2013sleep} state that slow oscillations (SOs), that travel as plane waves in an anterior-posterior direction (\cite{massimini2004sleep}), play a crucial role in memory consolidation during non-rapid eye movement (non-REM) sleep, and \cite{muller2016rotating} identified a dominant rotational temporal-parietal-frontal directionality of spindle oscillations that accompany SOs. Beyond spatiotemporal patterns during sleep, \cite{das2024planar} showed that spatial modes that regulate plane waves are absent in navigational memory tasks in humans while in verbal memory tasks, they observed different clusters of traveling waves depending on the letters that appear in words. Hence, an indicator of the functionality of a rhythm is its spatiotemporal organization (see further, \cite{breakspear2003modulation, mohan2024direction}). While reductionist approaches to the temporal dynamics of activity patterns have been widely researched to understand the functionality of the more local dynamics in the brain, most recently, neuroscientific research has shown an increasing interest to include the identification of the spatial dynamics, especially on a larger scale (see \cite{pessoa2022entangled, sporns2022complex}). 

In-silico methods can support these investigations by computational modeling of specific brain activity for the evaluation of candidate mechanisms. In-silico methods have been applied to surface EEG measurements (\cite{sanchez2017shaping, cakan2022spatiotemporal}), and to intracranially recorded activity in humans (\cite{deco2017dynamics, das2024planar, mohan2024direction, muller2016rotating}), rodents (see \cite{bhattacharya2022traveling, liang2021corticalmousewaves, dasilva2021modulation}), and other species (\cite{muller2014stimulus}). Intracranial recording methods measure activity of higher spatial and temporal resolutions, hence, in-silico methods require an adjustment to spatially denser models. On a smaller scale (i.e. not the whole brain), \cite{capone2023simulations} showed that different granularity of the recorded space changed the measured density of SO wave velocity in mice, where faster waves were neglected on a lower spatial resolution. On a larger scale,  \cite{popovych2021inter} found that the fit of simulated activity to empirical functional connectivity depends both on parcellation schemes and spatial resolution, and \cite{proix2016parcellation} shows that the parcellation size affects the dynamics of a whole-brain model whereas it was challenging to identify a consistent type of change.

Key to the emergence of different types of spatiotemporal patterns is the dynamical landscape of a computational model that can be decomposed into different regions of interest by the different types of stability a dynamical system experiences. \cite{sanchez2017shaping} showed that bistability is required for the organization of neocortical SOs both in-silico, as well as empirically. \cite{cakan2022spatiotemporal} identified a temporal destabilization of a stable high-activity state (up state) by a fatigue mechanism (spike-frequency adaptation) for transitioning into a low-activity state (down state) which is interrupted by noise to ultimately alternate at a low frequency ($<2Hz$). These SO wavefronts propagate as global plane waves. For the formation of more complex patterns, the presence of multi- or metastability is required (see \cite{kelso2012multistability}). These types of stability have been shown to play a crucial role in enabling elaborate spatiotemporal organizations in computational models (see, \cite{roberts2019metastable, kelso2012multistability}) with hallmarks of them being present in the human brain (see, \cite{freyer2009bistability, freyer2011biophysical}).

Different types of instability can also enable the formation of complex local patterns. \cite{townsend2018detection} applied methods from the analysis of turbulent flows to determine velocity vector fields over empirically recorded brain activity of mice. In those velocity vector fields, outward (sources) or inward (sinks) rotational waves emerge from unstable, or stable foci, respectively. Analogously for empirical data of humans, \cite{das2024planar} investigated the organization of sinks and sources and their role for different memory tasks, showing that in spatial tasks more sources, in verbal memory tasks more sinks were detected. Along that line, in the model study of \cite{breakspear2003modulation}, the authors emphasized the importance of balance between local short-range versus long-range connections\footnote{\cite{breakspear2003modulation} refer to excitatory couplings between cortical columns as long-range connections.} for the transition from independent, locally appearing oscillations to chaotic synchronization to global patterns. \cite{liang2021corticalmousewaves} supported this observation when investigating the spatiotemporal patterns in awake and anesthetized rodents. They not only emphasized the presence of complex local patterns during wakefulness but also showed, with computational modeling, that the coherence in low frequency bands is enhanced by stronger long-range connections between cortical areas further apart. Information processing has also been shown to be crucially affected by long-range connections by \cite{deco2021rare}, where the authors compared two whole-brain models, one with connections which exponentially decayed with distance and one with additional sparse long-range connections that deviated from that rule. They investigated complex brain activity that is functionally beneficial for the transmission of information between cortical regions and found the information cascade, i.e. the flow of brain activity across different spatial scales, to be significantly improved by the presence of these long-range connections. Studies such as the above, where brain activity is simulated with networks equipped with empirically informed structure, have been shown to reliably predict empirically observed patterns. \cite{cakan2022spatiotemporal}, for example, showed that the observed direction of SOs can be implicated by the antero-posterior structural connectivity gradient that decreases in connectivity strength from the anterior to the posterior direction.

Given the large number of computational modeling studies which investigate the spatiotemporal organization of neural activity on larger scales, we are left with the question in how far results generalize across the different whole-brain modeling approaches. Here, we specifically investigate whether, and how strongly, the specific choice of the dynamical system and of the spatial resolution changes the observed patterns, and how the connectivity profiles affect the emergent dynamics beyond empirically observed variability. We compare the emergent dynamics of whole-brain models based on the biophysically realistic adaptive linear-nonlinear cascade (aLN) model (\cite{augustin2017low, cakan2020biophysically, cakan2022spatiotemporal}) and the phenomenological Wilson-Cowan model (\cite{wilson1972excitatory}), both equipped with spike-frequency adaptation as a fatigue mechanism. To identify the role of spatial density in the models, we show the results for three network parcellations based on the Schaefer local-global parcellation schemes (\cite{schaefer2018local}) with 100, 200, and 500 nodes. We find that the coarse-grained dynamical landscape remains robust across models and network resolutions. However, results may not generalize when exploring specific dynamical states.

\section{Materials and Methods} \label{section:methods}
\subsection{Data}
\subsubsection{Participants}
We used diffusion tensor imaging (DTI) data and anatomical T1 scans which were acquired at the Universit\"atsmedizin Greifswald from 27 participants (15 females; age range = 50 - 78 years, mean age = 63.55 years). Prior to participating in the study, all participants gave a written informed consent and were subsequently reimbursed for participation. The study was approved by the local ethics committee at the Universit\"atsmedizin Greifswald and was conducted in accordance with the Declaration of Helsinki. 

\subsubsection{Data acquisition and preprocessing}\label{subsec:data_acqu}
The acquisition parameters and preprocessing of the DTI and anatomical T1 scans were identical to those described in \citet{cakan2022spatiotemporal}.

We defined the anatomical regions according to the Schaefer cortical parcellation scheme (\cite{schaefer2018local}) with 100, 200, and 500 nodes, respectively. We employed the same probabilistic tractography algorithm with 5,000 randomly sampled streamlines per voxel, which yielded one structural connectivity matrix and one fiber length matrix per participant. One participant was excluded because the tractography procedure at the highest network resolution failed. Following probabilistic tractography, we normalized the resulting structural connectivity matrix for each participant by dividing the connection probability $C_{ij}$ from seed region \textit{i} to target region \textit{j} by 5,000 (number of streamlines per voxel) \textit{x} \textit{n} (number of voxels in the seed region \textit{i}). As probabilistic tractography contains no directional information, we estimated $C_{ij}$ by averaging the connection probabilities from \textit{i} to \textit{j} and \textit{j} to \textit{i} (\cite{cabral2012functional}).

In addition to the individual connectomes, we constructed average structural connectivity matrices C and average fiber length matrices D for each parcellation.

\subsection{Whole-brain network models}\label{subsec:models}
We used whole-brain networks that consist of $N\in\{100, 200, 500\}$ nodes following the parcellation schemes described in Section \ref{subsec:data_acqu}. Each node represents a brain region and consists of an excitatory ($E$) and an inhibitory ($I$) population of model neurons. The nodes are connected by edges with the connections strengths given by the connectivity matrices C. Each excitatory population is equipped with an activity-dependent adaptation mechanism ($A$) that acts as a hyperpolarising feedback current.

\subsubsection{The aLN model}
The adaptive linear-nonlinear (aLN) model is a mean-field neural mass model of a network of coupled adaptive exponential integrate-and-fire (AdEx) neurons. It was developed in \cite{augustin2017low} and validated against simulations of spiking neural networks in \cite{cakan2020biophysically}. We used the \textit{neurolib} framework introduced in \cite{cakan2021neurolib} for the numerical simulations. The dynamics of each node (\cite{cakan2022spatiotemporal}) is summarized by the following equations:
\begin{equation}
    \begin{aligned}
    \tau_\alpha\frac{\text{d}\mu_\alpha}{\text{d}t}&=-\mu_\alpha^{syn}(t)+\mu_\alpha^{ext}(t)+\mu_\alpha^{ou}(t)-\mu_\alpha(t),\\
    \mu_\alpha^{syn}(t)&=J_{\alpha E}\bar{s}_{\alpha E}(t)+J_{\alpha I}\bar{s}_{\alpha I}(t),\\
    \sigma_{\alpha}^2(t)&=\sum_{\beta\in\{E,I\}}\frac{2J_{\alpha\beta}^2\sigma_{s,\alpha\beta}^2(t)\tau_{s,\beta}\tau_m}{(1+r_{\alpha\beta}(t))\tau_m+\tau_{s.\beta}}+\sigma_{ext,\alpha}^2\\
    \frac{\text{d}\bar{s}_{\alpha\beta}}{\text{d}t}&=\tau_{s,\beta}^{-1}\bigl(1-\bar{s}_{\alpha\beta(t)}(t)\bigr)\cdot r_{\alpha\beta}(t)-\bar{s}_{\alpha\beta}(t),\\
    \frac{\text{d}\sigma_{s,\alpha,\beta}^2}{\text{d}t}&=\tau_{s,\beta}^{-1}\bigl(1-\bar{s}_{\alpha\beta}(t)\bigr)^2\cdot\rho_{\alpha\beta}(t)+\bigl(\rho_{\alpha\beta}(t)-2\tau_{s,\beta}(r_{\alpha\beta}(t)+1)\bigr)\cdot\sigma_{s,\alpha\beta}^2(t),\ \ \ \text{ for }\alpha,\ \beta\in\{E,I\},
    \label{eq:aln}
    \end{aligned}
\end{equation}where $\bar{s}_{\alpha\beta}$ represents the mean and $\sigma_{s, \alpha\beta}^2$ the variance of the fraction of active synapses. Means and variances are computed across all neurons within each population. Given $\mu_\alpha$, the mean membrane current, its standard deviation $\sigma_\alpha$, and a set of nonlinear transfer functions $\Phi_\gamma(\mu_\alpha,\sigma_\alpha),\ \gamma\in\{\tau, V, r\}$, the mean membrane potentials $\bar{V}_\alpha=\Phi_V(\mu_\alpha,\sigma_\alpha)$ and the population firing rate $r_\alpha=\Phi_r(\mu_\alpha,\sigma_\alpha)$ can be calculated from the Fokker-Plank equations as in \cite{Richardson2007FiringrateRO}. The time constant $\tau_\alpha$ is input-dependent with $\tau_\alpha=\Phi_\tau(\mu_\alpha, \sigma_\alpha)$. The values for $\bar{V}_E, \tau_\alpha$, and $r_\alpha$ are evaluated at every time step with precomputed functions such that the effective input rate from population $\beta$ to $\alpha$ is determined by the mean $r_{\alpha\beta}$ and the variance $\rho_{\alpha\beta}$ with\begin{equation}
    \begin{aligned}
    r_{\alpha\beta}(t) &= \frac{c_{\alpha\beta}}{J_{\alpha\beta}}\tau_{s,\beta}\bigl(K_\beta\cdot r_{\beta}(t-d_{\alpha})+\delta_{\alpha\beta E}\cdot K_{gl}\sum_{j=0}^N C_{ij}\cdot r_{\beta}(t-D_{ij})\bigr)\\
    \rho_{\alpha\beta}(t) &= \frac{c_{\alpha\beta}^2}{J_{\alpha\beta}}\tau_{s,\beta}^2\bigl(K_\beta\cdot r_{\beta}(t-d_{\alpha})+\delta_{\alpha\beta E}\cdot K_{gl}\sum_{j=0}^N C_{ij}^2\cdot r_{\beta}(t-D_{ij})\bigr),
    \end{aligned}
\end{equation}\noindent 
The mean adaptation current  $\bar{I}_A$ is given by\begin{equation}
    \begin{aligned}
    \frac{\text{d}\bar{I}_A}{\text{d}t} = \tau_A^{-1}\bigl(a(\bar{V}_E(t)-E_A)-\bar{I}_A\bigr)+b\cdot r_E(t).
    \end{aligned}
\end{equation}All parameters not explained above are given and explained in Table \ref{tab:model_parameters_aln}. Values were chosen as in \cite{cakan2022spatiotemporal} with the global coupling strength $K_{gl}$ fixed to one value for all parcellations, see Table \ref{tab:model_parameters_aln}. For the determination of units for the parameters, see \cite{cakan2022spatiotemporal}.

\subsubsection{The Wilson-Cowan model}
The Wilson-Cowan model (\cite{wilson1972excitatory}) describes the dynamics of the proportions of excitatory ($r_E(t)$) and inhibitory ($r_I(t)$) neurons firing per unit time (\cite{Kilpatrick2013WC}). Even though the aLN and Wilson-Cowan models represent neuronal firing somewhat differently, we denote both dynamical variables with $r_k\ \in\{E,I\}$ for brevity. The framework in \cite{cakan2021neurolib} provides an implementation of the original model equations including a refractory term. Since the refractory time only rescales the solutions $r_E(t)$, and $r_I(t)$ but has no qualitative effect on the dynamics (\cite{Pinto1996refractory}), we omitted it for this study. Furthermore, a spike-frequency adaptation current is considered. The dynamics in each node is thus determined by the following equations:
\begin{equation}
    \begin{aligned}
    \tau_E\frac{\text{d}r_{E,j}}{\text{d}t}=&-r_{E,j}(t)+F_E\bigl(w_{EE}r_{E,j}(t)-w_{EI}r_{I,j}(t)+\mu^{ext}_{E}+I^{ext}_j(t)-a_j(t)+\mu^{ou}_E\bigr)\\
    \tau_I\frac{\text{d}r_{I,j}}{\text{d}t}=&-r_{I,j}(t)+F_I\bigl(w_{IE}r_{E,j}(t)-w_{II}r_{I,j}(t)+\mu^{ext}_{I}+\mu^{ou}_I\bigr)\\
    \tau_a\frac{\text{d}a_j}{\text{d}t}=&-a_j(t)+bF_A\bigl(r_{E,j}(t)\bigr).\label{eq:wc}
    \end{aligned}
\end{equation} $I^{ext}_j(t)$, the input from other nodes to the excitatory population of node $j$, is determined by the connectivity matrix $C=\{C_{jk}\}$ and the delay matrix $D=\{D_{jk}\}$, and scaled by a global coupling strength $K_{gl}\in\mathbb{R}_0^+$:
\begin{equation}
    \begin{aligned}
    I^{ext}_j(t)=K_{gl}\cdot\sum_{k=1}^N \bigl(C_{jk} \cdot r_{E,k}(t-D_{jk})\bigr).
    \end{aligned}
\end{equation}
To simplify the Equations \eqref{eq:wc}, we consider a mean external input $\mu^{ext}_\alpha$ for $\alpha\in\{E,I\}$ to each node, which is constant across nodes. The transfer functions $F_\alpha(\cdot),\ \alpha\in\{E,I,A\}$, are chosen to be sigmoidal: \begin{align*}
    F_\alpha(x)=\frac{1}{1+\exp{\bigl(-a_\alpha(x-\nu_\alpha)\bigr)}}. 
\end{align*}
A description for each parameter can be found in Table \ref{tab:model_parameters_wc}. These parameter values were chosen, because they give rise to a dynamical landscape which is similar to other systems that also reliably produce SOs (\cite{cakan2020biophysically, cakan2022spatiotemporal}). The parameter setting required minor adjustments compared to previous studies that used the Wilson-Cowan model to simulate various types of spatiotemporal patterns (\cite{Levenstein2019, Papadopoulos2020, Torao2021}).

\subsubsection{Noise}
For the investigation of simulated sleep SOs, shown in Figures \ref{fig:sc_gradient_change_1}, A18, \ref{fig:coherence_aln}, and A19, noise input to each population $\alpha\in\{E,I\}$ in both models was considered. Noise is described by an Ornstein-Uhlenbeck process \begin{equation*}
    \frac{\text{d}\mu^{ou}_\alpha(t)}{\text{d}t} = -\frac{\mu^{ou}_\alpha}{\tau_{ou}}+\sigma_{ou}\xi(t),
\end{equation*}where $\xi(t)$ is sampled from a normal distribution with zero mean and unit variance and $\tau_{ou}$ is the time constant set to $5$ ms for both models. The variance $\sigma_{ou}$, also referred to as noise strength, is different for each model and given in Tables \ref{tab:model_parameters_aln} and \ref{tab:model_parameters_wc}.

\subsection{Analysis}
\subsubsection{State space analysis} \label{subsection:methods_state_space}
The analysis of the state space was conducted numerically in the absence of noise. We randomly initialized and simulated the model for 101 x 101 parameter values (10,201 simulations in total) for the mean external inputs to the E and I populations for a duration of 30 s. The duration was extended to 1 min for the Wilson-Cowan model with adaptation, as in some cases $r_E$ needed a longer time to return to baseline after the application of the positive stimulus, see paragraph below. 

For every point in state space, we applied a negative, but increasing, followed by a positive, but decaying stimulus. Examples are shown in Figure \ref{fig:aln_example_traces}. Subsequently, we computed the difference between the average $r_E$ over the last 2 s of simulation and the 2 s prior to the application of the positive stimulus. As in \cite{cakan2022spatiotemporal}, the point was classified as bistable, if this difference was larger than 10 Hz for the aLN and larger than 0.1 for the Wilson-Cowan model for at least one node in the network. These thresholds were chosen because the bistable states displayed differences larger than these values across the entire state space in both models for the chosen parameterizations, detailed in Section \ref{subsec:models}.

Furthermore, we computed the difference between the maximum and minimum value of $r_E$ over the last 2 s of simulation. We classified each point as oscillating if this value was larger than 10 Hz for the aLN and 0.1 for the Wilson-Cowan model for at least one node in the network.

Figure A1 shows the single-node bifurcation diagrams for both models with and without adaptation obtained using the procedures described above.

\subsubsection{State classification} \label{subsection:methods_state_classification}
\begin{figure}[H]
    \centering
    \includegraphics[width=\textwidth]{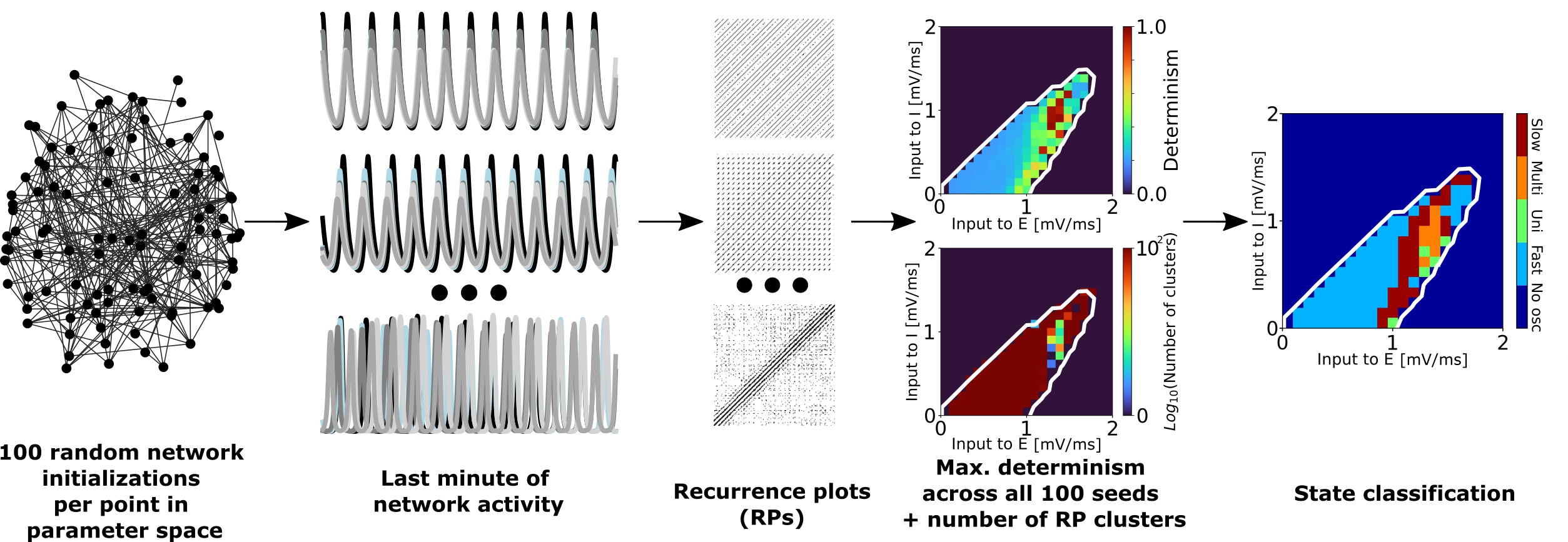}
    \caption{Summary of the procedure used to classify network states into unistable, multistable, fast metastable, and slow metastable. For each model (aLN or Wilson-Cowan) and each parcellation (100, 200, or 500 nodes) we conducted 100 randomly initialized simulations of 2 minutes duration for each point in the slice of parameter space spanned by varying the external excitatory ($\mu_E^{ext}$) and inhibitory ($\mu_I^{ext}$) input currents. We discarded the first minute of activity to eliminate transient effects and used the last minute of network activity to compute the recurrence plots. Based on these, we computed the maximum determinism value across all 100 seeds, and we clustered the recurrence plots using the DBSCAN algorithm. Combining the information from these two sources, we classified each point into one of the four states mentioned above.}
    \label{fig:summary_state_classification}
\end{figure}

To characterize the temporal dynamics of each point in the oscillatory regions (identified as described in Section \ref{subsection:methods_state_space}), we used the procedure summarized in Figure \ref{fig:summary_state_classification}. For each point in the slice of state space spanned by the external input currents to the excitatory and inhibitory populations, $\mu_E^{ext}$ and $\mu_I^{ext}$, we simulated network activity in the absence of noise over a period of 2 min and for 100 random initializations. The first minute of activity was discarded to account for transient effects. Next, for each initialization, we computed recurrence plots with entries:
\begin{equation}
    \begin{linenomath}
        R(t, t') = 
        \begin{cases}
            1, & \text{if $\lVert \overrightarrow{x}(t) - \overrightarrow{x}(t') \rVert \leq \epsilon$}\\
            0, & \text{otherwise,}
        \end{cases}   
    \end{linenomath}
\end{equation}
where $\overrightarrow{x}(t)$, $\overrightarrow{x}(t')$ contain the values of $r_E$ at time points \textit{t} and \textit{t'} across all nodes. $\epsilon$ is the recurrence threshold, and $\lVert \cdot \rVert$ denotes the Euclidean norm. To account for different amplitudes of $r_E$, which could lead to different results if a fixed threshold $\epsilon$ were to be used across initializations and parametrizations, we adjusted the recurrence threshold $\epsilon$ until the recurrence rate (defined as the proportion of non-zero entries in the resulting recurrence plot) of 0.1 (\cite{zbilut2002recurrence}) was reached.

For each parametrization, we clustered the resulting recurrence matrices using the DBSCAN algorithm (\cite{ester1996density}). Additionally, we computed the determinism value DET, 
\begin{equation}
    \begin{linenomath}
        DET = \frac{\sum_{l=l_{min}}^N l P(l)}{\sum_{l=1}^N l P(l)},
    \end{linenomath}
\end{equation}
for each initialization, where $P(l)$ is the fraction of the diagonal lines with length $l$ in the recurrence plot, and $l_{min}$ specifies a minimum diagonal length. The determinism value ranges between 0 and 1.

We used the number of clusters to classify each state in the limit cycle as either unistable (if the number of clusters was equal to 1), multistable (if the number of clusters was $\leq$30), or metastable (if the number of clusters was $>$30).  The thresholds were determined based on the visual inspection of the number of clusters per point in the oscillatory regimes, as exemplified in Figure \ref{fig:summary_state_classification} (panel in the fourth column, bottom plot, depicting the number of clusters in the oscillatory region of state space). This led to a clear boundary between metastable versus multi- and unistable regions (panel in the fourth column, bottom plot of Figure \ref{fig:summary_state_classification}, dark red versus multicolored regions) . We further distinguished between fast and slow metastable states by the maximum determinism value across the 100 initializations. Fast metastable states are characterized by values $\leq$0.35, with short state durations, while slow metastable states are characterized by values $>$0.35, with longer state durations. We opted for the maximum determinism value as this allowed us to identify the presence of at least one slow metastable state across the 100 initializations. The value of 0.35 was chosen as the threshold based on the visual inspection of the determinism values computed for all state space locations in the oscillatory region. This showed clusters of regimes with determinism values $> 0.35$, across the state space (see example in Figure \ref{fig:summary_state_classification} in the panel in the fourth column, top plot, showing the maximum determinism value in the oscillatory region of the state space). Additionally, the choice was confirmed through the visual inspection of the interhemispheric cross-correlation (see Section \ref{subsection:methods_interhem_cc} and examples in Figure \ref{fig:aln_slow_fast_example}) for several points in the state space. This allowed us to visually confirm the difference in state duration between slow and fast metastable states.

\subsubsection{Kuramoto order parameter}
Using the simulation data described in Section \ref{subsection:methods_state_space}, we computed the Kuramoto order parameter R(t),
\begin{equation}
    \begin{linenomath}
        R(t) = |\frac{1}{N}\sum_{n=1}^{N}e^{i\theta_n(t)}|,
    \end{linenomath} \label{eq:kuramoto}
\end{equation}
where $\theta_n(t)$ denotes the instantaneous phase obtained from the Hilbert transform of the time series $r_E$ for each node \textit{n}, and \textit{N} $\in$ \{100, 200, 500\} denotes the total number of nodes in the network.

Subsequently, we summarized the results for each model and each network resolution using the mean and the standard deviation of $R(t)$. High values of the mean indicate a synchronous solution, whereas low values indicate an asynchronous solution. With respect to the standard deviation, high values are indicative of metastability and low values correspond to solutions which remain stable over time.

\subsubsection{Interhemispheric cross-correlation} \label{subsection:methods_interhem_cc}
To investigate spatial properties of oscillatory states, we computed the sliding-window time-lagged cross-correlation as in \citet{roberts2019metastable}. We calculated the intrahemispheric Kuramoto order parameter for each hemisphere. Subsequently, the windowed time-lagged cross-correlation between the two parameters was determined with a window of length 100 ms and 90\% overlap between consecutive windows, and with a lag $l$ of 50 ms.

\subsubsection{Singular value decomposition}\label{subsec:SVD}
\label{subsec:singular_value_decomp}
To conduct a singular value decomposition (SVD), we firstly computed the velocity vector fields (cf. \cite{roberts2019metastable}). For each node \textit{n}, we used the instantaneous Hilbert transform of $r_E$ to determine the phase $\theta_n$, after which we calculated the velocity $v_n$ using its spatial and temporal derivatives:
\begin{equation}
    \begin{linenomath}
        v_{n} = -(|\frac{\partial \theta_{n}}{\partial t}| / ||\nabla \theta_{n}||^2)\nabla \theta_{n}.
    \end{linenomath}
\end{equation}
The spatial derivative was calculated using the constrained natural element method (\cite{illoul2011some}), as described in \citet{roberts2019metastable}. This method allows for the calculation of the components of the gradient vector without the need for interpolation to and from a 3D grid.

The SVD was then performed for the velocity vector fields $\textbf{v} = \{v_n\}_{n=1}^N$, according to the method used by \citet{liang2021corticalmousewaves} and introduced in \citet{townsend2018detection}. Briefly, for each of the two models and for every network resolution, we concatenated the time series $v(t)$ of the velocity vector fields across all four state types identified in Section \ref{subsection:methods_state_classification} to obtain a matrix W (time steps and state types in rows and nodes in columns). This matrix was decomposed using SVD as:
\begin{equation}
    %\begin{linenomath}
        W = U \Sigma V^T,
    %\end{linenomath}
\end{equation}
where the columns of \textit{U} represent the left and the rows of $V^T$ represent the right singular vectors of $W$. Hence, the rows of $V^T$ represent the spatial modes of W, the columns of U their time course and the diagonal elements of $\Sigma$ the eigenvalues $\sigma$ in descending order of magnitude. The variance explained by each mode is given by $\sigma_k^2/\sum_i \sigma_i^2$.

We then projected the spatial modes identified on the concatenated data onto the individual vector velocity fields of each parametrization and quantified the proportion of explained variance by each projected spatial mode $m$ onto the $n$-th velocity vector field as $M_{m,n}^2/\sum_i M_{i,n}^2$, where $M$ denotes the projection matrix.

\subsubsection{Structural gradient manipulation} \label{subsection:gradient_manipulation_methods}
To investigate the effect of the structural gradient on the propagation of SOs, we used the sleep model parametrization introduced in \citet{cakan2022spatiotemporal} for the aLN model, with minor adjustments of the adaptation parameters (see Table A5). The adjustment was necessary because the parcellations of higher resolution had stronger pairwise connectivity strengths compared to the 100 node case, which caused the model to be in the up state for prolonged intervals of time due to a shift in state boundaries. The manual increase of the adaptation parameters ensured that the model visually displayed SOs (\cite{cakan2022spatiotemporal}). For the Wilson-Cowan model with 100 nodes, we conducted an evolutionary optimization in \textit{neurolib}, with resting-state functional connectivity and functional connectivity dynamics and with power spectrum of EEG in sleep stage N3 as optimization objectives (full procedure described in \citet{cakan2022spatiotemporal}). As the evolutionary optimization was computationally notfeasible for the networks with 200 and 500 nodes, we manually adjusted the adaptation parameters obtained for the network with 100 nodes (see Table A6) in the same manner as described above for the aLN model. To compare our results with previous work, compare dynamical landscapes across models and resolutions, we used the parameters given in Tables \ref{tab:model_parameters_aln}, and \ref{tab:model_parameters_wc}. For the sleep models, we modified a small number of parameters to place the model in a regime, where realistic SOs are produced (Tables A5, and A6, for the aLN and Wilson-Cowan model respectively).

The antero-posterior structural connectivity gradient defined as the slope of the linear regression between the node degree and its coordinate along the antero-posterior axis of the brain (\cite{cakan2022spatiotemporal}) is shown in Figure A17 for the three parcellations.

We manipulated the strength of the gradient by defining $p$ - the maximum percentage by which the connection strengths of the most anterior node were modified - and creating an equally spaced set of N values in [-$p$, +$p$], where N is the number of nodes in the network. We rank-ordered all nodes according to their coordinate along the antero-posterior axis and changed the connection strengths of each node by the corresponding value from this interval. We modified the connection strengths based on percentages rather than absolute values to ensure that no negative values were introduced in the structural connectivity matrix.

Additionally, we constructed control models with gradients similar to the networks constructed before in which we preserved the total sum of connection strengths, but destroyed the relationship between the connection strength and the corresponding fiber length. This was achieved by permuting the entries of the structural connectivity matrix until the value of the antero-posterior gradient fell within a predefined range, while maintaining the fiber length matrix intact.

To determine the direction of propagation of SO up/down state transitions along the antero-posterior axis, we first computed the proportion of regions in the down state as a function of time. The down states were identified by thresholding the excitatory $r_E(t)$ $\leq$ $\theta$ $\cdot$ $max(r_E(t))$, with $\theta$ = 0.01 for the aLN and $\theta$ = 0.2 for the Wilson-Cowan model at every time step. Subsequently, we applied a 0.5 - 2 Hz bandpass filter to the resulting time series, computed the Hilbert transform, and identified the transition phase of a node as the phase of the Hilbert transform at the time point at which the node transitioned from the up (down) to the down (up) state. Phases were averaged across all transitions of each node. We then computed the Pearson correlation coefficient between the average transition phase and the node coordinate along the antero-posterior axis. Positive (negative) values of the correlation between the up-to-down transition phases and the node coordinates indicate a preferential antero-posterior (postero-anterior) direction of propagation, and vice-versa for the down-to-up transitions.

\subsection{Manipulation of short- vs. long-range connection strengths}\label{subsec:short-long-definition-methods}
\begin{figure}[H]
    \centering
    \includegraphics[width=\textwidth]{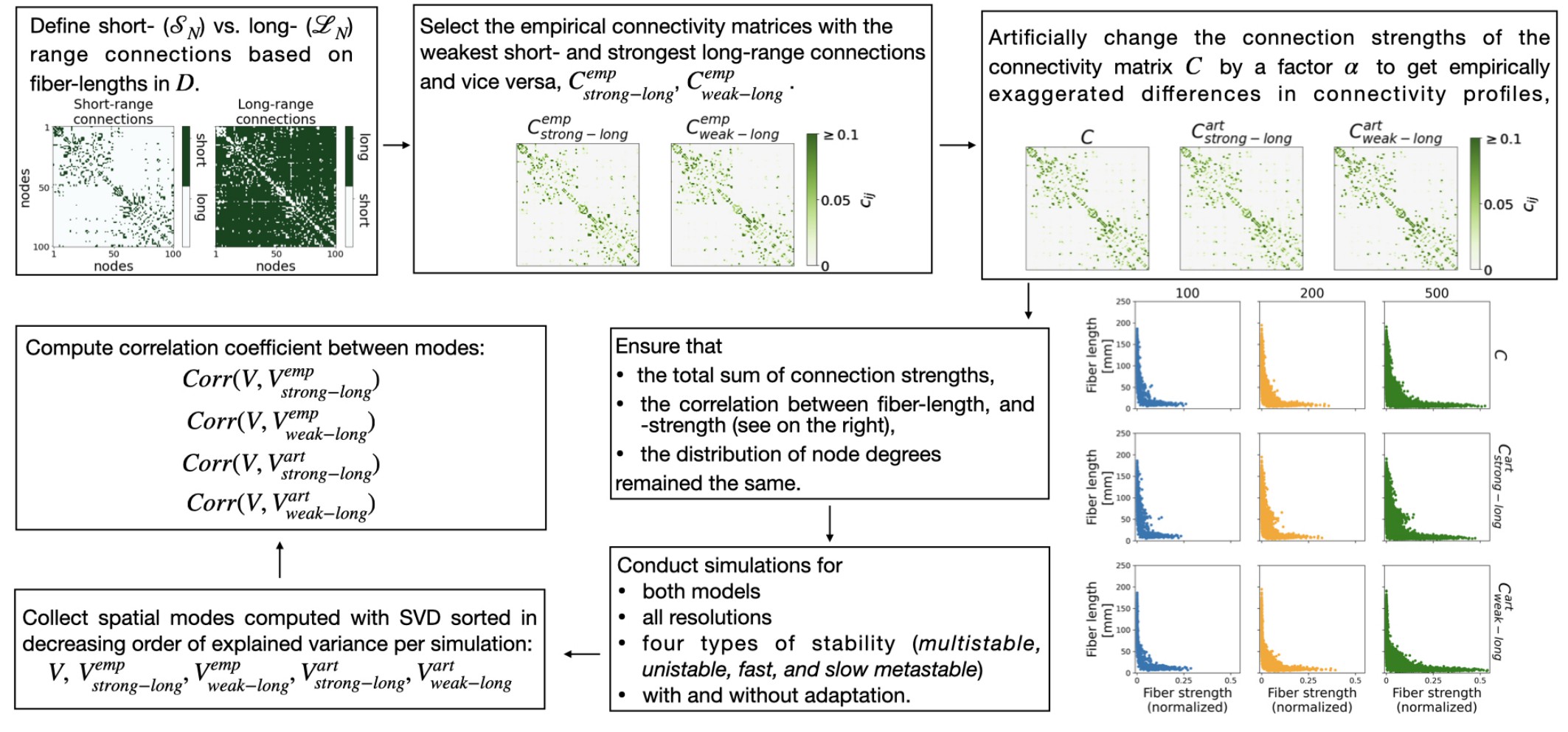}
    \caption{Summary of the procedure used to manipulate and investigate the effect of weaker versus stronger long-range connection strengths  on the network dynamics. For an explanation, see text.}
    \label{fig:long-vs-short-pipeline}
\end{figure}

We collected all pairs $(n,\tilde{n})$, $n,\tilde{n}=1,...,N,\ N\in\{100,200,500\}$, of indices of nodes connected by short-range connections in set $\mathcal{S}_N$, and of nodes connected by long-range connections in set $\mathcal{L}_N$ (see Figure \ref{fig:long-vs-short-pipeline},panel on the top left). A connection was marked as short range, if the corresponding element $D_{n\tilde{n}}$ of the delay matrix $D$ was smaller than $50\ \rm{mm.}$

We identified the subjects with the weakest short- and the strongest long-range connections \begin{align*}
    min_{subject}\bigl(\sum_{(n,\tilde{n})\in\mathcal{S}_N} C^{subject}_{n\tilde{n}}\bigr), \\
    max_{subject}\bigl(\sum_{(n,\tilde{n})\in\mathcal{L}_N} C^{subject}_{n\tilde{n}}\bigr),
\end{align*} and retained the corresponding connectivity matrices $C^{emp}_{weak-long}$, $C^{emp}_{strong-long}$ (see Figure \ref{fig:long-vs-short-pipeline}, panel in the middle of the top row). 

To artificially manipulate two matrices beyond the empirically observed variability (see Figure \ref{fig:long-vs-short-pipeline}, panel in the top right), we used the factors $\alpha=0.1$ and $\gamma=\alpha\frac{|\mathcal{S}_N|}{|\mathcal{L}_N|}$, where $|\cdot|$ denotes the cardinality, to manipulate the connectivity strengths into a biophysically exaggerated disproportion by \begin{align*}
    C^{art}_{strong-long} = C -\alpha C_{short} + \gamma C_{long} \\
    C^{art}_{weak-long} = C + \alpha C_{short} - \gamma C_{long}.
\end{align*}$C_{short}$ ($C_{long}$) denotes the connectivity matrix between nodes connected by short-range (long-range) connections and with the strength for nodes connected by long-range (short-range) connections set to zero. For the non-zero entries, we used the corresponding elements of the averaged connectivity matrix $C_{n\tilde{n}}$, i.e.,\begin{align*}
    C_{short}=\begin{cases}
    C_{n\tilde{n}}, & \text{for } (n, \tilde{n}) \in \mathcal{S}_N \\
    0, & \text{otherwise,}
\end{cases}\\\
    C_{long}=\begin{cases}
    C_{n\tilde{n}}, & \text{for } (n, \tilde{n}) \in \mathcal{L}_N \\
    0, & \text{otherwise.}
\end{cases}
\end{align*}
Thus we ensured that the total sum of connections strengths remained constant, i.e. $\sum_{n,\tilde{n}=1}^NC_{n\tilde{n}} = \sum_{n,\tilde{n}=1}^NC^{art}_{n\tilde{n}}$. Figure \ref{fig:long-vs-short-pipeline} (plots on the bottom right) shows the correlations between fiber-length and -strength for the empirical and for the manipulated connectivity matrices. There was no qualitative change. We also ensured that there was no qualitative change in the distribution of node degrees (not shown).\\
Furthermore, we individually inspected the total sum over the short- and long-range connections of $C^{art}_{strong-long}$ to confirm that long-range connectioned were strengthened, that short-range connections were weakened, and that ther difference between the two sums was enhanced (i.e. $\bigl|\sum_{(n,\tilde{n})\in\mathcal{S}_N}^N C^{art}_{n\tilde{n},strong-long}-\sum_{(n,\tilde{n})\in\mathcal{L}_N}^N C^{art}_{n\tilde{n},strong-long}\bigr| > \bigl|\sum_{(n,\tilde{n})\in\mathcal{S}_N}^N C^{emp}_{n\tilde{n},strong-long}-\sum_{n,\tilde{n}\in\mathcal{L}_N}^N C^{emp}_{n\tilde{n},strong-long}\bigr|$, where $|\cdot|$ denotes the absolute value). A similar construction was conducted for $C^{art}_{weak-long}$.

\subsubsection{Correlation coefficient between spatial modes}
We conducted numerical simulations for four locations in the state space covering unistability, multistability, fast and slow metastable patterns (see Figure A13). Simulations wer performed for both models with and without adaptation, for the averaged connectivity matrix $C$, for the four connectivity matrices $C^{emp}_{strong-long}$, $C^{emp}_{weak-long}$, $C^{art}_{strong-long}$, and $C^{art}_{weak-long}$, and for all resolutions. Then, we computed the velocity vector fields for each resultant activity, concatenated them per setting, and applied SVD as described in Section \ref{subsec:SVD}. This resulted in five matrices $V, V^{emp}_{strong-long}, V^{emp}_{weak-long}, V^{art}_{strong-long}$, and $V^{art}_{weak-long}$ of the spatial modes per setting. To identify the similarity between spatial modes, we computed the Person correlation coefficient between each row of $V^T$ and each row of the matrices $V^{emp}_{strong-long}, V^{emp}_{weak-long}, V^{art}_{strong-long}$, and $V^{art}_{weak-long}$:\begin{align*}
    Corr(V,V^{type}_{strength})\ \text{ for } type\in\{emp,art\},\ strength\in\{strong-long, weak-long\}.
\end{align*}This was done for each selected state, with and without adaptation, for the aLN and the Wilson-Cowan models, and for all three parcellations. The resulting correlation coefficient matrices have values ranging between $-1$ and $1$. Values close to zero indicate little to no similarity, while values closer to $1$, $-1$ indicate high similarity.

\subsubsection{Coherence values}
As in Section \ref{subsection:gradient_manipulation_methods}, we conducted numerical simulations of SOs for the aLN (parameters, see Table A5) and the Wilson-Cowan model (parameters, see Table A6) using the average connectivity matrix $C$, as well as the modified matrices $C^{type}_{strength}$ for  $type\in\{emp,art\},\ strength\in\{strong-long, weak-long\}$.

For each numerical simulation, we computed, analogously to \cite{liang2021corticalmousewaves}, the magnitude-squared coherence 
\begin{align*}
coh_{n\tilde{n}}(f)=\frac{P_{n\tilde{n}}(f)^2}{P_{n}(f)P_{\tilde{n}}(f)},\ \text{ with }\  coh_{n\tilde{n}}(f)\in[0,1],
 \end{align*}where $P_{n}(f)$ and $P_{\tilde{n}}(f)$ are the power spectra over temporal frequencies of the firing rates of the excitatory population for the nodes $n$ and $\tilde{n}$ and $P_{n\tilde{n}}(f)$ is the corresponding cross-power spectrum. A value close to one indicates high correspondence between nodes (i.e., the nodes are highly correlated) for frequency $f$ and vice versa for values close to zero.

We separately consider the coherence between nodes connected with a short range (i.e. all pairs $(n,\tilde{n})$ of nodes from $\mathcal{S}_N$) versus nodes connected with a long range (i.e. all pairs $(n,\tilde{n})$ of nodes from $\mathcal{L}_N$) connection.

\section{Results} \label{section:results}
\subsection{State space}
\begin{figure}[H]
    \centering
    \includegraphics[width=\textwidth]{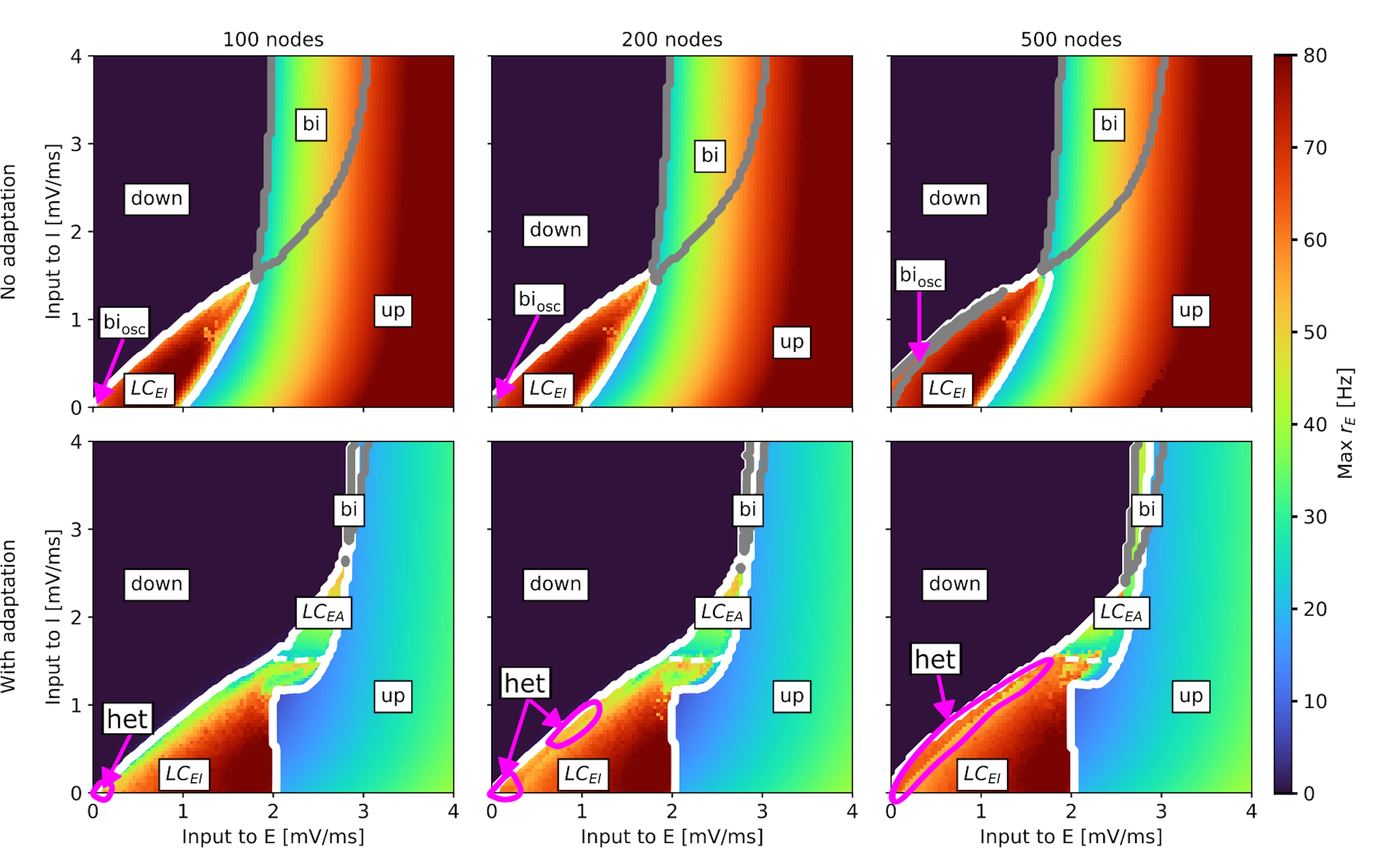}
    \caption{Slice of state space of the whole-brain aLN model without (\textit{b} = 0 pA; top row) and with (\textit{b} = 20 pA; bottom row) adaptation for a brain network with 100 (left column), 200 (middle column), and 500 (right column) nodes spanned by the external input currents to the E and I populations. In every panel, the horizontal axis shows the external input current to the excitatory population ($\mu_E^{ext}$) and the vertical axis shows the external input current to the inhibitory population ($\mu_I^{ext}$). The heatmap shows the maximum excitatory firing rate $r_E$ (Hz) across all nodes in the network. State transition boundaries are indicated by solid white lines for the fast ($LC_{EI}$) and slow ($LC_{EA}$) oscillatory regions and by solid grey lines for the bistable regimes (\textit{bi} - bistability between up and down states; $bi_{osc}$ - bistability between $LC_{EI}$ and the down state). The white dashed lines indicate the border between the two oscillatory regions. Up state (up) and down state (down) regions are also marked. $het$ indicates the areas where we identified heterogeneous slow-fast oscillations (for $b$ = 20 pA). Model parameters are given in Table \ref{tab:model_parameters_aln}.}
    \label{fig:aln_state_space}
\end{figure}

\begin{figure}[H]
    \centering
    \includegraphics[width=\textwidth]{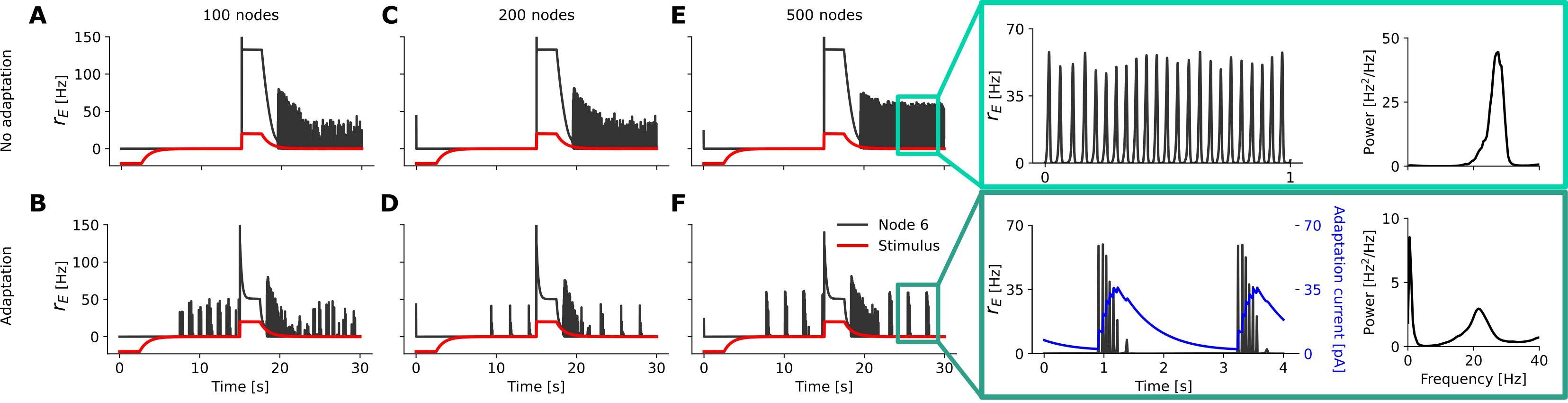}
    \caption{Example time series of the firing rate $r_E$ of one randomly chosen node (black line) of the whole-brain aLN network at several points in the state space: (A), (C), and (E) illustrate bistability between the down state and the fast oscillatory region $LC_{EI}$ using a decaying stimulus (red) delivered to all nodes in the network ($\mu_E^{ext}$ = $\mu_I^{ext}$ = 0.0 mV/ms, b = 0 pA for all three parcellations); (B), (D), and (F) illustrate coexisting slow and fast oscillations for the case of adaptation (b = 20 pA for all three parcellations, $\mu_E^{ext}$ = 0.08 mV/ms for the 100 node resolution, $\mu_E^{ext}$ = 0.04 mV/ms for 200 and 500 nodes, $\mu_I^{ext}$ = 0.0 mV/ms for all three parcellations). All other model parameters are given in Table \ref{tab:model_parameters_aln}. The light (top) and dark green (bottom) insets display enlarged intervals of the time series of the firing rate $r_E$ (black) and, in case of finite adaptation, the current $I_A$ (blue) for the chosen node, and also show the power spectrum for the brain network with 500 nodes averaged across all nodes.}
    \label{fig:aln_example_traces}
\end{figure}

\begin{figure}[H]
    \centering
    \includegraphics[width=\textwidth]{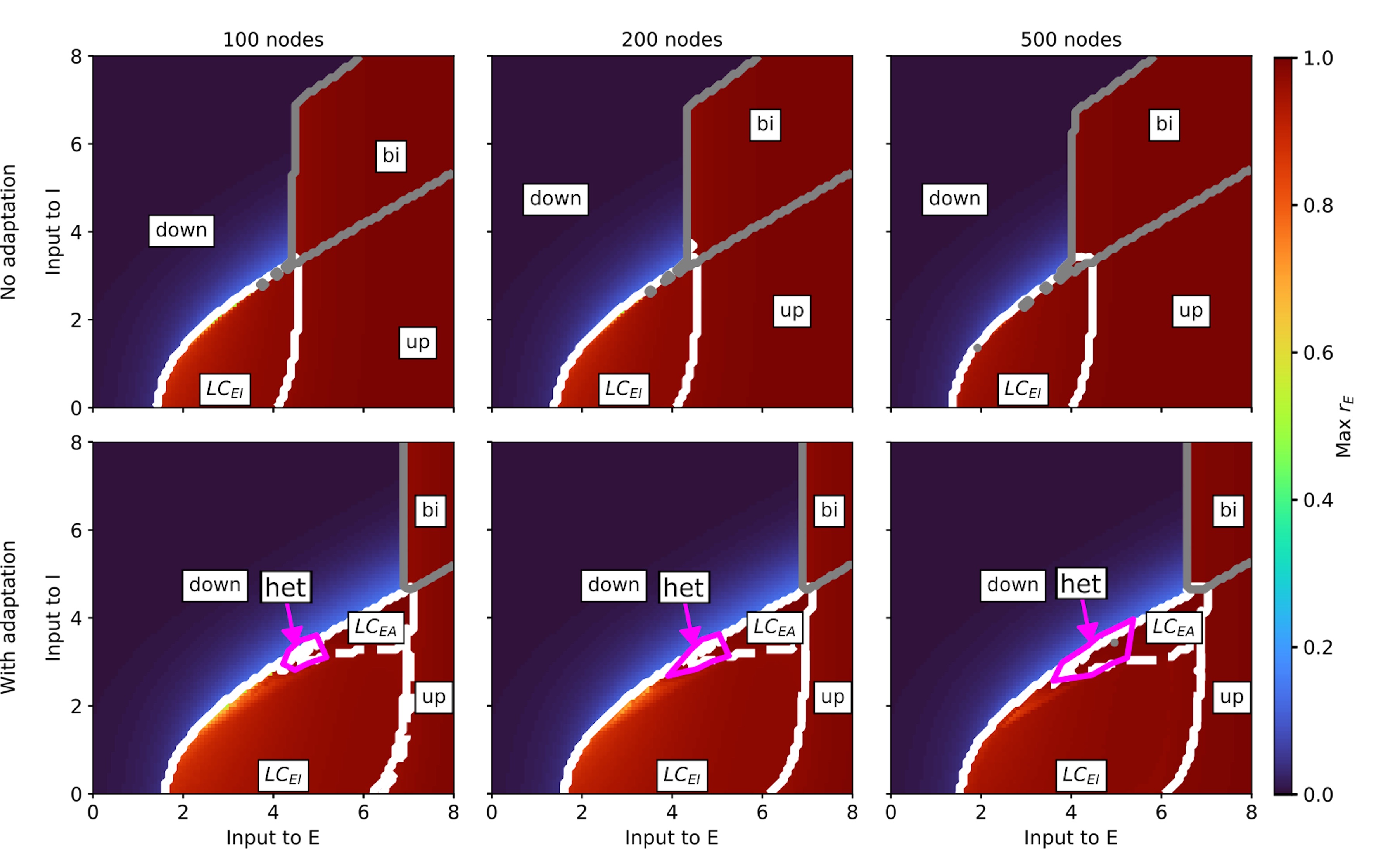}
    \caption{Slice of state space of the whole-brain Wilson-Cowan model without (\textit{b} = 0; top row) and with (\textit{b} = 60; bottom row) spike-triggered adaptation for a brain network with 100 (left column), 200 (middle column), and 500 (right column) nodes spanned by the external input currents to the E and I populations. In every panel, the horizontal axis shows the external input current to the excitatory population ($\mu_E^{ext}$), and the vertical axis shows the external input current to the inhibitory population ($\mu_I^{ext}$). The heatmap shows the maximum value of $r_E$ across all nodes in the network. State boundaries are indicated by solid white lines for the fast ($LC_{EI}$) and by dotted white lines for the regimes of slow ($LC_{EA}$) oscillations. Solid grey lines denote the boundary of the regime of bistability between up and down states (\textit{bi}). $het$ indicate the areas where we identified heterogeneous slow-fast oscillations. Up state (up) and down state (down) regions are also marked. All model parameters are given in Table \ref{tab:model_parameters_wc}.}
    \label{fig:wc_state_space}
\end{figure}

Figures \ref{fig:aln_state_space} and \ref{fig:wc_state_space} show the results of the state space analysis for the whole-brain aLN and Wilson-Cowan models. In line with previous results for the aLN model (\cite{cakan2022spatiotemporal}), we identify several dynamical regimes: a down-state, where all network nodes display no or low activity; an up-state, characterized by constant high firing rate; an oscillatory region $LC_{EI}$, where the activity oscillates between a minimum and a maximum value with frequencies $>$10 Hz (see dominant frequencies in Figure A3 for the aLN and Figure A4 for the Wilson-Cowan model); a bistable regime between up- and down-states; and a slow oscillatory region $LC_{EA}$ with frequencies $<$2 Hz (see bottom panels in Figure A3 for the aLN and Figure A4 for the Wilson-Cowan model) in the case with adaptation. Similar to \citet{cakan2022spatiotemporal}, we observe a very small bistable region for the aLN model where an up-state and the fast $LC_{EI}$ coexist. We also find a small bistable region where an up-state and the slow $LC_{EA}$ coexist. For both whole-brain models, these states are "inherited" from the single-node models (shown in Figure A1), although only very few points displaying bistability between oscillatory and up states can be identified here (purple arrow in Figure A1).

Our results show that, for both models, the state space remains generally robust to changes in network resolution, but there are some differences between the aLN and the Wilson-Cowan implementations. For the aLN model, we observe a region of bistability between the down-state and the $LC_{EI}$ in the case without adaptation, respectively a heterogeneous oscillation (different oscillation frequencies either within the same node or across nodes) in the case with adaptation for the network model with 100 nodes (see Figure \ref{fig:aln_example_traces} for an example time series of a nested slow-fast oscillation). Examining the top row in Figure \ref{fig:aln_state_space} reveals that the region of bistability between the down state and the $LC_{EI}$ in the case of no adaptation expands as the number of nodes in the network increases. Inspecting the average dominant frequency, as well as the standard deviation of the dominant frequency of each node (bottom panels in Figures A3 and A5) confirms that, for the case with adaptation, this region corresponds to an expanding regime of heterogeneous slow-fast oscillations across nodes. 

For the Wilson-Cowan model, we also find a region of heterogeneous oscillations in the case with adaptation (example time series in Figure A2), which expands with increasing network resolution (bottom row in Figures A4 and A6). However, in contrast to the aLN model, this region emerges at the border between the $LC_{EI}$ and $LC_{EA}$ and no regime of bistability between the down state and the $LC_{EI}$ appears.

\subsection{State classification} \label{section:results_state_classification}
\begin{figure}[H]
    \centering
    \includegraphics[width=0.95\textwidth]{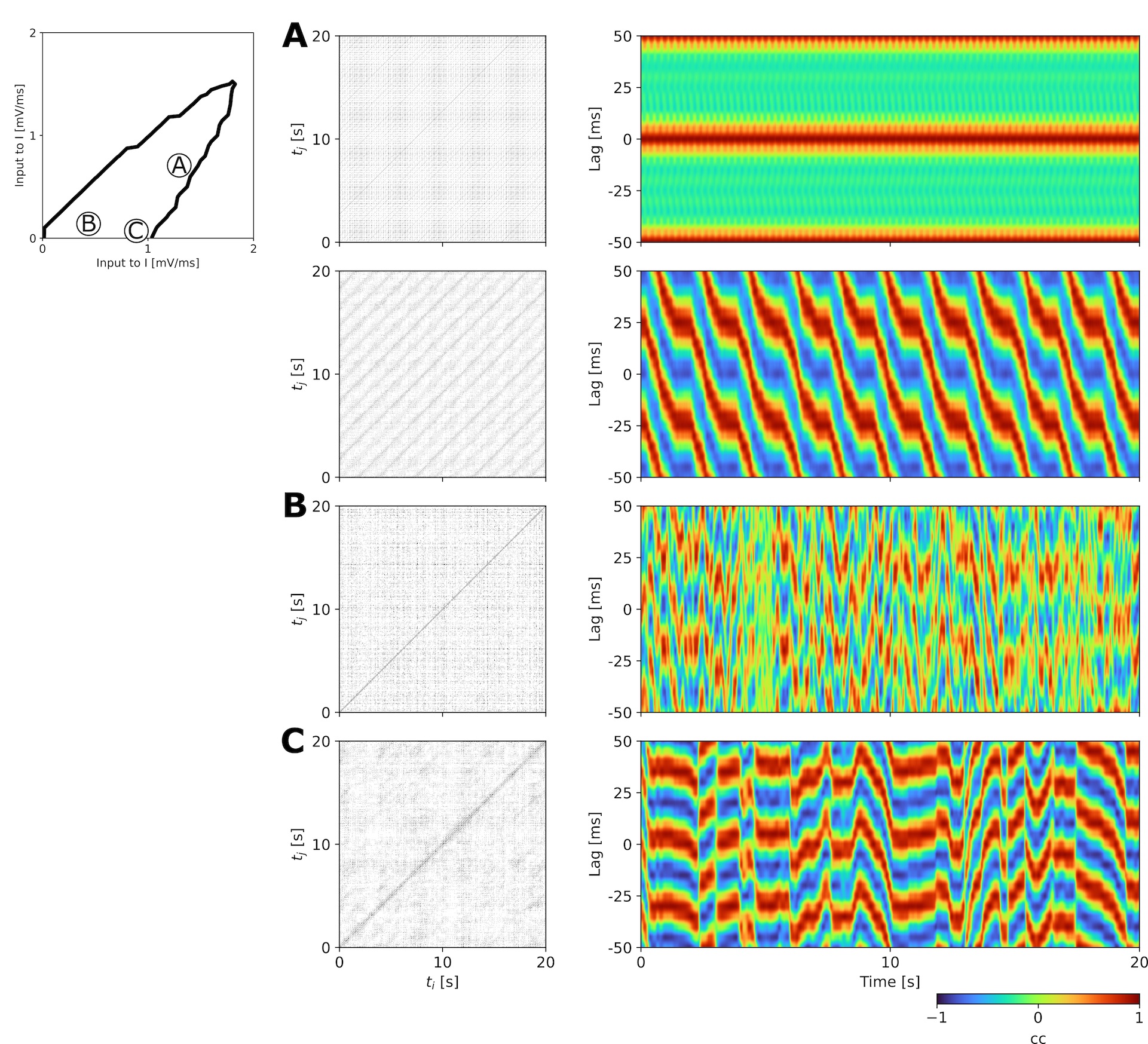}
    \caption{Examples of multistable (A), fast metastable (B), and slow metastable (C) states of the aLN model with 100 nodes and without adaptation ($b = 0$ pA). In each subplot, the left panel shows the recurrence plots, and the right panel the corresponding cross-correlograms. The interhemispheric cross-correlations (cc, see Section \ref{subsection:methods_interhem_cc}) range from -1 (blue) to 1 (red). For the multistable example (A), results are shown for two different random initializations of the network (top and bottom rows). Parameters (positions in state space are shown in the inset on the top left): (A) - ($\mu_E^{ext}$ = 1.3 mV/ms, $\mu_I^{ext}$ = 0.8 mV/ms), (B) - ($\mu_E^{ext}$ = 0.4 mV/ms, $\mu_I^{ext}$ = 0.1 mV/ms), (C) - ($\mu_E^{ext}$ = 0.9 mV/ms, $\mu_I^{ext}$ = 0.0 mV/ms). The simulation time was 20 s. All other parameters are given in Table \ref{tab:model_parameters_aln}.}
    \label{fig:aln_slow_fast_example}
\end{figure}

The analysis methods presented in Section \ref{subsection:methods_state_classification} allowed us to identify four types of states (unistable, multistable, fast metastable, and slow metastable) across the $LC_{EI}$ and $LC_{EA}$ regions. Figure \ref{fig:aln_slow_fast_example} shows examples of recurrence plots and interhemispheric cross-correlograms for a multistable, a fast metastable, and a slow metastable state. The recurrence plots allow us to identify the temporal structure of these states, with the multistable state displaying a clear repetitive pattern over the 20 s of activity shown here (Figure \ref{fig:aln_slow_fast_example}a), the fast metastable state displaying rapid state switches, as evidenced by the noisy recurrence plot in Figure \ref{fig:aln_slow_fast_example}b, and the slow metastable state showing states which persist for a longer duration, as demonstrated by the appearance of more defined clusters (Figure \ref{fig:aln_slow_fast_example}c). The cross-correlograms additionally allow us to highlight the spatiotemporal properties of these states. As mentioned in \citet{roberts2019metastable}, if short incoherent waves dominate, we would expect the interhemispheric coherence to be close to zero across all explored time lags and time points, whereas waves with longer wavelengths would display specific signatures composed of alternating high and low correlation values as a function of the time lag that would persist for a longer time. In the example highlighted here, the multistable state shows repeating spatiotemporal patterns for both initializations. In the fast and slow metastable cases (r.h.s. in Figure \ref{fig:aln_slow_fast_example}b and c), we observe signatures of wave patterns which remain stable for a few hundred miliseconds (in the fast metastable case) up to a few seconds (in the slow metastable case), before rapidly desynchronizing for brief periods of time and transitioning into other wave patterns. To further highlight the difference in state durations between the fast and the slow metastable case, we computed the distribution of state durations identified for the fast and slow metastable points shown in Figure \ref{fig:aln_slow_fast_example} (Figure A9), where we observe longer state durations (up to a few seconds) in the slow compared to the fast metastable case.

\begin{figure}[H]
    \centering
    \includegraphics[width=\textwidth]{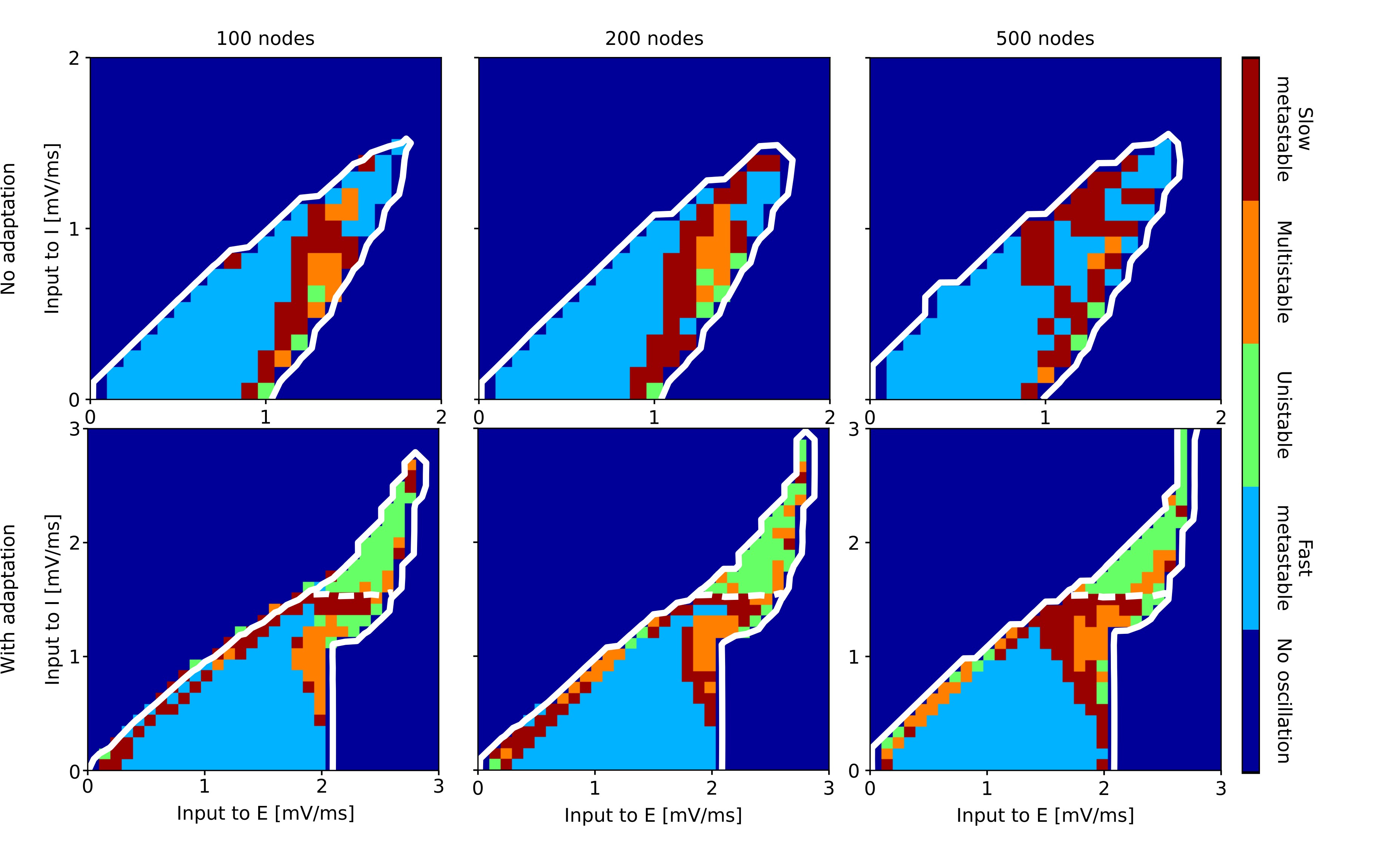}
    \caption{Classification of states inside the oscillatory regions for the aLN whole-brain model in the case without (b = 0 pA; top row) and with (b = 20 pA; bottom row) adaptation for a parcellation with 100 (left column), 200 (middle column), and 500 (right column) nodes. The slice of state space is spanned by the external input current to the E and I populations. The white solid contour marks the two oscillatory regions, and the white dashed lines indicate the approximate border between them.}
    \label{fig:aln_state_classification}
\end{figure}

\begin{figure}[H]
    \centering
    \includegraphics[width=\textwidth]{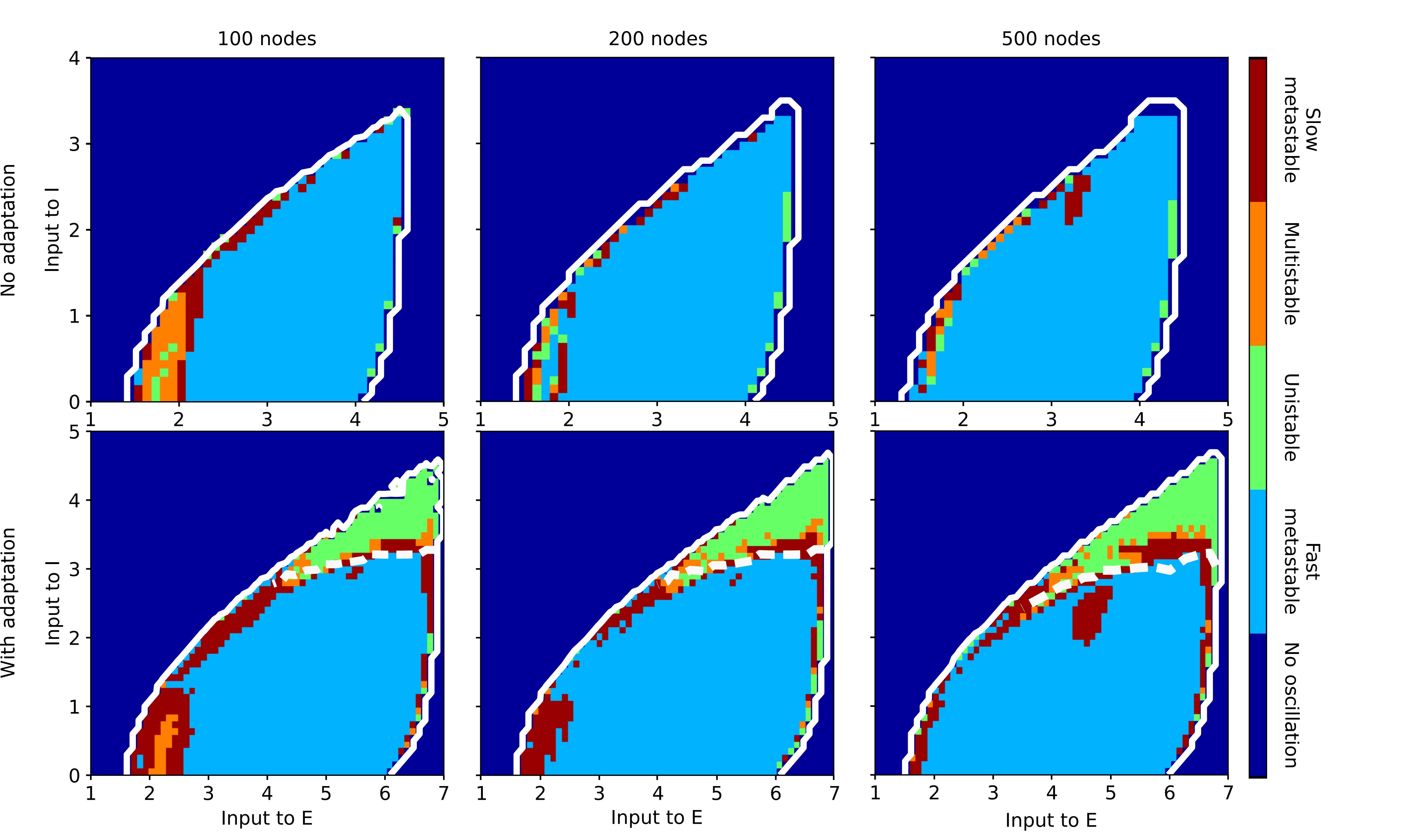}
    \caption{Classification of states inside the oscillatory regions for the Wilson-Cowan whole-brain model in the case without (b = 0; top row) and with (b = 60; bottom row) adaptation for a parcellation with 100 (left column), 200 (middle column), and 500 (right column) nodes. The slice of state space is spanned by the external input current to the E and I populations. The white solid contour marks the two oscillatory regions, and the white dashed lines indicate the approximate border between them.}
    \label{fig:wc_state_classification}
\end{figure}

Results of the state classification for the entire slice of state space are summarized in Figures \ref{fig:aln_state_classification} (aLN model) and \ref{fig:wc_state_classification} (Wilson-Cowan model). Qualitatively, the results are similar across models and resolutions, with all four regimes being present in all cases, and with the fast metastable regime occupying the largest portion of the $LC_{EI}$, while being absent from the $LC_{EA}$ region (which is dominated by unistable patterns). However, some quantitative differences are apparent. For the aLN model without adaptation, both the multistable and slow metastable regimes emerge on the right side of the $LC_{EI}$ region close to the up state. For the Wilson-Cowan model, however, they appear on the left side of this region close to the down state.

As metastability is usually identified based on the mean and the standard deviation (SD) of the Kuramoto order parameter (metastability corresponds to a high standard deviation of the Kuramoto order parameter), we report these results for completeness in Figures A7 and A8. The results show high synchrony (mean Kuramoto $\sim$1) and low metastability (SD of Kuramoto $\leq$0.1) in the areas identified above as uni/multistable, lower synchrony (mean $\sim$0.4 - 0.7) and higher metastability (SD $\sim$0.1 - 0.2) for the corresponding slow metastable points, and lowest values for the corresponding fast metastable points (mean and SD $<$0.1). Given that the Kuramoto order parameter is only sensitive to global states and misses local synchrony and that the fast metastable dynamics are also more local, these results are not surprising.

\subsection{Spatial modes of activity}
Our analysis of the spatial modes of activity reveals that, in general, the modes which explain a larger proportion of variance of the activity (percentages given in Tables A1 and A2) in the concatenated data (obtained by concatenating the velocity vector fields computed for each point in the oscillatory regions, with time steps in rows and nodes in columns) consist of large-scale waves traveling mainly along the horizontal and dorso-ventral axes. The results are summarized in Figures \ref{fig:aln_spatial_modes}a,b for the aLN model and in Figures A12a,b in the appendix for the Wilson-Cowan model. For example, modes 1 and 4 in the aLN model (Figure \ref{fig:aln_spatial_modes}a) and modes 2 and 4 in the Wilson-Cowan model (Figure A12a) exemplify large-scale waves with coherent horizontal and dorso-ventral directions of propagation encompassing approximately three quarters of the brain. Another example of a large-scale wave pattern is represented by the hemispheric-segregated pattern present in the Wilson-Cowan model (mode 3 in Figure A12a) and in the aLN model (mode 9 in Figure \ref{fig:aln_spatial_modes}). Interestingly, these modes explain similar proportions of variance (1.78\% in the aLN vs. 1.44\% in the Wilson-Cowan model). In contrast, modes explaining less variance within each model and each resolution usually capture more complex patterns of propagation. For example, in both models, mode 13 (Figures \ref{fig:aln_spatial_modes}a and A12b) displays smaller clusters of arrows with the same color and direction (i.e. same horizontal and dorso-ventral directions), as well as more neighboring arrows with different colors and directions compared to the large-scale modes indicated above. While we identify similar modes in both models (see above), the overall proportion of variance explained by the 15 first modes differs (30.28\% for the aLN vs. 9.19\% for the Wilson-Cowan model with adaptation). There is also a tendency towards decreased explained variance per mode with increasing model resolution, as well as differences in the percentages of variance explained by the dominant modes between the models with and without adaptation (Tables A1 and A2).

To verify whether the modes obtained from the decomposition of the concatenated data can be reliably identified in the individual velocity vector fields computed for each parametrization in the oscillatory regions $LC_{EI}$ and $LC_{EA}$, we projected these modes and investigated the explained proportion of variance for the state types identified in Section \ref{section:results_state_classification} (i.e. fast metastable, slow metastable, uni/multistable). Figures \ref{fig:aln_spatial_modes}c and A12c show that, in general, the most dominant five modes, representing global propagation patterns, explain the largest proportion of variance in individual states regardless of state type. Nevertheless, the largest proportion of variance is explained in the stable states ($>$25\% explained by the first five modes), followed by the slow ($>10\%$), and the fast metastable states ($<10\%$). We also observe that the first dominant mode identified in the concatenated data does not necessarily capture the largest proportion of variance in individual states (Figure \ref{fig:aln_spatial_modes}c in contrast with Figure A12c), suggesting that while this pattern of activity is consistently present across states, it may not be dominant in all of them.

As a further example, we examined the spatial modes of activity in the $LC_{EA}$ region, obtained from the data concatenated over all points identified as unistable and with an average dominant frequency $\leq$2 Hz, for the aLN and Wilson-Cowan models with 100 nodes and adaptation. Figures A10 and A11 confirm the presence of large-scale activity patterns traveling along the horizontal and dorso-ventral directions similar to the ones described above. For example, mode 1 in the aLN model and mode 7 in the Wilson-Cowan model are similar to modes 9, respectively 3, described above, whereas modes 2, 3, and 4 in both models are similar to modes 1 and 4, respectively 2 and 4, described above. Furthermore, we also observe that most spatial modes contain a small component propagating along the antero-posterior direction (for example, the arrows pointing anteriorly/posteriorly in the first two modes of both models, which is in agreement with previous reports regarding the antero-posterior direction of SO propagation (\cite{cakan2022spatiotemporal, massimini2004sleep}). In both cases, the modes obtained from the decomposition of the unistable patterns in the $LC_{EA}$ region of slow oscillations explain a significantly higher proportion of variance compared to those obtained from the decomposition of the data concatenated over all state types in both oscillatory regions: 73.52\% vs. 30.28\% for the aLN and 58.99\% vs. 9.19\% for the Wilson-Cowan model, with the first mode explaining 26.71\% of the variance (aLN) and 24.33\% (Wilson-Cowan) vs. 9.31\% and 3.72\%.

\subsection{Similarity of spatial modes of imbalanced short- versus long-range connection strengths}
To identify the impact of the balance between short- and long-range connection strength, we compared the $10\%$-most dominant spatial modes (i.e. the spatial modes that explain the largest amount of variance in the spatial organization of activity patterns) of the activity induced by empirically informed and artificially manipulated connectivity matrices. We simulated both models for parameters corresponding to all four types of stability per resolution (see Figure A13 for the corresponding locations in state space). We used the average connectivity matrix $C$ whose resulting spatial modes are collected in the columns of $V$, and compared results obtained to the results for the empirical and the artificially enhanced matrices with weaker vs. stronger long-range connections: $C^{emp}_{weak-long}, C^{art}_{weak-long}$, $C^{emp}_{strong-long}, C^{art}_{strong-long}$, whose resulting spatial modes are collected in $V^{emp}_{weak-long}, V^{art}_{weak-long}$, $V^{emp}_{strong-long}, V^{art}_{strong-long}$, respectively. Then we estimated the distribution of the values of the correlation coefficients $Corr(V,V^{type}_{strength})$ for $type\in\{emp, art\},\ strength\in\{weak-long, strong-long\}$ where we normalized each distribution by its' maximum value to ensure the option of visual comparability. Results are shown in Figure \ref{fig:long_short_density_aln} for the aLN, and in Figure A16 for the Wilson-Cowan model. Means and standard deviations of the distributions are given in Table A3 for the aLN and in in Table A4 for the Wilson-Cowan model. Additionally, we show the resulting correlation coefficient matrices for all settings without adaptation in Figures A14 (aLN model) and A15 (Wilson-Cowan model).

All distributions are centered around a value of zero. However, we notice that the distributions in Figure \ref{fig:long_short_density_aln} for the fast metastable states appear visually the broadest (indicating higher similarity between spatial modes). The computed standard deviations (Table A3) agree with this observation except for the cases of unistability with adaptation at resolutions $N\in\{100,200\}$. This is because the activity for those settings converges to a spatially homogeneous unistable state for all matrices, having a diagonal of $Corr(V, V^{type}_{strength})_{nn}\approx 1$, and $Corr(V, V^{type}_{strength})_{n\tilde{n}}\approx 0$, for ${n\neq\tilde{n}}$. Therefore, the density is distributed around values close to zero and values close to one, broadening the width. Note, that the correlation coefficients between $V$ and $V^{emp}_{strong-long}$ in the unistable state with adaptation are all close to zero, causing a peaky distribution for that case (see Figure \ref{fig:long_short_density_aln}, second column, fourth row, dashed dark blue line). This is a result of the model with $C^{emp}_{strong-long}$ converging to not only spatially but also temporally homogeneous, i.e., constant, activity. Furthermore, we observe overall lower absolute values for the correlation coefficients between spatial modes, which indicates a loss of similarity in the spatial organization of the patterns induced by the average connectivity matrix compared to the activity caused by the connectivity matrices with weaker, and stronger long-range connections. The Wilson-Cowan model does not align with the highest similarity between spatial modes in fast metastable states but rather shows the broadest width in unistable states (see Figure A16, and Table A4). It agrees with the same findings that overall the similarity is comparably low for all settings, but lower on average than for the aLN model ($avg(\sigma)_{aLN,100}=0.040750,\ avg(\sigma)_{aLN,200}=0.018125,\ avg(\sigma)_{aLN,500}=0.005$,\ $avg(\sigma)_{wc,100}=0.027375,\ avg(\sigma)_{wc,200}=0.011250,\ avg(\sigma)_{wc,500}=0.004875$).

The above observations lead to three main conclusions. Firstly, we see a higher similarity of spatial organization in states of stability that promote more local, complex activity patterns rather than the global, synchronized patterns that appear in unistable or multistable states. Exceptions occur if a state of spatially homogeneous activity is reached. Secondly, while the states showing the broadest widths differ between both models (multistable states for the Wilson-Cowan model vs. unistable or fast metastable states for the aLN model), the overall low similarity in the spatial organization between activity patterns caused by the average connectivity matrix vs. by the connectivity matrices with weaker and stronger long-range connections generalizes across all resolutions, both model types and all settings. Finally, we see that the results of the comparison between the spatial organization of activity patterns induced by the different connectivity matrices are predominantly the same for the artificial versus empirical connectivity matrices for both models and all resolutions. 

\subsection{Effect of the antero-posterior gradient of structural connectivity strengths on sleep SO propagation}\label{subsec:application}
The results presented above show that for both the aLN and Wilson-Cowan models dynamical features remain generally robust to changes in the parcellation. Also, the phenomenological Wilson-Cowan model is capable of producing qualitatively similarly complex spatiotemporal dynamics as the biophysically realistic aLN model. In the current section, we explore whether this remains to be the case when both models are applied to the phenomenon of sleep SO propagation (\cite{cakan2022spatiotemporal}). In particular, we examine whether the relation between the antero-posterior structural connectivity gradient and the propagation of sleep SOs as waves of silence from anterior to posterior brain areas remains present in both models and for all parcellations. Furthermore, we test whether changes in the strength of this connectivity gradient have a causal effect on the direction of propagation of SOs.

Figure \ref{fig:sc_gradient_change_1} shows that the relation reported in \citet{cakan2022spatiotemporal} is present in both the aLN and Wilson-Cowan models for all three network resolutions. Furthermore, decreasing the gradient strength along the antero-posterior axis causes a reversal of the direction of SO propagation, with down states being initiated preferentially in posterior areas and traveling towards the front of the brain. Increasing the gradient strength increases this preference to propagate from anterior to posterior areas. In the Wilson-Cowan model, however, the relation between node degree and the transition phase decreases with the increase in resolution, as the magnitude of the correlation coefficients decreases at higher resolutions. This could potentially be caused by the fact that in the Wilson-Cowan model the adaptation strength $b$ and adaptation time constant $\tau_A$ had to be drastically increased at higher resolutions in order to observe SOs.

To ensure that the results presented in Figure \ref{fig:sc_gradient_change_1} are not due to changes in the underlying network topology induced by the specific gradient manipulation method, we employ a control model in which we preserve the total sum of connection strengths in the network and destroy the relation between fiber length and connection strength (cf. Section \ref{subsection:gradient_manipulation_methods}). Figure A18 shows that the relationship described above remains present in the aLN model at all three network resolutions. In the Wilson-Cowan model, destroying the relation between the distance and connectivity strength destroys and even reverts the propagation direction of SOs, suggesting that the model is more sensitive to changes in the particular structure of the connectome.

\subsection{Stronger long-range connections lead to an increase in coherence as observed  empirically}
Motivated by the findings that show that rare long-range connections play an effective role in the cascade of information processing (see \cite{deco2021rare}) and that stronger long-range connections correlate with enhanced coherence between cortical regions over lower frequency ranges (\cite{liang2021corticalmousewaves}), we investigated how changes in the strength of long- versus short-range connections influence waves of SOs.

Since long-range connections are assumed to play a crucial role in the propagation of global patterns, we assume that the stronger the long-range connections, the higher the coherence over lower frequency values induced by slow oscillations. We therefore compared results obtained using the matrices $C_{strong-long}$, $C_{weak-long}$, and $C$. 

Figures \ref{fig:coherence_aln} and A19 show the average power spectra and coherence values for the aLN and the Wilson-Cowan models for three different parcellations. In Figure \ref{fig:coherence_aln}a we see that for all parcellations the dominant temporal frequencies are $<1\rm{Hz}$. Small differences between the power spectra for the different parcellations caused by the three different connectivity matrices $C_{strong-long}$, $C_{weak-long}$, and $C$ are more pronounced for the empirical matrices, which is confirmed by values for the dominant temporal frequency, given in Table A7. Furthermore, the power for lower frequencies decreases with increasing resolution, in particular for the stronger long-range connections (blue line), see Table A7. The decrease in power is less pronounced in the artificial compared to the empirical case. We argue that this is caused by the matrix $C$ being an average, hence the connection strengths are more evenly distributed rather than promoting sparse connectivity profiles, unlike for the empirical matrices $C^{emp}$. According to our method of manipulation, the connection strengths in the artificially manipulated matrices are also more evenly distributed than for the empirical matrices.

In the artificial case, we see that the change of coherence over frequency for the aLN (see Figure \ref{fig:coherence_aln}b) and for the Wilson-Cowan model (see Figure A19b) agrees with our expectation. The coherence over low frequencies is higher for SOs induced by $C_{strong-long}$ (blue lines) than $C_{weak-long}$ (green lines), both between nodes connected with short-range (solid lines) and long-range (dotted lines) connections. This is also observable in the corresponding coherence values given in Table \ref{tab:aln_coherence_table}, where we can see that, in the artificial case, the coherence values are higher at $f=0.5\ \rm{Hz}$ for the SOs induced by $C_{strong-long}$ compared to $C_{weak-long}$. With an increase in resolution, we see an alignment of coherence values over the entire frequency range between nodes connected with short- and long-range connections due to an overall decrease of coherence values between nodes connected  but short-range connections (see Figures \ref{fig:coherence_aln}b, and A19b).

The results of the Wilson-Cowan model agree mostly with the results of the aLN, however, the dominant temporal frequency varies more strongly depending on the parcellation and the used connectivity matrix, see Table A8. The coherence values are consistently larger for waves of SOs induced by $C_{strong-long}$ in the artificial case, see Table A9.

We observe the expected effect in neither model for the empirical matrices. We argue that this is due to the fitting process applied to the averaged matrix $C$ whose distribution of connection strengths is more similar to the artificially manipulated connectivity matrices than to the empirical matrices.

Overall, models and resolutions agree with the expected increase in coherence values over low frequencies for the artificially manipulated matrices, but do not display the same effect for the empirically selected matrices.

\section{Discussion} \label{section:discussion}
In this work, we investigated whether we can employ generalized whole-brain models for the study of complex brain dynamics or whether the latter are significantly influenced by the choice details of the dynamical system and the parcellation. To that end, we compared a biophysically realistic model (aLN) and a phenomenological model (Wilson-Cowan) with similar state spaces and bifurcations at three network resolutions (the Schaefer parcellation scheme with 100, 200, 500 nodes). Overall, we found that the results remain relatively robust to changes in both model and parcellation, but dynamics at detail appear sensitive to these changes, indicating the need for careful model adjustment depending on the application.

We started our analysis with the exploration of the coarse-grained structure of the dynamical landscape. We found that both the aLN and the Wilson-Cowan model display a down state of no or low activity, an up state of constant high activity, a fast limit cycle, where the activity oscillates between low and high values with frequencies $>$ 10 Hz, a bistable regime, where the activity remains either in a stable up or a stable down state depending on the initial condition in the case with and without adaptation, and a slow limit cycle, where the activity oscillates at low frequencies ($<2$ Hz) in the case of finite adaptation. The state boundaries remained relatively robust to changes in network resolution and are in agreement with those previously reported in the literature (\cite{cakan2022spatiotemporal}). Nevertheless, we reported the emergence of a region of bistability between the down state and the $LC_{EI}$ in the case without adaptation in the aLN model, respectively of heterogeneous oscillations in the case with adaptation for both models. This is not present for a single node and it enlarged with increasing network resolution. We hypothesize that this is due to the fact that in the parcellations with higher resolutions we observe stronger local connection strengths (\cite{roberts2019metastable}), which in turn favor the emergence of more complex dynamics, such as heterogeneous oscillations.

In a second step, we classified the oscillatory network states. We identified four types of states, namely unistable, multistable, slow, and fast metastable states, in both models and at all resolutions, and observed quantitative changes with respect to the distribution of each type of state in the oscillatory regimes both across models and across resolutions (Figures \ref{fig:aln_state_classification} and \ref{fig:wc_state_classification}). Our detailled analysis of the types of oscillatory network states revealed that complex wave dynamics emerge even at low network resolutions and in relatively simple phenomenological models. Furthermore, the detailed mapping of the oscillatory regimes presented here can provide useful information for further studies aiming to explore induced state transitions, such as, for example, through the application of electrical stimulation (for example, see \citet{ladenbauer2017promoting, ladenbauer2023towards}).

We explored large-scale patterns through the spatial modes obtained from singular value decomposition. We found that results are qualitatively similar across models and resolutions, but that specific patterns emerge depending on either model or resolution. Given that recent work (\cite{das2024planar, mohan2024direction}) investigating the relation between spatiotemporal wave patterns and cognitive function has shown an association between specific patterns and specific behavioral processes, future modeling work in this direction should take into account the variability introduced by model and parcellation when exploring such phenomena.

We showed that changes in the balance of connectivity strengths between short- and long-range connections alter the spatial organization in states exhibiting global patterns (multi- and unistable) as well as complex patterns (fast and slow metastable), a result which stays predominantly consistent across models, resolutions, parametrizations, and states (see Figures \ref{fig:long_short_density_aln} and A16). Artificially manipulating the long- versus short-range connection strengths beyond empirically observed variability had no significantly different effect to the loss of similarity between the spatial organization of activity patterns induced by the artificially manipulated and the empirical connectivity matrices. Furthermore, we noticed that the strongest similarity in the spatial modes collected from the activity patterns caused by the different connectivity matrices was observable in the fast and slow metastable states in which complex local activity patterns emerge (see \cite{kelso2012multistability}).

Furthermore, in the specific case of sleep SOs, we have shown that the aLN model is robust to changes in network resolution and even in parcellation scheme (as we used the original parametrization introduced in \citet{cakan2022spatiotemporal} with only minimal parameter adjustments). In this case, we were also able to demonstrate that changes in the antero-posterior structural connectivity gradient have a causal effect on the propagation of SOs. In contrast, the Wilson-Cowan model required optimization for the Schaefer parcellation scheme with 100 nodes and an additional adjustment of its parameters for higher resolutions. Here, manipulating the antero-posterior gradient of node degrees showed a robust causal effect only in the case where the model parameters were explicitly fitted to data rather than adjusted to support SO activity. The model also displayed high sensitivity to the changes in of the relationship between connection strength and distance. 

For understanding the impact of changes in the strength of short- vs. long-range connections on SOs, we investigated power spectra and coherence values (see Figures \ref{fig:coherence_aln}, and A19). For the case of artificially manipulated connectivity matrices we found the coherence in lower frequency bands ($<2\ \rm{Hz}$) to be higher in value for matrices with stronger long-range connections, than for the averaged $C$ matrix that was used for the fitting process as well as for $C^{art}_{weak-long}$. This agrees with the results of \cite{liang2021corticalmousewaves} who also observed an increase in coherence between cortical regions in mice connected by stronger long-range connections. Our results are consistent across models and resolutions. For the empirical connectivity matrices $C^{emp}$ we found the opposite effect (see Figures \ref{fig:coherence_aln}b, A19b). This could be due to the fitting process being conducted with the averaged $C$ matrix. Since the artificially manipulated connectivity profiles are based on the averaged $C$ matrix, they are more similar in the distribution of the connection strengths, unlike the empirical connectivity matrices that are characterised by rather sparse connectivity profiles.

We thus conclude that the deployment of whole-brain models for the investigation of the coarse-grained dynamics provides results which are fairly independent of model type and resolution. All model variants enable the same dynamical landscape with qualitatively similar changes in dynamical features with resolution and with the manipulation of the connectivity profiles. In the specific application to sleep SOs, both the phenomenological and the biophysically realistic model show similar changes in the temporal dynamics. While the antero-posterior directionality of simulated SOs by the aLN corresponds to the expected changes induced by the manipulation of the underlying antero-posterior structural connectivity gradient, the phenomenological Wilson-Cowan model requires a much more careful handling to demonstrate the empirically observed directionality. In total, this indicates that both model types are fairly robust to the simulation of empirically realistic temporal features, but not so for propagation dynamics. Nonetheless, for the investigation of quantitative features, detailed dynamics or specific application cases, the phenomenological Wilson-Cowan model requires a much more careful handling and finer tuning, while the biophysically realistic aLN model allows the investigation of specific features in a more reliable way.

\section*{Conflict of Interest Statement}
%All financial, commercial or other relationships that might be perceived by the academic community as representing a potential conflict of interest must be disclosed. If no such relationship exists, authors will be asked to confirm the following statement: 

The authors declare that the research was conducted in the absence of any commercial or financial relationships that could be construed as a potential conflict of interest.

\section*{Author Contributions}
CD - Conceptualization, Formal analysis, Methodology, Visualization, Writing - original draft, Writing - review \& editing; RS - Conceptualization, Formal analysis, Methodology, Visualization, Writing - original draft, Writing -review \& editing; AF - Experiments \& Data Collection, Writing - review \& editing; KO - Funding acquisition, Supervision, Writing - review \& editing.

\section*{Funding}
The author(s) declare that financial support was received for the research, authorship, and/or publication of this article. This work was funded by the Deutsche Forschungsgemeinschaft (DFG, German Research Foundation) Project number 327654276, SFB1315, project number B03 (KO and AF) and project grants to AF: Research Unit 5429/1 (467143400), FL 379/34-1, FL 379/35-1.

\section*{Data Availability Statement}
All simulations for the aLN model were conducted using the neurolib framework (\cite{cakan2021neurolib}), available at \href{https://github.com/neurolib-dev/neurolib.git}{https://github.com/neurolib-dev/neurolib.git}. All simulations for the Wilson-Cowan model were conducted using an expanded version of the neurolib framework, available in a different github-repository \href{https://github.com/ronja-roevardotter/WC-model_withAdaptation.git}{https://github.com/ronja-roevardotter/WC-model\_withAdaptation.git}. This repository also contains the connectome matrices $C$ and $D$ for all three parcellations.

\newpage
\section*{Figures}

\let\cleardoublepage\clearpage
\begin{sidewaysfigure}
    \vspace*{1cm} 
    \centering
    \includegraphics[width=\textwidth]{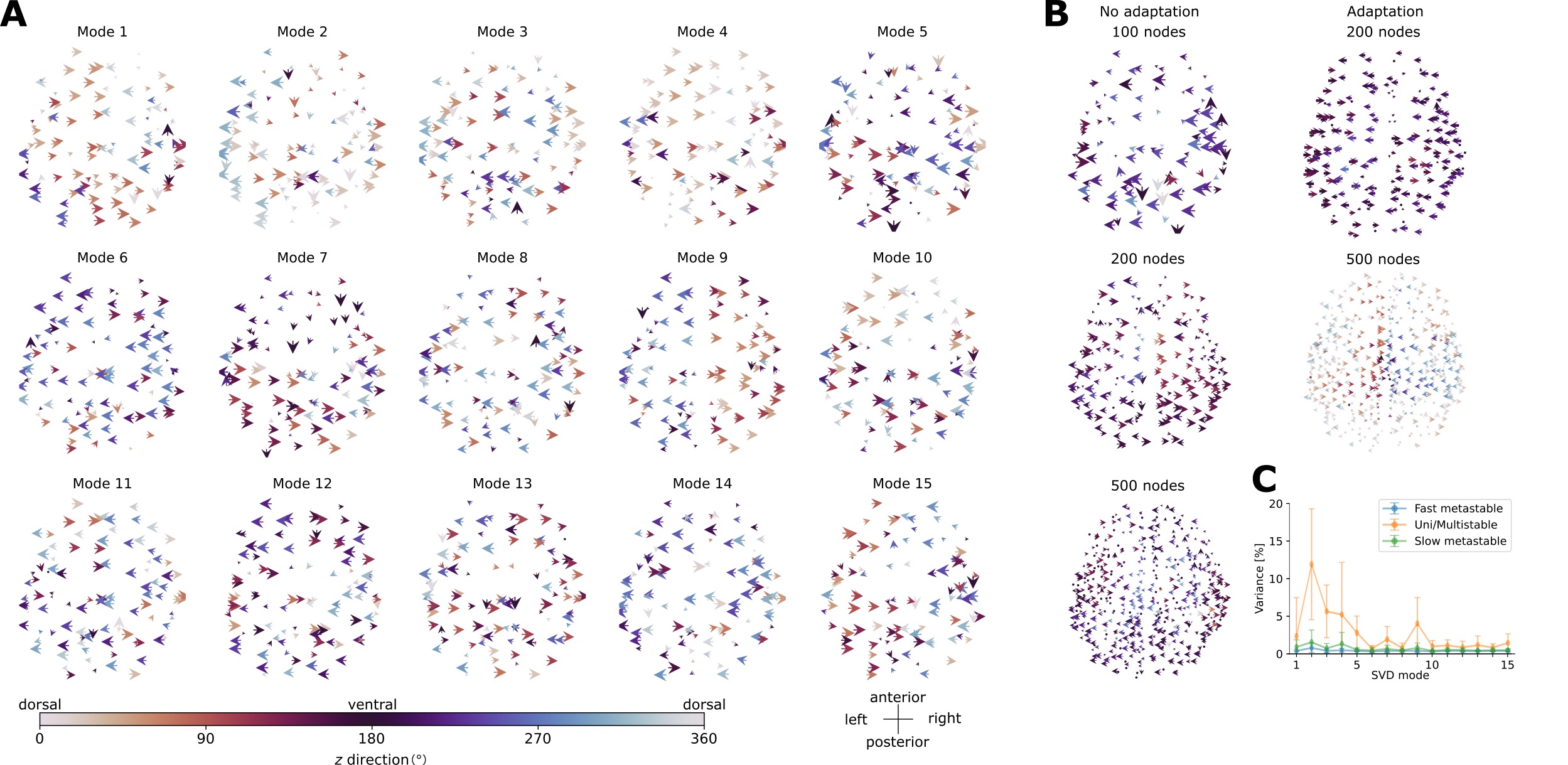}
    \caption{(A) First 15 modes obtained from the singular value decomposition of the velocity vector fields in the whole-brain aLN model with 100 nodes and spike-triggered adaptation (b = 20 pA). Modes are ordered in decreasing order of explained variance. (B) Left panels: Modes explaining the largest proportion of variance for the whole-brain aLN model without spike-triggered adaptation (b = 0 pA) with a parcellation of 100, 200 and 500 nodes. Right panels: Same as before, but with spike-triggered adaptation (b = 20 pA) and with a parcellation of 200 and 500 nodes. The arrows represent the orientation in the $xy$ plane (left-right and antero-posterior directions) and are color-coded according to the direction along the \textit{z}-axis (dorso-ventral direction). (C) Percentage of explained variance (mean $\pm$ standard deviation across points in the parameter space) of the first 15 modes identified in (A) for the aLN model with 100 nodes and spike-triggered adaptation (b = 20 pA). The percentage is shown for the different pattern types identified in Section \ref{section:results_state_classification}: uni/multistable (orange), fast metastable (blue), and slow metastable (green).}
    \label{fig:aln_spatial_modes}
\end{sidewaysfigure}

\begin{figure}[H]
    \centering
    \includegraphics[width=0.94\textwidth]{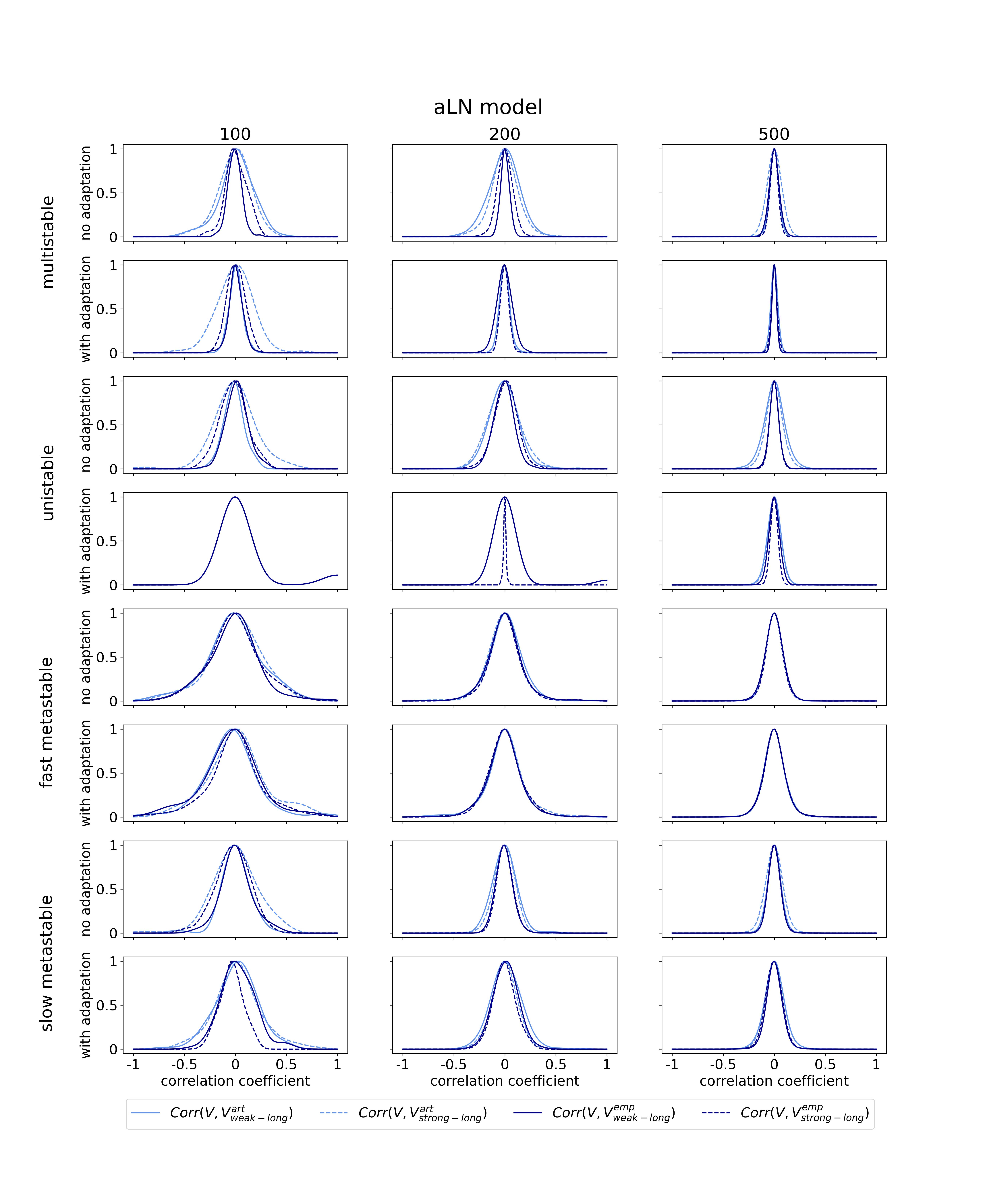}
    \caption{Distribution of the values from the matrices $Corr(V,V^{type}_{strengh})$ of correlation coefficients, each normalized to its maximum value. Correlation coefficients are computed between the spatial modes obtained with the averaged connectivity matrix $C$ and with the spatial modes of the empirically derived (darker colors, $V^{emp}$) and the artificially manipulated (lighter colors, $V^{art}$) matrices, with stronger (dashed, $V_{strong-long}$) and weaker (solid, $V_{weak-long}$) long-range connections. Distributions are estimated using kernel density estimation. Each column corresponds to one parcellation, each pair of rows (upper row without, lower row with adaptation) to the type of stability (multistable, unistable, fast, and slow metastable). Means and standard deviations are provided in Table A3. For parameters, see A13.}
    \label{fig:long_short_density_aln}
\end{figure}

\begin{figure}[H]
    \centering
    \includegraphics[width=\textwidth]{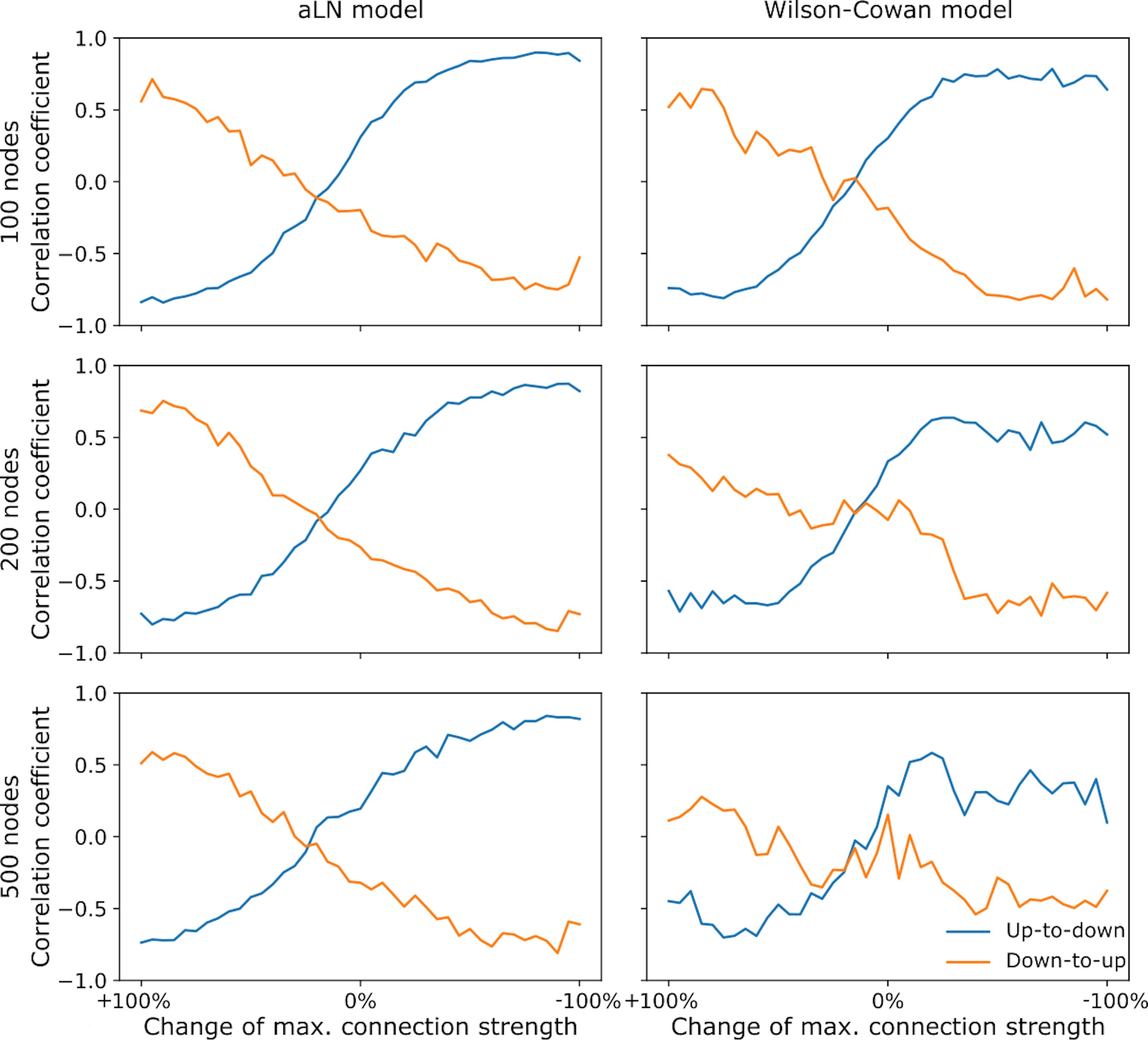}
    \caption{Correlation coefficient between the mean transition phases of the nodes from the up to the down state (blue) and vice-versa (orange) and the node coordinates along the antero-posterior axis as a function of the percentage by which the connection strengths of the most anterior node were changed. The left (right) column shows results for the aLN (Wilson-Cowan) models with 100 (top row), 200 (middle row), and 500 nodes (bottom row). 0\% corresponds to the unchanged structural antero-posterior gradient where the value of the $y$-slope was not changed, -100\% indicates that the gradient was enhanced, whereas +100\% indicates that the gradient was reversed. Model parameters are given in Tables A5 and A6.}
    \label{fig:sc_gradient_change_1}
\end{figure}

\begin{figure}[H]
    \centering
    \includegraphics[width=\textwidth]{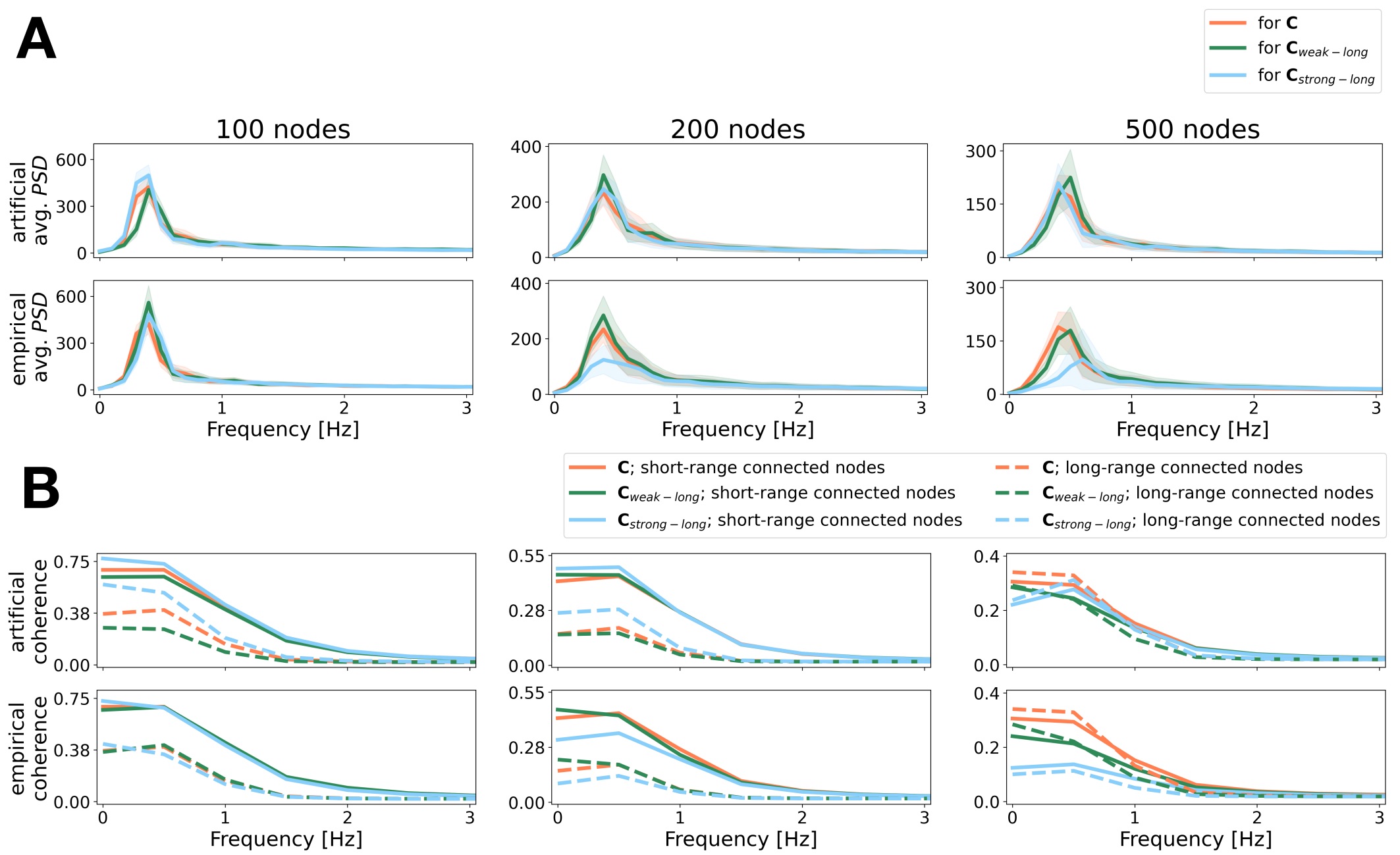}
    \caption{Power and coherence as a function of frequency for SO activity generated by the aLN model. Results are shown for the average connectivity matrix, $C$, (coral), and the connectivity matrices with weaker, $C_{weak-long}$, (green) and stronger, $C_{strong-long}$, (blue) long-range connections. Every column corresponds to one parcellation. (A) Averaged power spectra with standard deviation for each activity induced by the three connectivity matrices. The top (bottom) row shows the results for the artificially changed (empirically selected) connections. (B) Corresponding coherence values plotted separately for nodes that are connected through short- (solid lines) or long-range (dashed lines) connections. Model parameters are given in Table A5.}
    \label{fig:coherence_aln}
\end{figure}

\section*{Tables}
\begin{table}[H]
    \centering
    \caption{Parameter values used for the aLN model. Values are taken from \citet{cakan2022spatiotemporal}.}
    \begin{tabular}{||c|c|c||}
         \hline
         Parameter & Value & Description \\
         \hline
         $\mu_e^{ext}$ & [0 - 4]mV/ms & Mean external input to E\\
         $\mu_I^{ext}$ & [0 - 4]mV/ms & Mean external input to I\\
         $\sigma_{ou}$ & 0 or 0.37 mV/ms$^{3/2}$ & Noise strength\\
         $\tau_{ou}$ & 5 ms & Noise time constant\\
         $K_e$ & 800 & Number of excitatory inputs per neuron\\
         $K_i$ & 200 & Number of inhibitory inputs per neuron\\
         $c_{EE}$, $c_{EI}$ & 0.3 mV/ms & Maximum AMPA PSC amplitude\\
         $c_{EI}$, $c_{II}$ & 0.5 mV/ms & Maximum GABA PSC amplitude\\
         $J_{EE}$ & 2.4 mV/ms & Maximum synaptic current from E to E\\
         $J_{IE}$ & 2.6 mV/ms & Maximum synaptic current from I to E\\
         $J_{EI}$ & -3.3 mV/ms & Maximum synaptic current from I to E\\
         $J_{II}$ & -1.6 mV/ms & Maximum synaptic current from I to I\\
         $\tau_{s,E}$ & 2 ms & Excitatory synaptic time constant\\
         $\tau_{s,I}$ & 5 ms & Inhibitory synaptic time constant\\
         $d_E$ & 4 ms & Synaptic delay to excitatory neurons\\
         $d_I$ & 2 ms & Synaptic delay to inhibitory neurons\\
         C & 200 pF & Membrane capacitance\\
         $g_L$ & 10 nS & Leak conductance\\
         $\tau_m$ & C/$g_L$ & Membrane time constant\\
         $E_L$ & -65 mV & Leak reversal potential\\
         $\delta_T$ & 1.5 mV & Threshold slope factor\\
         $V_T$ & -50 mV & Threshold voltage\\
         $V_s$ & -40 mV & Spike voltage threshold\\
         $V_r$ & -70 mV & Reset voltage \\
         $T_{ref}$ & 1.5 ms & Refractory time\\
         $\sigma^{ext}$ & 1.5 mV/$\sqrt{\textnormal{ms}}$ & Standard deviation of external input\\
         $E_A$ & -80 mV & Adaptation reversal potential\\
         a & 0 nS & Subthreshold adaptation conductance\\
         b & 0, 20 pA & Spike-triggered adaptation incremenent\\
         $\tau_A$ & 600 ms & Adaptation time constant\\
         $K_{gl}$ & 265 & Global coupling strength\\
         $v_{gl}$ & 20 m/s & Global signal speed \\
         \hline
    \end{tabular}    
    \label{tab:model_parameters_aln}
\end{table}

\begin{table}[H]
    \centering
    \caption{Parameter values used for the Wilson-Cowan model.}
    \begin{tabular}{||c|c|c||}
         \hline
         Parameter & Value & Description \\
         \hline
         $\mu_e^{ext}$ & [0 - 8] & Mean external input to E\\
         $\mu_I^{ext}$ & [0 - 8] & Mean external input to I\\
         $\sigma_{ou}$ & 0 or 0.49 & Noise strength\\
         $\tau_{ou}$ & 5 & Time constant of the Ornstein-Uhlenbeck process\\
         $\tau_E$ & 2.5 & Excitatory membrane time constant\\
         $\tau_I$ & 3.75 & Inhibitory membrane time constant\\
         $w_{EE}$ & 16 & Excitatory-excitatory coupling strength\\
         $w_{EI}$ & 12 & Inhibitory-excitatory coupling strength\\
         $w_{IE}$ & 12 & Excitatory-inhibitory coupling strength\\
         $w_{II}$ & 3 & Inhibitory-inhibitory coupling strength\\
         $a_E$ & 1 & Gain factor of the excitatory population\\
         $a_I$ & 1 & Gain factor of the inhibitory population\\
         $\nu_E$ & 5 & Threshold of the excitatory population\\
         $\nu_I$ & 5 & Threshold of the inhibitory population\\
         $a_A$ & 3 & Adaptation gain factor\\
         $\nu_A$ & 2 & Adaptation threshold\\
         b & 0, 60 & Adaptation strength\\
         $\tau_A$ & 4625 & Adaptation time constant\\
         $K_{gl}$ & 0.5& Global coupling strength\\
         $v_{gl}$ & 80 & Global signal speed \\
         \hline
    \end{tabular}    
    \label{tab:model_parameters_wc}
\end{table}

\begin{table}[H]
    \centering
    \caption{Maximum coherence values for non-zero frequencies for the metastable states of the aLN model for all settings shown in Figure \ref{fig:coherence_aln}B. Both for the artificial and the empirical case, values in \textbf{bold} indicate the highest coherence values per parcellation, per set of nodes connected on a short-range (long-range). The corresponding frequencies were 0.5 Hz for all settings. Parameters are as for Figure \ref{fig:coherence_aln}.}
    \begin{tabular}{||ll|rrrrrr||}
\hline
          & Property & \multicolumn{6}{c||}{$coh(f_{max})$} \\
          & Resolution & \multicolumn{2}{l}{100} & \multicolumn{2}{l}{200} & \multicolumn{2}{l||}{500} \\
          & Distance &            short &  long & short &  long & short &  long \\
\hline
artificial & $\textbf{C}$ &       0.69 &  0.40 &  0.44 &  0.19 &  \textbf{0.29} & \textbf{0.33} \\
          & $\textbf{C}_{weak-long}$ &       0.64 &  0.26 &  0.45 &  0.16 &  0.24 &  0.24 \\
          & $\textbf{C}_{strong-long}$ &   \textbf{0.73} & \textbf{0.52} & \textbf{0.49} & \textbf{0.28} &  0.28 &  0.31 \\
empirical & $\textbf{C}$ &   \textbf{0.69} & 0.40 & \textbf{0.44} & \textbf{0.19} & \textbf{0.29} & \textbf{0.33} \\
          & $\textbf{C}_{weak-long}$ &   \textbf{0.69} & \textbf{0.41} &  0.43 & \textbf{0.19} &  0.21 &  0.22 \\
          & $\textbf{C}_{strong-long}$ &       0.68 &  0.34 &  0.35 &  0.13 &  0.14 &  0.11 \\
\hline
\end{tabular}
    \label{tab:aln_coherence_table}
\end{table}

\newpage
\bibliographystyle{apalike}
\bibliography{references.bib}  %%% Uncomment this line and comment out the ``thebibliography'' section below to use the external .bib file (using bibtex) .

\begin{thebibliography}{}

\bibitem[Augustin et~al., 2017]{augustin2017low}
Augustin, M., Ladenbauer, J., Baumann, F., and Obermayer, K. (2017).
\newblock Low-dimensional spike rate models derived from networks of adaptive
  integrate-and-fire neurons: comparison and implementation.
\newblock {\em PLoS Computational Biology}, 13(6):e1005545.

\bibitem[Bhattacharya et~al., 2022]{bhattacharya2022traveling}
Bhattacharya, S., Brincat, S.~L., Lundqvist, M., and Miller, E.~K. (2022).
\newblock Traveling waves in the prefrontal cortex during working memory.
\newblock {\em PLoS Computational Biology}, 18(1):e1009827.

\bibitem[Breakspear et~al., 2003]{breakspear2003modulation}
Breakspear, M., Terry, J.~R., and Friston, K.~J. (2003).
\newblock Modulation of excitatory synaptic coupling facilitates
  synchronization and complex dynamics in a biophysical model of neuronal
  dynamics.
\newblock {\em Network: Computation in Neural Systems}, 14(4):703.

\bibitem[Cabral et~al., 2012]{cabral2012functional}
Cabral, J., Kringelbach, M., and Deco, G. (2012).
\newblock Functional graph alterations in schizophrenia: a result from a global
  anatomic decoupling?
\newblock {\em Pharmacopsychiatry}, 45(S 01):S57--S64.

\bibitem[Cakan et~al., 2022]{cakan2022spatiotemporal}
Cakan, C., Dimulescu, C., Khakimova, L., Obst, D., Fl{\"o}el, A., and
  Obermayer, K. (2022).
\newblock Spatiotemporal patterns of adaptation-induced slow oscillations in a
  whole-brain model of slow-wave sleep.
\newblock {\em Frontiers in Computational Neuroscience}, 15:800101.

\bibitem[Cakan et~al., 2021]{cakan2021neurolib}
Cakan, C., Jajcay, N., and Obermayer, K. (2021).
\newblock neurolib: A simulation framework for whole-brain neural mass
  modeling.
\newblock {\em Cognitive Computation}.

\bibitem[Cakan and Obermayer, 2020]{cakan2020biophysically}
Cakan, C. and Obermayer, K. (2020).
\newblock Biophysically grounded mean-field models of neural populations under
  electrical stimulation.
\newblock {\em PLoS Computational Biology}, 16(4):e1007822.

\bibitem[Capone et~al., 2023]{capone2023simulations}
Capone, C., De~Luca, C., De~Bonis, G., Gutzen, R., Bernava, I., Pastorelli, E.,
  Simula, F., Lupo, C., Tonielli, L., Resta, F., et~al. (2023).
\newblock Simulations approaching data: cortical slow waves in inferred models
  of the whole hemisphere of mouse.
\newblock {\em Communications Biology}, 6(1):266.

\bibitem[Das et~al., 2024]{das2024planar}
Das, A., Zabeh, E., Ermentrout, B., and Jacobs, J. (2024).
\newblock Planar, spiral, and concentric traveling waves distinguish cognitive
  states in human memory.
\newblock {\em bioRxiv}, pages 2024--01.

\bibitem[Dasilva et~al., 2021]{dasilva2021modulation}
Dasilva, M., Camassa, A., Navarro-Guzman, A., Pazienti, A., Perez-Mendez, L.,
  Zamora-L{\'o}pez, G., Mattia, M., and Sanchez-Vives, M.~V. (2021).
\newblock Modulation of cortical slow oscillations and complexity across
  anesthesia levels.
\newblock {\em NeuroImage}, 224:117415.

\bibitem[Deco et~al., 2017]{deco2017dynamics}
Deco, G., Kringelbach, M.~L., Jirsa, V.~K., and Ritter, P. (2017).
\newblock The dynamics of resting fluctuations in the brain: metastability and
  its dynamical cortical core.
\newblock {\em Scientific Reports}, 7(1):3095.

\bibitem[Deco et~al., 2021]{deco2021rare}
Deco, G., Perl, Y.~S., Vuust, P., Tagliazucchi, E., Kennedy, H., and
  Kringelbach, M.~L. (2021).
\newblock Rare long-range cortical connections enhance human information
  processing.
\newblock {\em Current Biology}, 31(20):4436--4448.

\bibitem[Ester et~al., 1996]{ester1996density}
Ester, M., Kriegel, H.-P., Sander, J., Xu, X., et~al. (1996).
\newblock A density-based algorithm for discovering clusters in large spatial
  databases with noise.
\newblock In {\em Knowledge Discovery and Data Mining}, volume~96, pages
  226--231.

\bibitem[Freyer et~al., 2009]{freyer2009bistability}
Freyer, F., Aquino, K., Robinson, P.~A., Ritter, P., and Breakspear, M. (2009).
\newblock Bistability and non-gaussian fluctuations in spontaneous cortical
  activity.
\newblock {\em Journal of Neuroscience}, 29(26):8512--8524.

\bibitem[Freyer et~al., 2011]{freyer2011biophysical}
Freyer, F., Roberts, J.~A., Becker, R., Robinson, P.~A., Ritter, P., and
  Breakspear, M. (2011).
\newblock Biophysical mechanisms of multistability in resting-state cortical
  rhythms.
\newblock {\em Journal of Neuroscience}, 31(17):6353--6361.

\bibitem[Illoul and Lorong, 2011]{illoul2011some}
Illoul, L. and Lorong, P. (2011).
\newblock On some aspects of the cnem implementation in 3d in order to simulate
  high speed machining or shearing.
\newblock {\em Computers \& Structures}, 89(11-12):940--958.

\bibitem[Kelso, 2012]{kelso2012multistability}
Kelso, J.~S. (2012).
\newblock Multistability and metastability: understanding dynamic coordination
  in the brain.
\newblock {\em Philosophical Transactions of the Royal Society B: Biological
  Sciences}, 367(1591):906--918.

\bibitem[Kilpatrick, 2013]{Kilpatrick2013WC}
Kilpatrick, Z.~P. (2013).
\newblock {\em Wilson-Cowan Model}, pages 1--5.
\newblock Springer New York, New York, NY.

\bibitem[Ladenbauer et~al., 2023]{ladenbauer2023towards}
Ladenbauer, J., Khakimova, L., Malinowski, R., Obst, D., T{\"o}nnies, E.,
  Antonenko, D., Obermayer, K., Hanna, J., and Fl{\"o}el, A. (2023).
\newblock Towards optimization of oscillatory stimulation during sleep.
\newblock {\em Neuromodulation: Technology at the Neural Interface},
  26(8):1592--1601.

\bibitem[Ladenbauer et~al., 2017]{ladenbauer2017promoting}
Ladenbauer, J., Ladenbauer, J., K{\"u}lzow, N., de~Boor, R., Avramova, E.,
  Grittner, U., and Fl{\"o}el, A. (2017).
\newblock Promoting sleep oscillations and their functional coupling by
  transcranial stimulation enhances memory consolidation in mild cognitive
  impairment.
\newblock {\em Journal of Neuroscience}, 37(30):7111--7124.

\bibitem[Levenstein et~al., 2019]{Levenstein2019}
Levenstein, D., Buzs{\'a}ki, G., and Rinzel, J. (2019).
\newblock Nrem sleep in the rodent neocortex and hippocampus reflects excitable
  dynamics.
\newblock {\em Nature Communications}, 10(1):2478.

\bibitem[Liang et~al., 2023]{liang2021corticalmousewaves}
Liang, Y., Liang, J., Song, C., Liu, M., Knöpfel, T., Gong, P., and Zhou, C.
  (2023).
\newblock Complexity of cortical wave patterns of the wake mouse cortex.
\newblock {\em Nature Communications}, 14(1):1434.

\bibitem[Massimini et~al., 2004]{massimini2004sleep}
Massimini, M., Huber, R., Ferrarelli, F., Hill, S., and Tononi, G. (2004).
\newblock The sleep slow oscillation as a traveling wave.
\newblock {\em Journal of Neuroscience}, 24(31):6862--6870.

\bibitem[Mohan et~al., 2024]{mohan2024direction}
Mohan, U.~R., Zhang, H., Ermentrout, B., and Jacobs, J. (2024).
\newblock The direction of theta and alpha travelling waves modulates human
  memory processing.
\newblock {\em Nature Human Behaviour}, pages 1--12.

\bibitem[Muller et~al., 2016]{muller2016rotating}
Muller, L., Piantoni, G., Koller, D., Cash, S.~S., Halgren, E., and Sejnowski,
  T.~J. (2016).
\newblock Rotating waves during human sleep spindles organize global patterns
  of activity that repeat precisely through the night.
\newblock {\em eLife}, 5:e17267.

\bibitem[Muller et~al., 2014]{muller2014stimulus}
Muller, L., Reynaud, A., Chavane, F., and Destexhe, A. (2014).
\newblock The stimulus-evoked population response in visual cortex of awake
  monkey is a propagating wave.
\newblock {\em Nature Communications}, 5(1):3675.

\bibitem[Papadopoulos et~al., 2020]{Papadopoulos2020}
Papadopoulos, L., Lynn, C.~W., Battaglia, D., and Bassett, D.~S. (2020).
\newblock Relations between large-scale brain connectivity and effects of
  regional stimulation depend on collective dynamical state.
\newblock {\em PLOS Computational Biology}, 16(9):1--43.

\bibitem[Pessoa, 2022]{pessoa2022entangled}
Pessoa, L. (2022).
\newblock {\em The entangled brain: How perception, cognition, and emotion are
  woven together}.
\newblock MIT Press.

\bibitem[Pinto et~al., 1996]{Pinto1996refractory}
Pinto, D.~J., Brumberg, J.~C., Simons, D.~J., Ermentrout, G.~B., and Traub, R.
  (1996).
\newblock A quantitative population model of whisker barrels: Re-examining the
  wilson-cowan equations.
\newblock {\em Journal of Computational Neuroscience}, 3(3):247--264.

\bibitem[Popovych et~al., 2021]{popovych2021inter}
Popovych, O.~V., Jung, K., Manos, T., Diaz-Pier, S., Hoffstaedter, F.,
  Schreiber, J., Yeo, B.~T., and Eickhoff, S.~B. (2021).
\newblock Inter-subject and inter-parcellation variability of resting-state
  whole-brain dynamical modeling.
\newblock {\em NeuroImage}, 236:118201.

\bibitem[Proix et~al., 2016]{proix2016parcellation}
Proix, T., Spiegler, A., Schirner, M., Rothmeier, S., Ritter, P., and Jirsa,
  V.~K. (2016).
\newblock How do parcellation size and short-range connectivity affect dynamics
  in large-scale brain network models?
\newblock {\em NeuroImage}, 142:135--149.

\bibitem[Rasch and Born, 2013]{rasch2013sleep}
Rasch, B. and Born, J. (2013).
\newblock About sleep's role in memory.
\newblock {\em Physiological Reviews.}

\bibitem[Richardson, 2007]{Richardson2007FiringrateRO}
Richardson, M. J.~E. (2007).
\newblock Firing-rate response of linear and nonlinear integrate-and-fire
  neurons to modulated current-based and conductance-based synaptic drive.
\newblock {\em Physical review. E - Statistical, nonlinear, and soft matter
  physics}, 76 2 Pt 1:021919.

\bibitem[Roberts et~al., 2019]{roberts2019metastable}
Roberts, J.~A., Gollo, L.~L., Abeysuriya, R.~G., Roberts, G., Mitchell, P.~B.,
  Woolrich, M.~W., and Breakspear, M. (2019).
\newblock Metastable brain waves.
\newblock {\em Nature Communications}, 10(1):1056.

\bibitem[Sanchez-Vives et~al., 2017]{sanchez2017shaping}
Sanchez-Vives, M.~V., Massimini, M., and Mattia, M. (2017).
\newblock Shaping the default activity pattern of the cortical network.
\newblock {\em Neuron}, 94(5):993--1001.

\bibitem[Schaefer et~al., 2018]{schaefer2018local}
Schaefer, A., Kong, R., Gordon, E.~M., Laumann, T.~O., Zuo, X.-N., Holmes,
  A.~J., Eickhoff, S.~B., and Yeo, B.~T. (2018).
\newblock Local-global parcellation of the human cerebral cortex from intrinsic
  functional connectivity mri.
\newblock {\em Cerebral Cortex}, 28(9):3095--3114.

\bibitem[Sporns, 2022]{sporns2022complex}
Sporns, O. (2022).
\newblock The complex brain: connectivity, dynamics, information.
\newblock {\em Trends in Cognitive Sciences}, 26(12):1066--1067.

\bibitem[Torao-Angosto et~al., 2021]{Torao2021}
Torao-Angosto, M., Manasanch, A., Mattia, M., and Sanchez-Vives, M.~V. (2021).
\newblock Up and down states during slow oscillations in slow-wave sleep and
  different levels of anesthesia.
\newblock {\em Frontiers in Systems Neuroscience}, 15.

\bibitem[Townsend and Gong, 2018]{townsend2018detection}
Townsend, R.~G. and Gong, P. (2018).
\newblock Detection and analysis of spatiotemporal patterns in brain activity.
\newblock {\em PLoS Computational Biology}, 14(12):e1006643.

\bibitem[Wilson and Cowan, 1972]{wilson1972excitatory}
Wilson, H.~R. and Cowan, J.~D. (1972).
\newblock Excitatory and inhibitory interactions in localized populations of
  model neurons.
\newblock {\em Biophysical Journal}, 12(1):1--24.

\bibitem[Zbilut et~al., 2002]{zbilut2002recurrence}
Zbilut, J.~P., Zaldivar-Comenges, J.-M., and Strozzi, F. (2002).
\newblock Recurrence quantification based liapunov exponents for monitoring
  divergence in experimental data.
\newblock {\em Physics Letters A}, 297(3-4):173--181.

\end{thebibliography}

%\begin{thebibliography}{1}




% \end{thebibliography}

%\begin{document}
\onecolumn

\title\textbf{Supplementary Material}

\maketitle

\setcounter{figure}{0}
\setcounter{table}{0}
\renewcommand{\figurename}{\textbf{Figure}}
\renewcommand{\tablename}{\textbf{Table}}
\renewcommand\thefigure{\textbf{A\arabic{figure}}}    
\renewcommand\thetable{\textbf{A\arabic{table}}}    

\begin{figure}[H]
    \centering
    \includegraphics[width=0.8\textwidth]{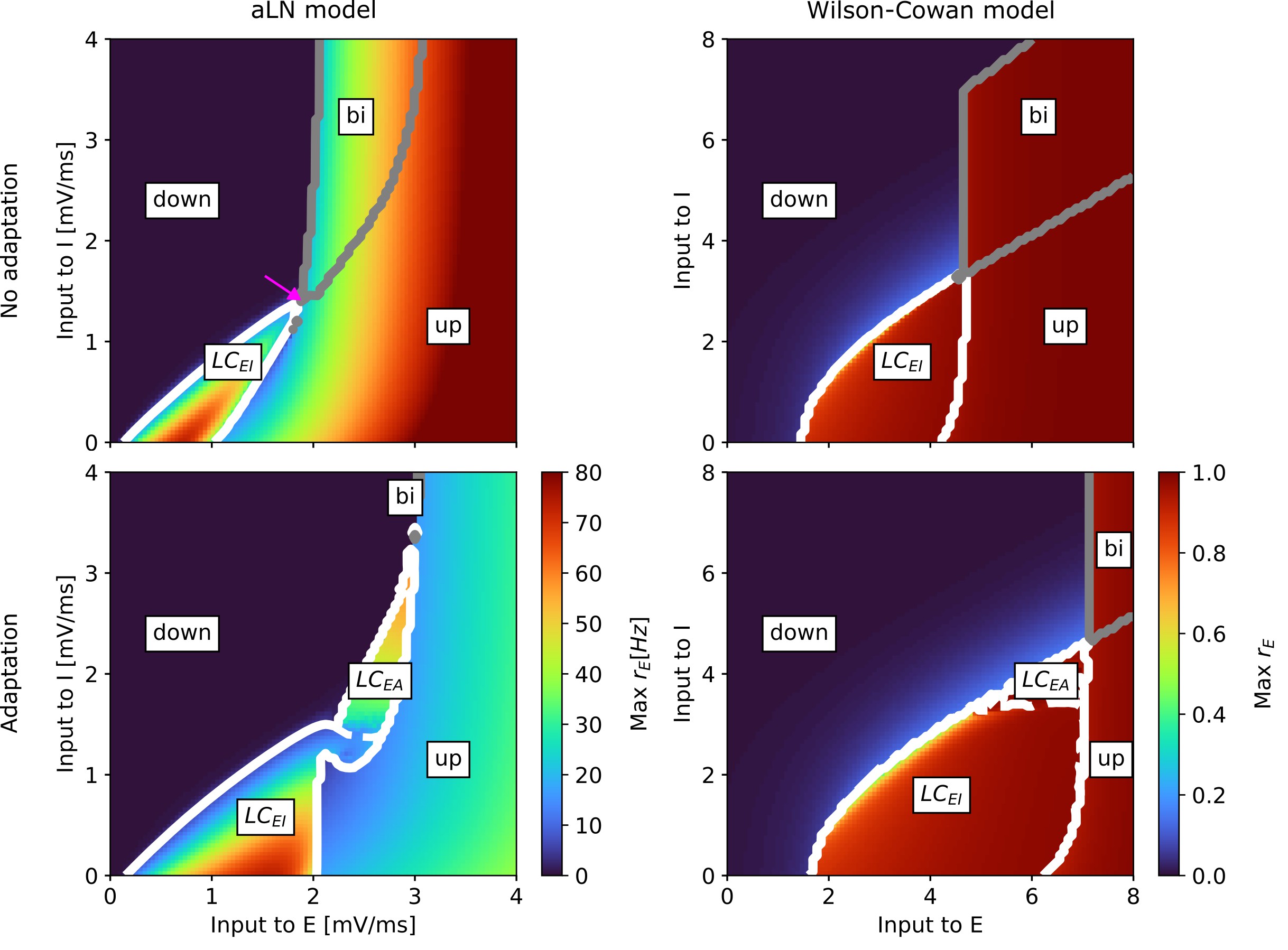}
    \caption{Slice of state space of the single-node aLN (left column) and Wilson-Cowan (right column) models without (b = 0; top row) and with adaptation (b = 20 pA, 60; bottom row) spanned by the external input currents to the E and I populations. In every panel, the horizontal (vertical) axis denotes the external input current $\mu_E^{ext}$ ($\mu_I^{ext}$) to the excitatory (inhibitory) population. The heatmap shows the maximum excitatory firing rate $r_E$  of the model. State boundaries are indicated by solid white lines for the fast ($LC_{EI}$) and slow ($LC_{EA}$) oscillatory regions, and by solid grey lines for the regime of bistability between up and down states (\textit{bi}). The magenta arrow indicates the region where few points with bistability between the up and the fast ($LC_{EI}$) state are found. Unistable up (up) and down state (down) regions are also marked. Model parameters are given in Tables 1 and 2.}
    \label{fig:single_node_bif}
\end{figure}

\begin{figure}[H]
    \centering
    \includegraphics[width=\textwidth]{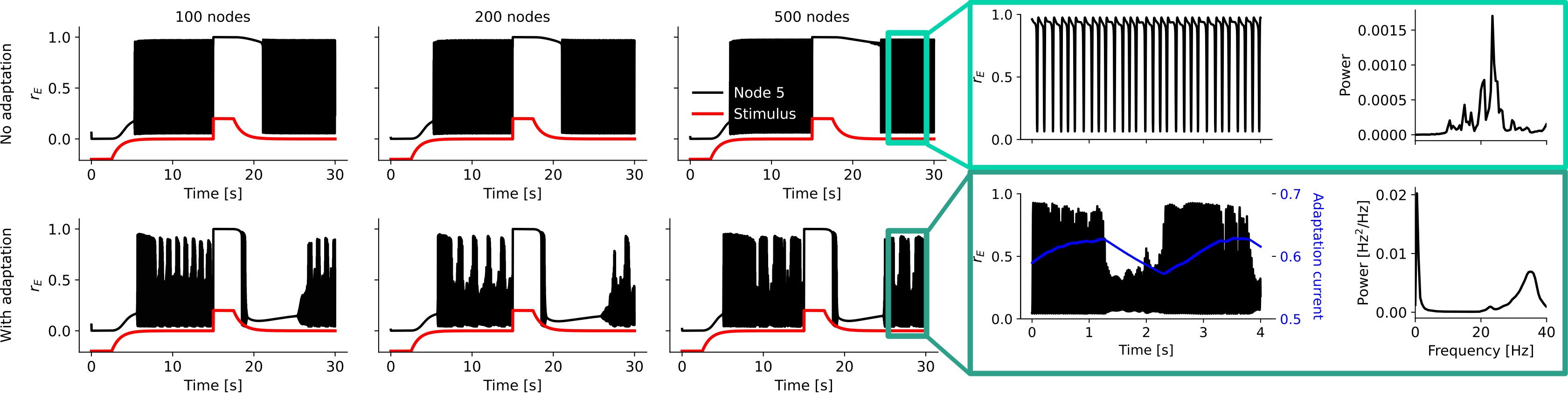}
    \caption{Example time series of the proportion $r_E$ of active neurons per unit of time of one randomly chosen node (black line) of the whole-brain Wilson-Cowan model at several points in state space, without ($b$ = 0; top row) and with ($b$ = 60; bottom row) adaptation for a network with 100 (left column), 200 (middle column), and 500 nodes (right column). Parameters are as follows: 100 nodes - $\mu_E^{ext}$ = 4.3, $\mu_I^{ext}$ = 2.8; 200 nodes - $\mu_E^{ext}$ = 4.3, $\mu_I^{ext}$ = 2.85; 500 nodes -  $\mu_E^{ext}$ = 4.2, $\mu_I^{ext}$ = 2.7. The light (top) and dark green (bottom) insets display enlarged intervals of the time series of the proportion $r_E$ of active neurons per unit of time and, in case of finite adaptation, the current $I_A$ for the chosen node, and also show the power spectrum for the brain network with 500 nodes averaged across all nodes. All other model parameter values are given in Table 2.}
    \label{fig:wc_example_traces}
\end{figure}

\begin{figure}[H]
    \centering
    \includegraphics[width=\textwidth]{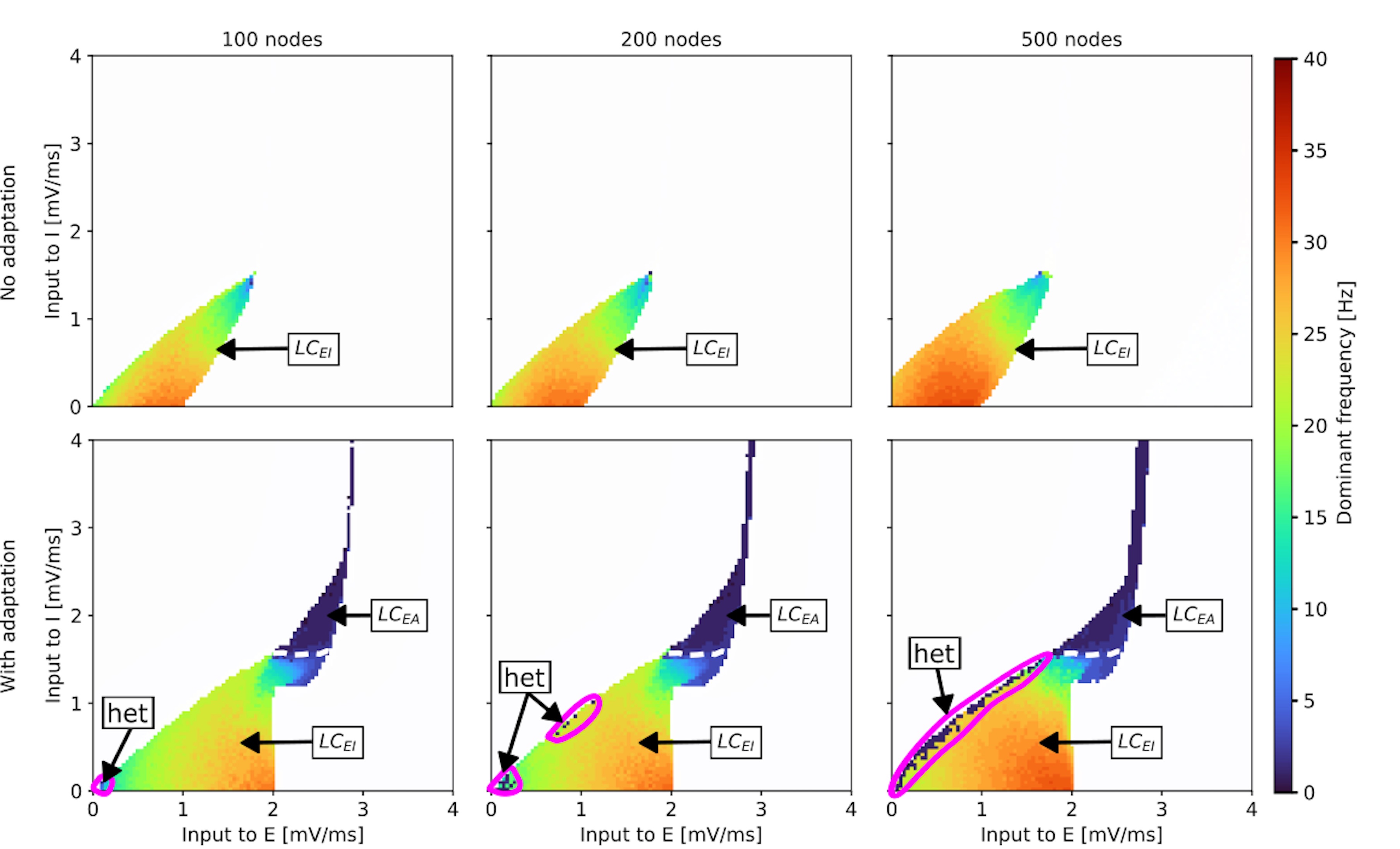}
    \caption{Frequency of the peak of the power spectrum averaged over all nodes of the whole-brain aLN model without (\textit{b} = 0 pA; top row) and with (\textit{b} = 20 pA; bottom row) adaptation for a brain network with 100 (left column), 200 (middle column), and 500 (right column) nodes as a function of the external input current to the E and I populations. In every panel, the horizontal axis shows the external input current to the excitatory population ($\mu_E^{ext}$) and the vertical axis shows the external input current to the inhibitory population ($\mu_I^{ext}$). The average dominant frequency (Hz) across all nodes in the network is color-coded. The white dashed line indicates the approximate border between the fast $LC_{EI}$ and slow $LC_{EA}$ oscillating regions. The magenta solid lines indicate the areas where heterogeneous (het) slow-fast oscillations were identified (for $b$ = 20 pA). All model parameters are summarized in Table 1.}
    \label{fig:aln_dominant_freq}
\end{figure}

\begin{figure}[H]
    \centering
    \includegraphics[width=\textwidth]{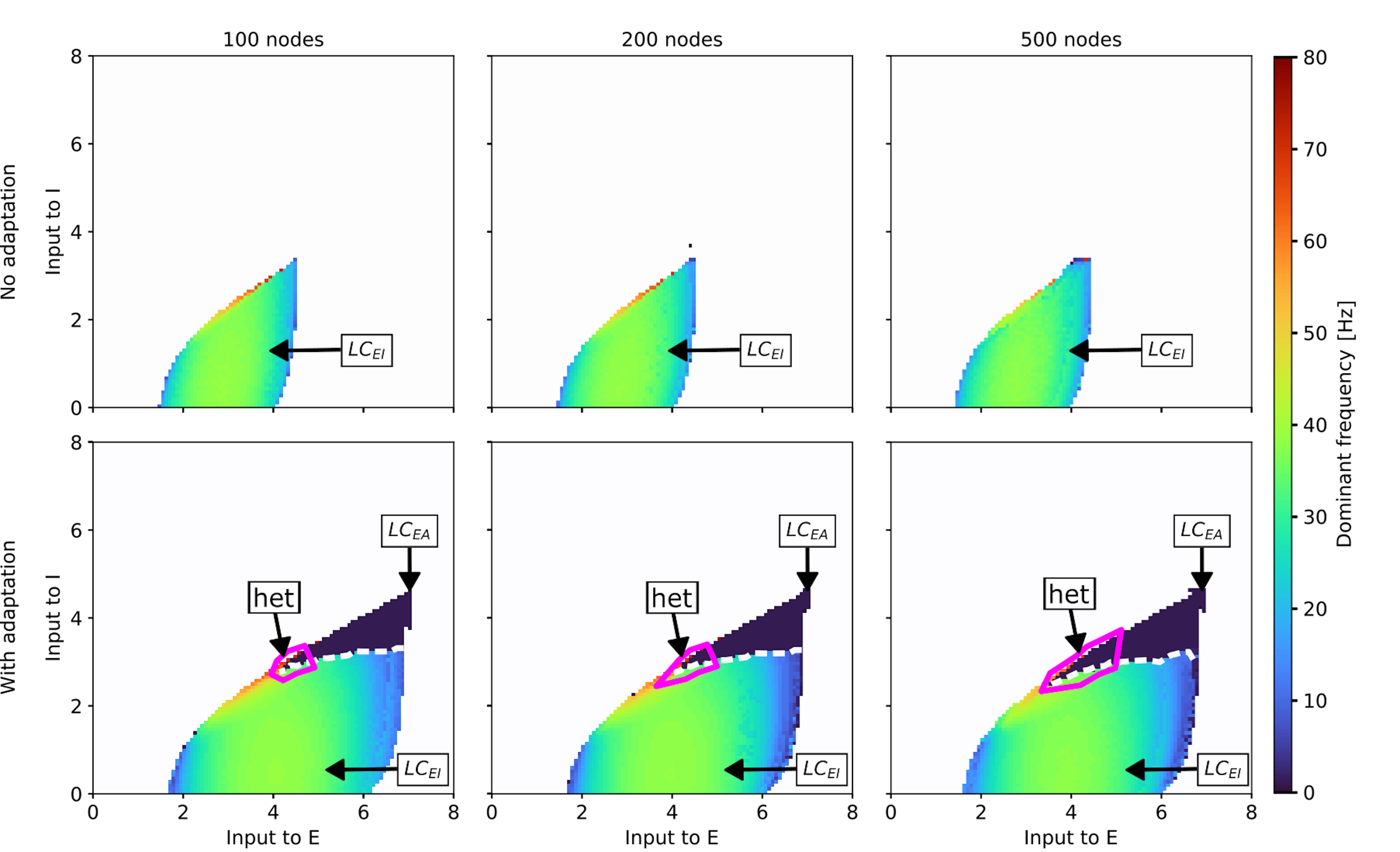}
    \caption{Frequency of the peak of the power spectrum averaged over all nodes of the whole-brain Wilson-Cowan model without (\textit{b} = 0; top row) and with (\textit{b} = 60; bottom row) adaptation for a brain network with 100 (left column), 200 (middle column), and 500 (right column) nodes as a function of the external input current to the E and I populations. In every panel, the horizontal axis shows the external input current to the excitatory population ($\mu_E^{ext}$) and the vertical axis shows the external input current to the inhibitory population ($\mu_I^{ext}$). The average dominant frequency (Hz) across all nodes in the network is color-coded. The white dashed line indicates the approximate border between the fast $LC_{EI}$ and slow $LC_{EA}$ oscillating regions. The magenta solid lines indicate the areas where heterogeneous (het) slow-fast oscillations were identified (for $b$ = 60). All model parameters are summarized in Table 2.}
    \label{fig:wc_dominant_freq}
\end{figure}

\begin{figure}[H]
    \centering
    \includegraphics[width=\textwidth]{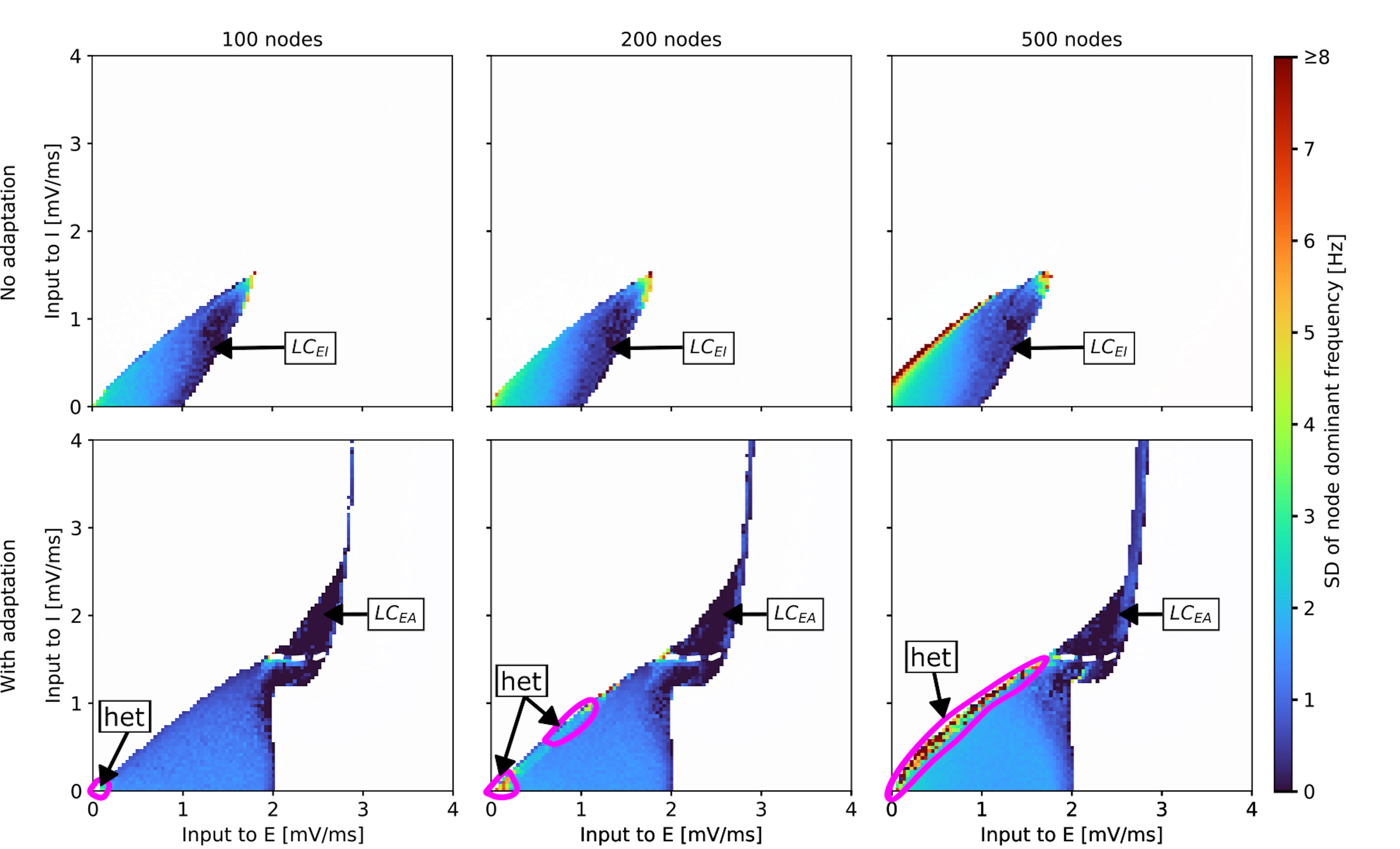}
    \caption{Standard deviation (SD) of the node dominant frequency of the whole-brain aLN model without (\textit{b} = 0 pA; top row) and with (\textit{b} = 20 pA; bottom row) adaptation for a brain network with 100 (left column), 200 (middle column), and 500 (right column) nodes as a function of the external input current to the E and I populations. In every panel, the horizontal axis shows the external input current to the excitatory population ($\mu_E^{ext}$) and the vertical axis shows the external input current to the inhibitory population ($\mu_I^{ext}$). The standard deviation of the node dominant frequency (Hz) across all nodes in the network is color-coded. The white dashed line indicates the approximate border between the fast $LC_{EI}$ and slow $LC_{EA}$ oscillating regions. The magenta solid lines indicate the areas where heterogeneous (het) slow-fast oscillations were identified (for $b$ = 20 pA). All model parameters are summarized in Table 1.}
    \label{fig:aln_SD_freq}
\end{figure}

\begin{figure}[H]
    \centering
    \includegraphics[width=\textwidth]{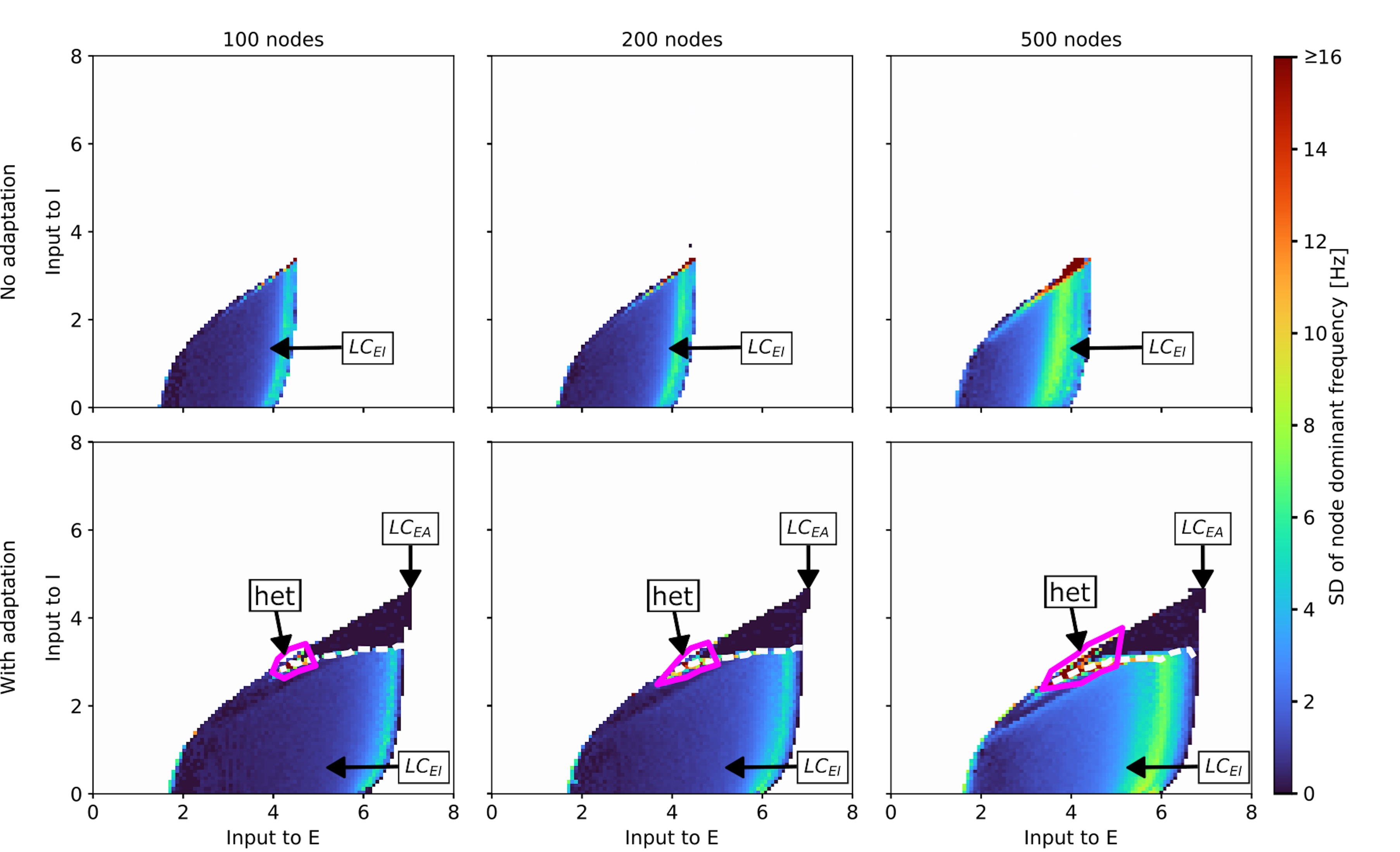}
    \caption{Standard deviation (SD) of the node dominant frequency of the whole-brain Wilson-Cowan model without (\textit{b} = 0; top row) and with (\textit{b} = 60; bottom row) adaptation for a brain network with 100 (left column), 200 (middle column), and 500 (right column) nodes as a function of the external input current to the E and I populations. In every panel, the horizontal axis shows the external input current to the excitatory population ($\mu_E^{ext}$) and the vertical axis shows the external input current to the inhibitory population ($\mu_I^{ext}$). The standard deviation of the node dominant frequency (Hz) across all nodes in the network is color-coded. The white dashed line indicates the approximate border between the fast $LC_{EI}$ and slow $LC_{EA}$ oscillating regions. The magenta solid lines indicate the areas where heterogeneous (het) slow-fast oscillations were identified (for $b$ = 60). All model parameters are summarized in Table 2.}
    \label{fig:wc_SD_freq}
\end{figure}

\begin{figure}[H]
    \centering
\includegraphics[width=0.95\textwidth]{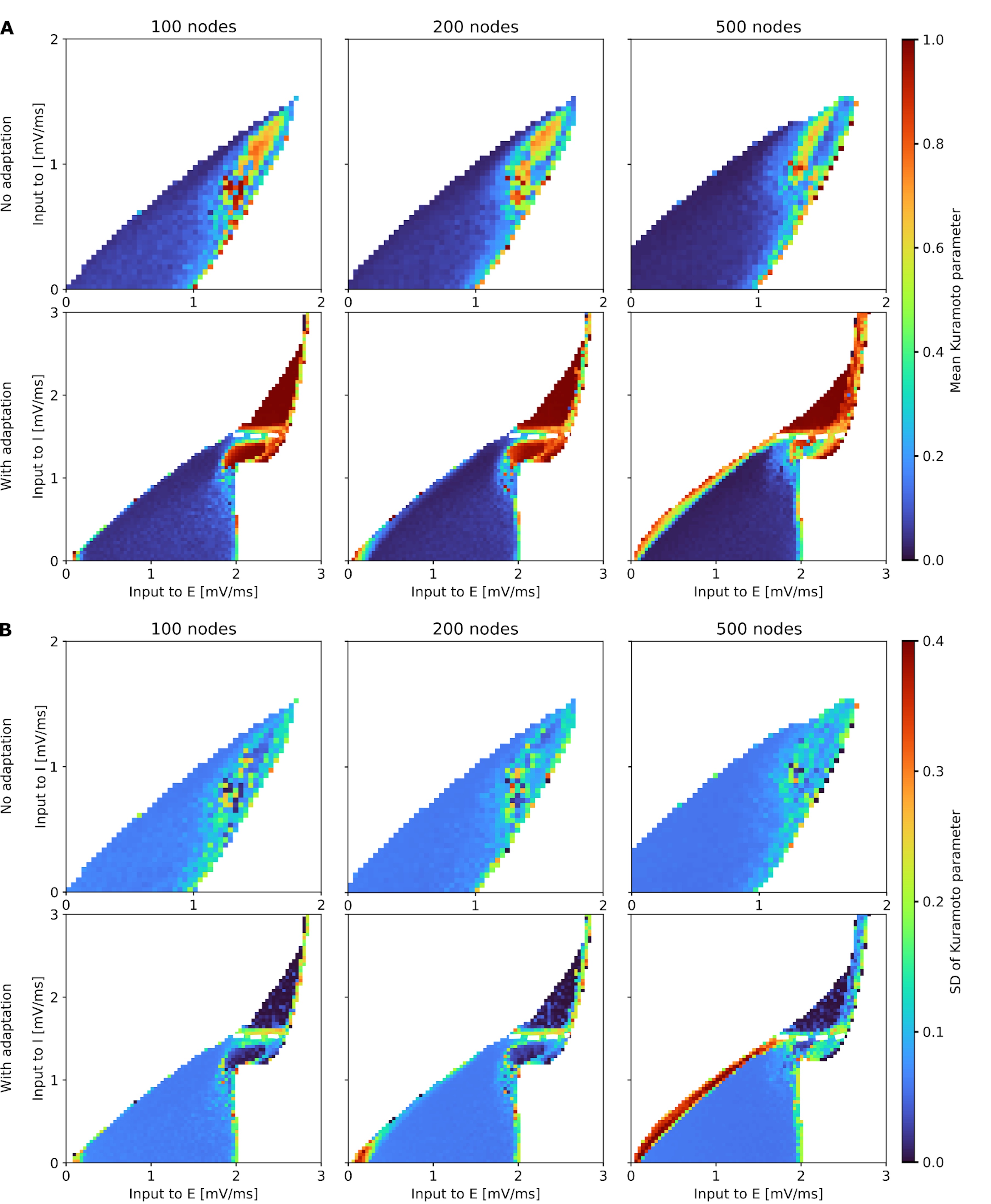}
    \caption{Mean (A) and standard deviation SD (B) of the Kuramoto order parameter for the aLN whole-brain model in the case without (b = 0 pA; top rows) and with (b = 20 pA; bottom rows) adaptation for 100 (left column), 200 (middle column), and 500 (right column) nodes. The slice of state space is spanned by the external input current to the E and I populations. The white dashed lines indicates the approximate border between the fast $LC_{EI}$ and slow $LC_{EA}$ oscillating regions.}
    \label{fig:aln_kuramoto}
\end{figure}

\begin{figure}[H]
    \centering
    \includegraphics[width=0.95\textwidth]{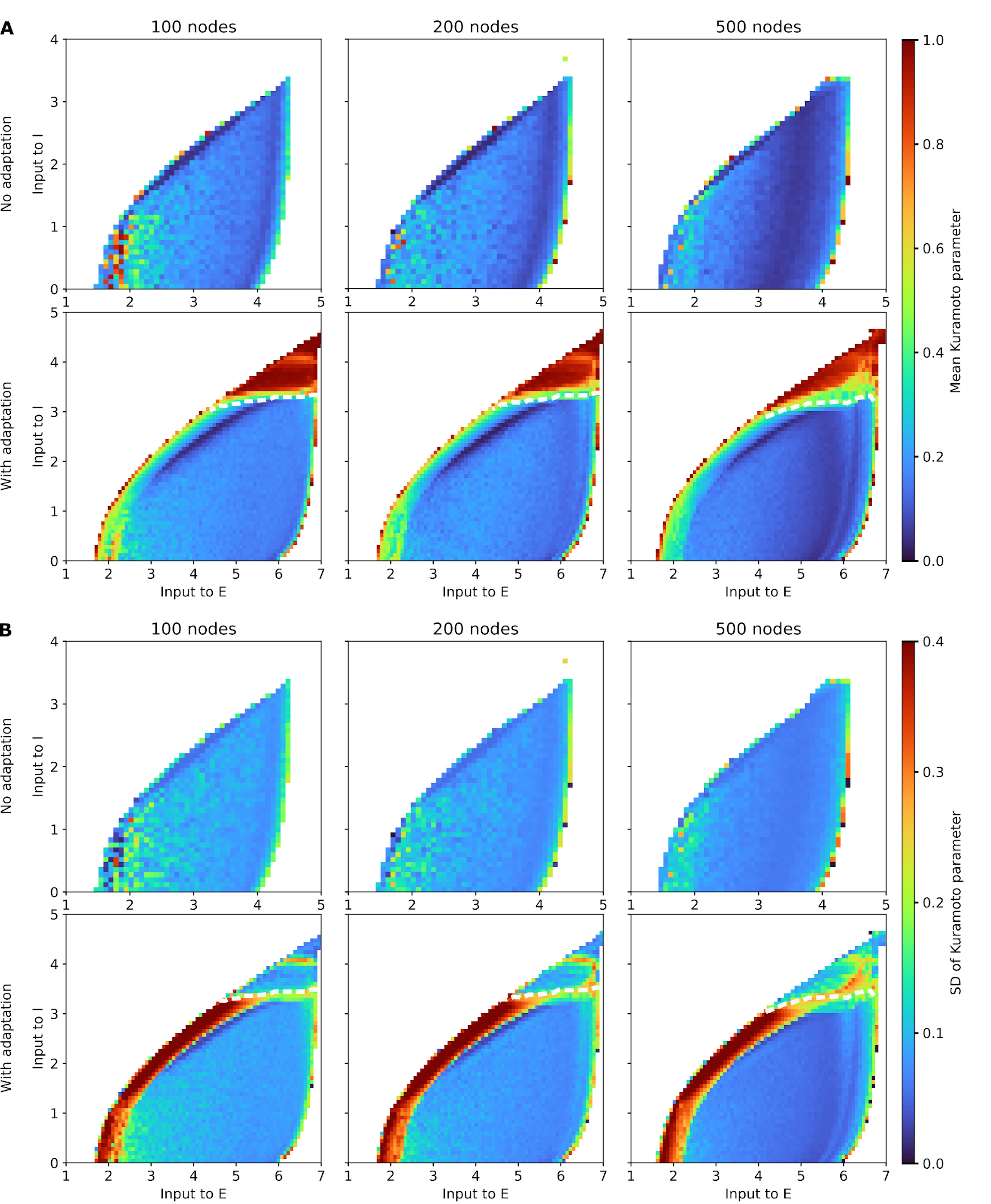}
    \caption{Mean (A) and standard deviation SD (B) of the Kuramoto order parameter for the Wilson-Cowan whole-brain model in the case without (b = 0; top rows) and with (b = 60; bottom rows) spike-triggered adaptation for 100 (left column), 200 (middle column), and 500 (right column) nodes. The slice of state space is spanned by the external input current to the E and I populations. The white dashed lines indicates the approximate border between the fast $LC_{EI}$ and slow $LC_{EA}$ oscillating regions.}
    \label{fig:wc_kuramoto}
\end{figure}

\begin{figure}[H]
    \centering
    \includegraphics[width=0.5\linewidth]{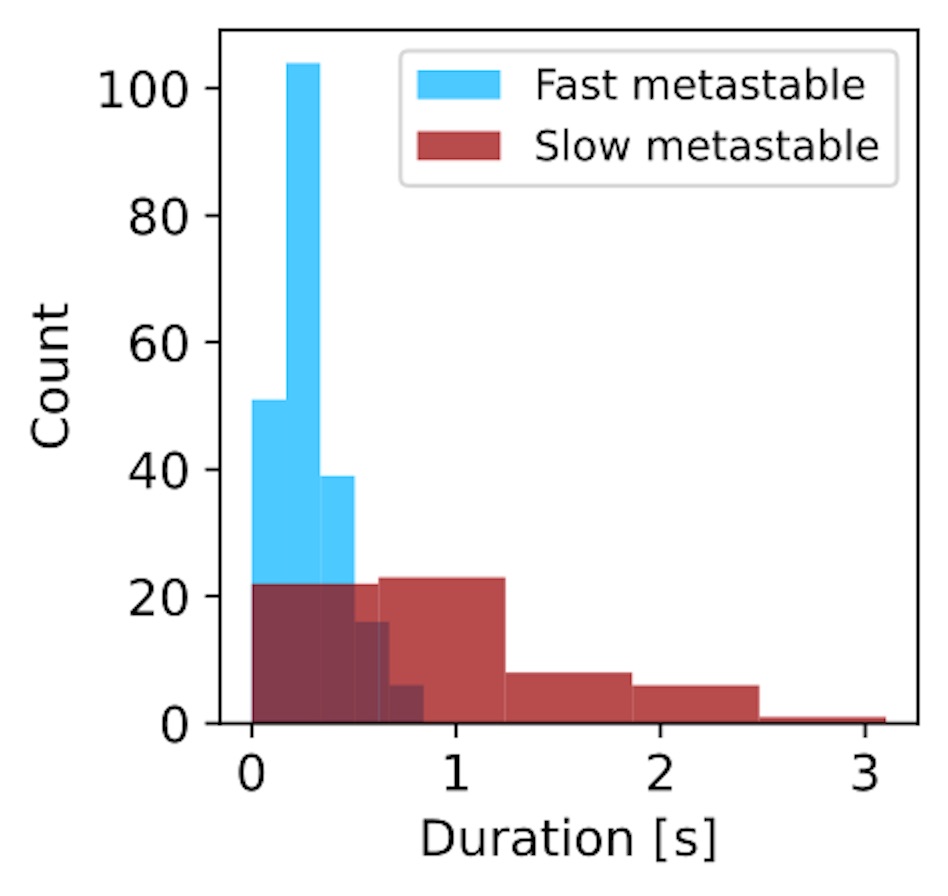}
    \caption{Histogram of the state durations for one slow metastable (red; location C in Figure 6) and one fast metastable point (blue; location B in Figure 6) in state space for the aLN model with 100 nodes and without adaptation ($b = 0$ pA). All other parameters are given in Table 1.}
    \label{fig:example_state_duration}
\end{figure}

\begin{figure}[H]
    \centering
    \includegraphics[width=\textwidth]{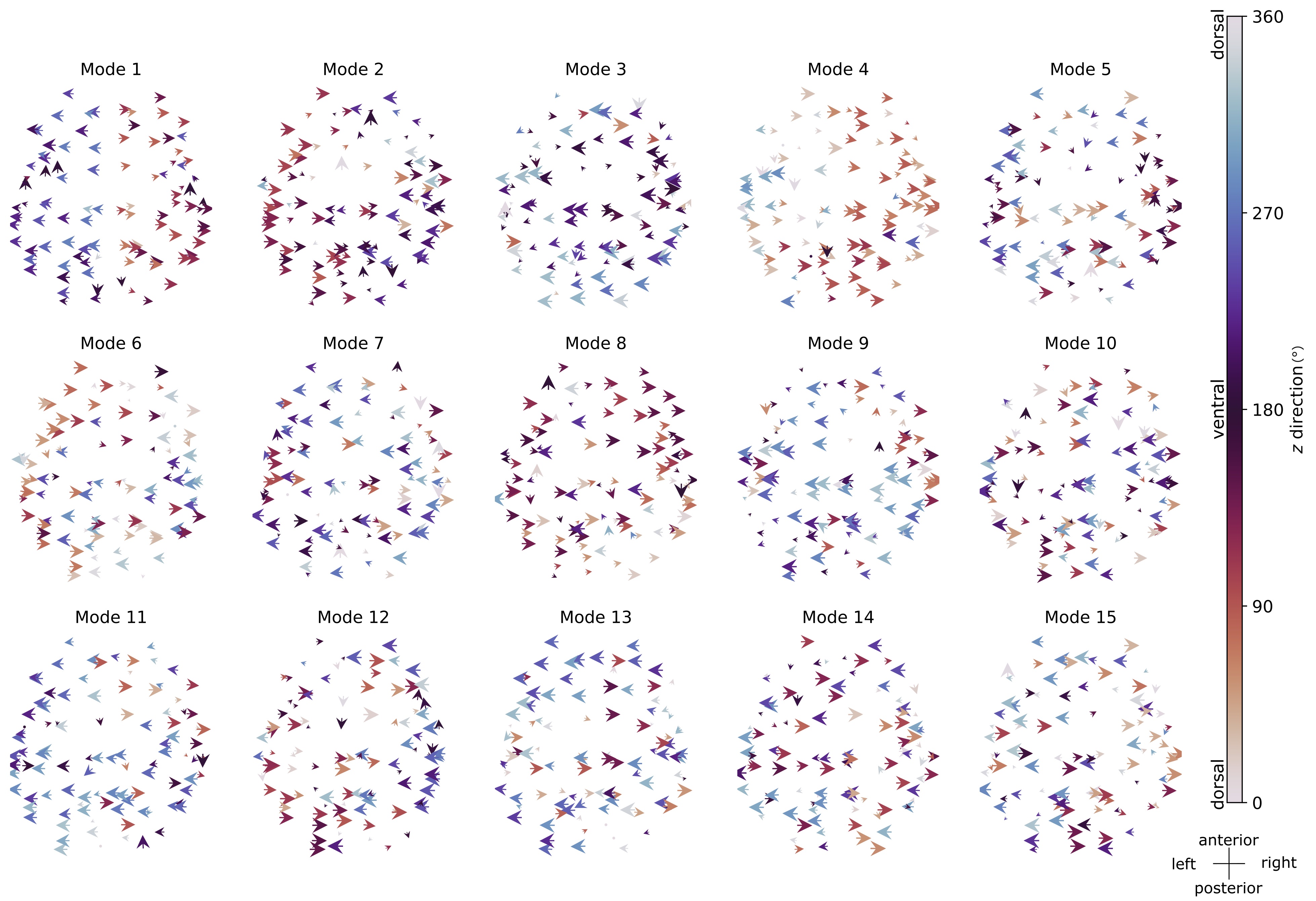}
    \caption{First 15 modes obtained from the singular value decomposition of the velocity vector fields in the whole-brain aLN model with 100 nodes and adaptation (b = 20 pA) for the unistable states in the $LC_{EA}$ region. Modes are ordered in decreasing order of explained variance.}
    \label{fig:aln_modes_100_slowLC}
\end{figure}

\begin{figure}[H]
    \centering
    \includegraphics[width=\textwidth]{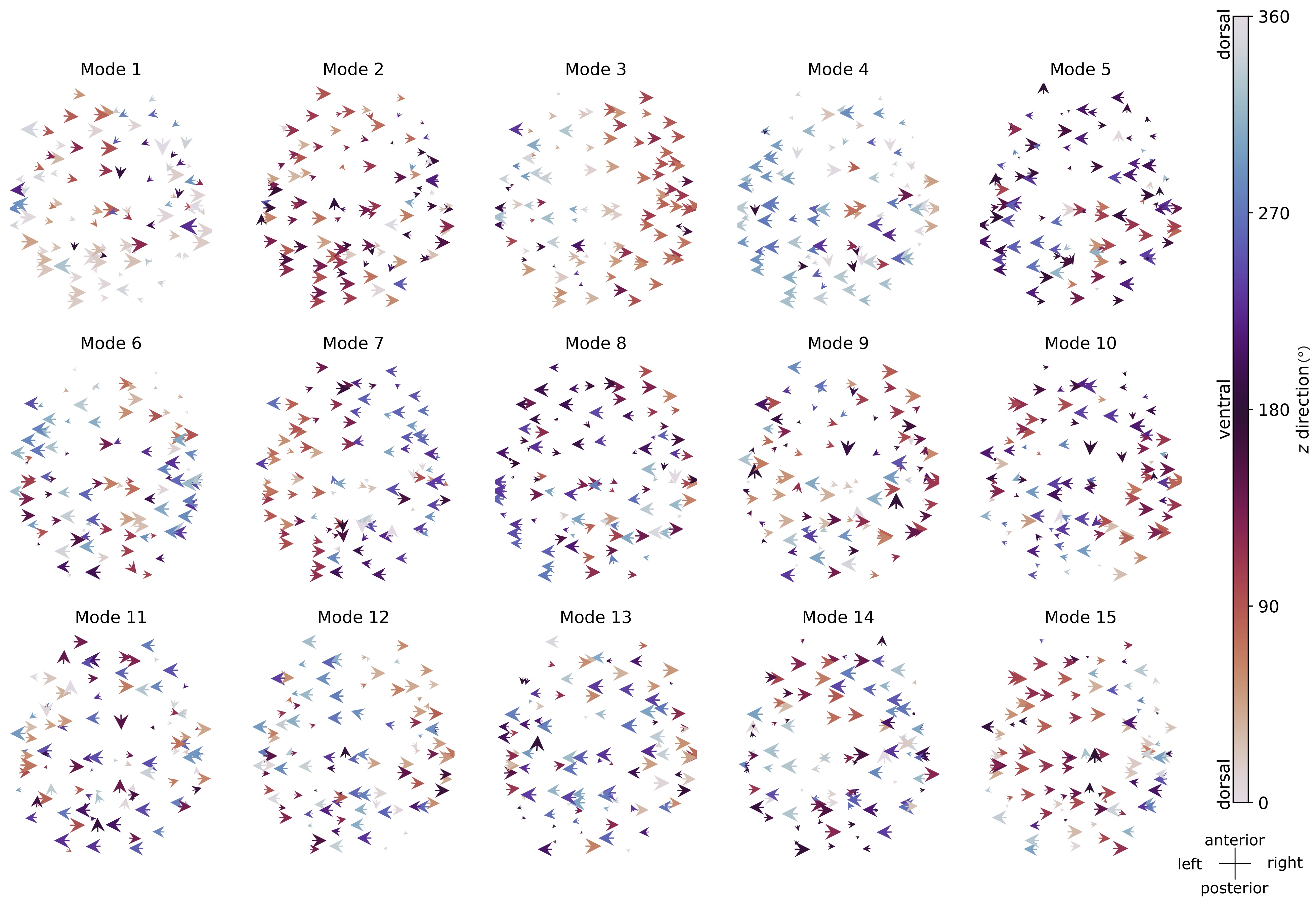}
    \caption{First 15 modes obtained from the singular value decomposition of the velocity vector fields in the whole-brain Wilson-Cowan model with 100 nodes and adaptation (b = 60) for the unistable states in the $LC_{EA}$ region. Modes are ordered in decreasing order of explained variance.}
    \label{fig:wc_modes_100_slowLC}
\end{figure}

\begin{sidewaysfigure}
%\begin{figure}[h!]
    \vspace*{1cm} % Adjust to move the figure up or down
    \centering
    \includegraphics[width=\textwidth]{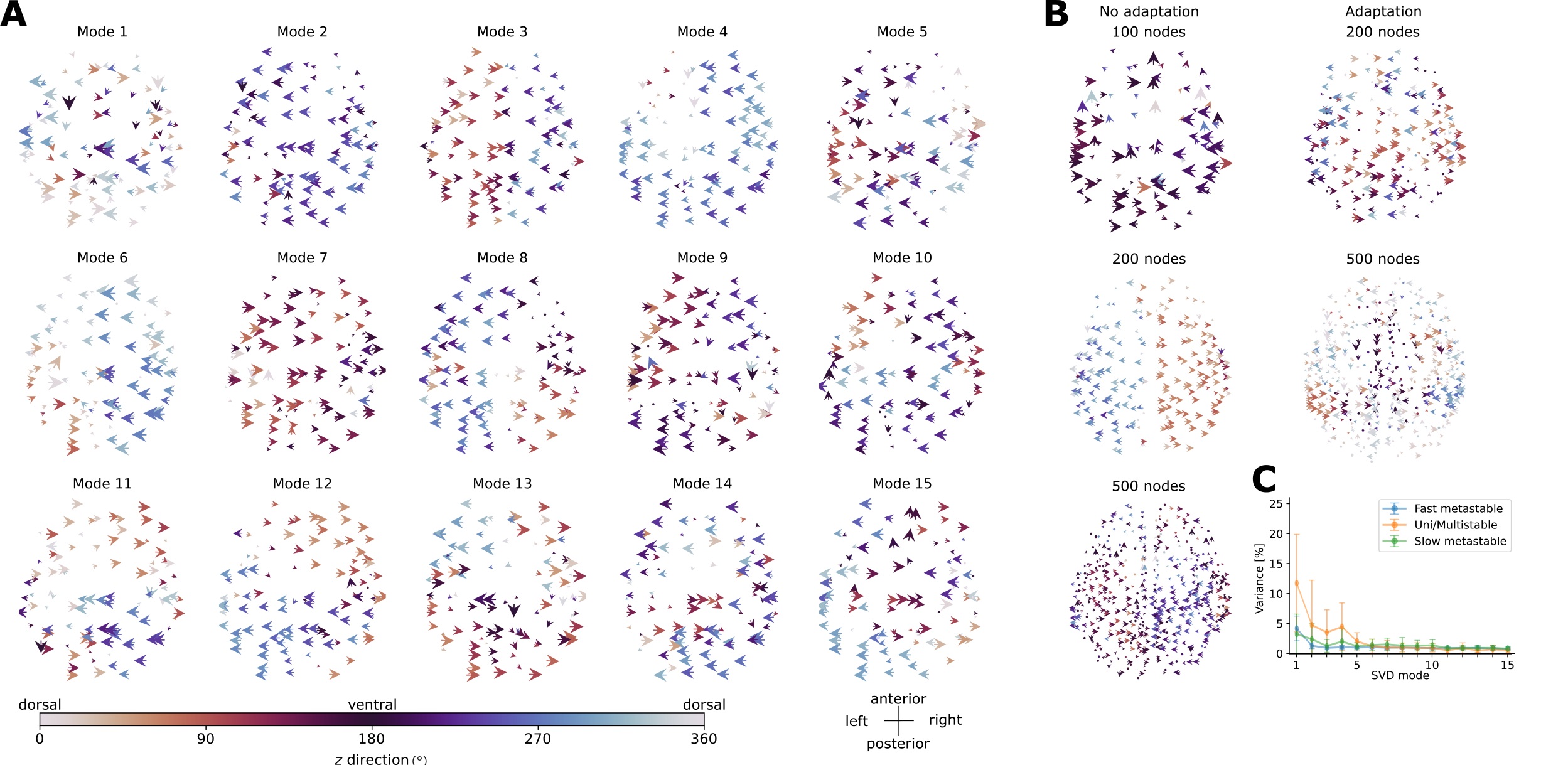}
    \caption{(A) First 15 modes obtained from the singular value decomposition of the velocity vector fields in the whole-brain Wilson-Cowan model with 100 nodes and adaptation (b = 60). Modes are ordered in decreasing order of explained variance. (B) Left panels: modes explaining the largest proportion of variance for the whole-brain Wilson-Cowan model without adaptation (b = 0) with a parcellation of 100, 200, and 500 nodes. Right panels: same as before, but with adaptation (b = 60) and with a parcellation of 200, and 500 nodes. The arrows represent the orientation in the $xy$ plane (left-right and antero-posterior directions) and are color-coded according to the direction along the \textit{z}-axis (dorso-ventral direction). (C) Percentage of explained variance (mean $\pm$ standard deviation across points in the parameter space) of the first 15 modes identified in (A) for the Wilson-Cowan model with 100 nodes and adaptation (b = 60). The percentage is shown for the different pattern types identified in Section 3.2: uni/multistable (orange), fast metastable (blue), and slow metastable (green).}
    \label{fig:wc_spatial_modes}
%\end{figure}
\end{sidewaysfigure}

\begin{table}[H]
\centering
\caption{Percentage of explained variance for the dominant 15 modes identified for the aLN model without ($b$ = 0 pA) and with ($b$ = 20 pA) adaptation for the whole-brain network with 100, 200, and 500 nodes. The column for 100 nodes and $b$ = 20 pA corresponds to the modes shown in Figure 9A.}
\begin{tabular}{||c|ccc|ccc||}
\hline
\textbf{}     & \multicolumn{3}{c|}{b = 0 pA}                       & \multicolumn{3}{c||}{b = 20 pA}                      \\ \hline
Mode & 100 nodes & 200 nodes & 500 nodes & 100 nodes & 200 nodes & 500 nodes \\ \hline
1    & 8.42               & 5.26               & 5.84               & 9.31               & 6.82               & 3.55               \\
2    & 6.99               & 2.10               & 2.33               & 7.43               & 2.76               & 3.06               \\
3    & 3.63               & 1.76               & 1.70               & 5.67               & 1.85               & 1.63               \\
4    & 3.31               & 1.18               & 1.13               & 4.74               & 1.68               & 1.21               \\
5    & 2.88               & 1.04               & 1.10               & 3.13               & 1.26               & 1.16               \\
6    & 2.65               & 1.01               & 0.83               & 2.74               & 1.19               & 0.99               \\
7    & 2.20               & 0.89               & 0.70               & 2.57               & 1.00               & 0.89               \\
8    & 1.84               & 0.76               & 0.62               & 1.85               & 0.88               & 0.82               \\
9    & 1.79               & 0.67               & 0.53               & 1.78               & 0.76               & 0.70               \\
10   & 1.58               & 0.65               & 0.50               & 1.63               & 0.74               & 0.66               \\
11   & 1.40               & 0.61               & 0.43               & 1.56               & 0.69               & 0.63               \\
12   & 1.29               & 0.59               & 0.40               & 1.46               & 0.67               & 0.62               \\
13   & 1.19               & 0.58               & 0.39               & 1.32               & 0.66               & 0.58               \\
14   & 1.13               & 0.55               & 0.37               & 1.18               & 0.64               & 0.55               \\
15   & 1.01               & 0.53               & 0.37               & 1.11               & 0.59               & 0.49               \\ \hline
\end{tabular}
\label{tab:aln_modes_variance}
\end{table}

\begin{table}[H]
\centering
\caption{Percentage of explained variance for the dominant 15 modes identified for the Wilson-Cowan model without ($b$ = 0) and with ($b$ = 60) adaptation for the whole-brain network with 100, 200, and 500 nodes. The column for 100 nodes and $b$ = 60 corresponds to the modes shown in Figure \ref{fig:wc_spatial_modes}A.}
\begin{tabular}{||c|ccc|ccc||}
\hline
\textbf{}     & \multicolumn{3}{c|}{b = 0}                          & \multicolumn{3}{c||}{b = 60}                         \\ \hline
Mode & 100 nodes & 200 nodes & 500 nodes & 100 nodes & 200 nodes & 500 nodes \\ \hline
1    & 14.91              & 13.21              & 6.64               & 3.72               & 11.61              & 5.14               \\
2    & 7.60               & 9.35               & 5.68               & 1.58               & 3.17               & 0.97               \\
3    & 3.59               & 2.86               & 3.05               & 1.44               & 0.97               & 0.57               \\
4    & 2.89               & 2.40               & 2.32               & 1.26               & 0.75               & 0.49               \\
5    & 2.31               & 1.64               & 1.87               & 1.19               & 0.71               & 0.39               \\
6    & 1.94               & 1.49               & 1.56               & 1.14               & 0.65               & 0.38               \\
7    & 1.75               & 1.17               & 1.44               & 1.10               & 0.64               & 0.34               \\
8    & 1.59               & 1.01               & 1.09               & 1.06               & 0.62               & 0.31               \\
9    & 1.48               & 0.91               & 1.04               & 1.04               & 0.55               & 0.31               \\
10   & 1.40               & 0.88               & 0.92               & 1.02               & 0.52               & 0.30               \\
11   & 1.15               & 0.83               & 0.88               & 0.95               & 0.49               & 0.29               \\
12   & 1.09               & 0.76               & 0.83               & 0.91               & 0.47               & 0.28               \\
13   & 1.01               & 0.69               & 0.71               & 0.87               & 0.46               & 0.27               \\
14   & 0.98               & 0.61               & 0.70               & 0.84               & 0.45               & 0.27               \\
15   & 0.94               & 0.59               & 0.64               & 0.80               & 0.45               & 0.27               \\ \hline
\end{tabular}
\label{tab:wc_modes_variance}
\end{table}

\begin{figure}[H]
    \centering
    \includegraphics[width=0.67\textwidth]{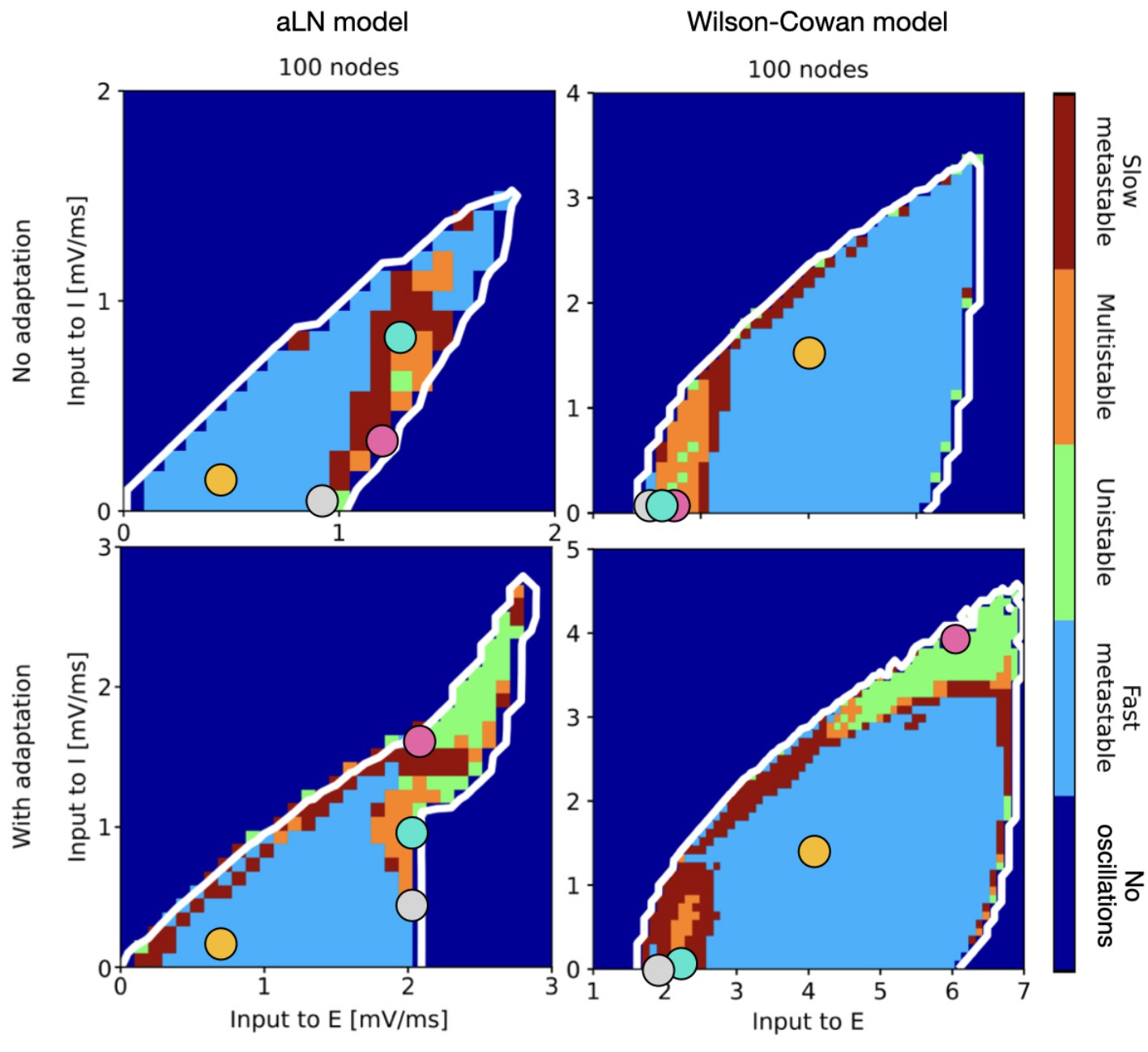}
    \caption{Locations in state space chosen for the results shown in Figures 10 and \ref{fig:long_short_density_wc} for the aLN and Wilson-Cowan (wc) models. Multistable state: turquoise, aLN, without adaptation $(\mu^{ext}_e, \mu^{ext}_i)=(1.3, 0.8)$, with adaptation $(\mu^{ext}_e, \mu^{ext}_i)=(2.0, 1.0)$; wc, without adaptation $(\mu^{ext}_e, \mu^{ext}_i)=(1.6, 0.1)$, with adaptation $(\mu^{ext}_e, \mu^{ext}_i)=(2.1, 0.1)$, unistable state: pink, aLN, without adaptation $(\mu^{ext}_e, \mu^{ext}_i)=(1.2, 0.3)$, with adaptation $(\mu^{ext}_e, \mu^{ext}_i)=(2.0, 1.7)$; wc, without adaptation $(\mu^{ext}_e, \mu^{ext}_i)=(1.7, 0.1)$, with adaptation $(\mu^{ext}_e, \mu^{ext}_i)=(6.0, 4.0)$, fast metastable state: yellow, aLN, without adaptation $(\mu^{ext}_e, \mu^{ext}_i)=(0.4, 0.1)$, with adaptation $(\mu^{ext}_e, \mu^{ext}_i)=(0.4, 0.1)$; wc, without adaptation $(\mu^{ext}_e, \mu^{ext}_i)=(3.0, 1.5)$, with adaptation $(\mu^{ext}_e, \mu^{ext}_i)=(4.0, 1.5)$, slow metastable state: grey, aLN, without adaptation $(\mu^{ext}_e, \mu^{ext}_i)=(0.9, 0.0)$, with adaptation $(\mu^{ext}_e, \mu^{ext}_i)=(2.0, 0.4)$; wc, without adaptation $(\mu^{ext}_e, \mu^{ext}_i)=(1.5, 0.0)$, with adaptation $(\mu^{ext}_e, \mu^{ext}_i)=(1.9, 0.0)$. For all simulations and models $b=0$ in the case of no and $b=20$ ($b=60$) for the aLN (wc) model in the case of finite adaptation.}
    \label{fig:stability-state-markers}
\end{figure}

\begin{figure}[H]
    \centering
    \includegraphics[width=\textwidth]{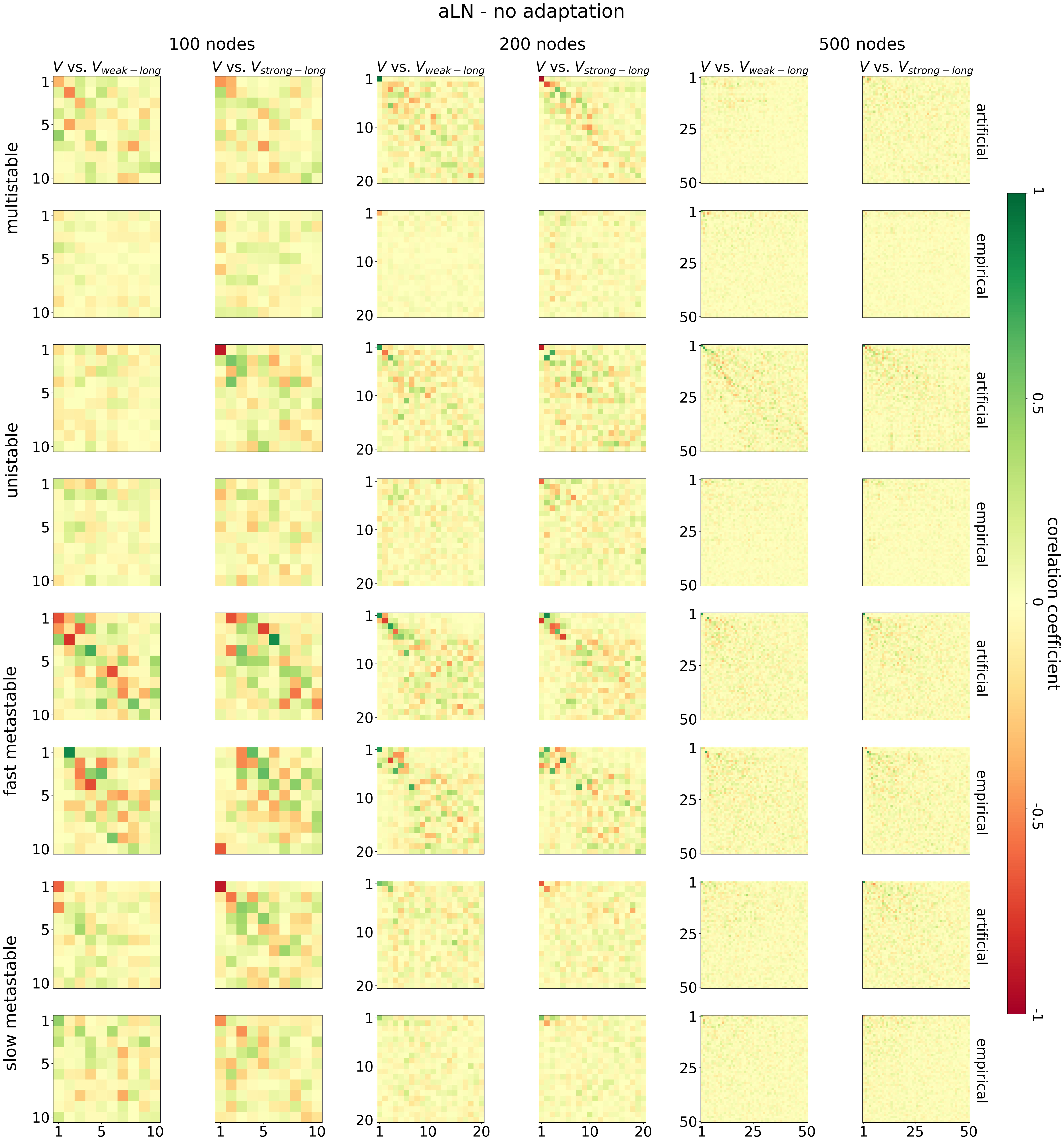}
    \caption{Matrices of correlation coefficients between spatial modes for the aLN model without adaptation. The left (right) column for each parcellation shows the comparison to the case of weak (strong) long-range connections. The upper (lower) rows for each class of states show the comparisons to the case of artificial (empirical) matrices. Colors denote the values of the correlation coefficients.}
    \label{fig:covariance-matrices-aln}
\end{figure}

\begin{figure}[H]
    \centering
    \includegraphics[width=\textwidth]{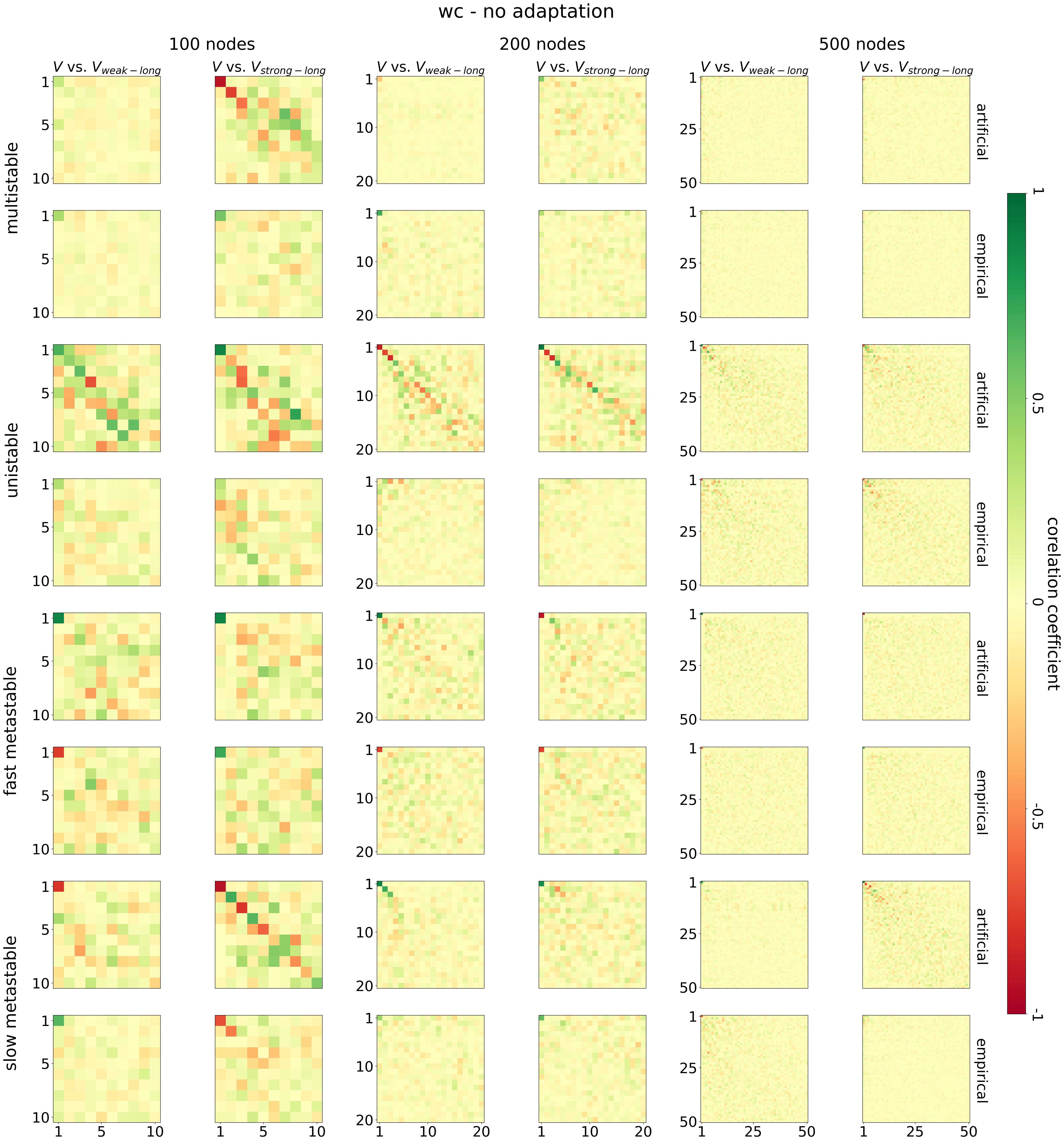}
    \caption{Matrices of correlation coefficients between spatial modes for the Wilson-Cowan model without adaptation. The left (right) column for each parcellation shows the comparison to the case of weak (strong) long-range connections. The upper (lower) rows for each class of states show the comparisons to the case of artificial (empirical) matrices. Colors denote the values of the correlation coefficients.}
    \label{fig:covariance-matrices-wc}
\end{figure}

\begin{table}[H]
    \centering
\caption{Average standard deviation ($\sigma$) and mean ($\mu$) of the density estimates of Figure 10 for the aLN model, per type of stability, with and without adaptation, and per resolution. Density estimates of broadest width per resolution are highlighted with and without adaptation in \textbf{bold}.}
\label{tab:density_params_table_aln}
\begin{tabular}{||ll|ccc|ccc||}
\hline
                &   & \multicolumn{3}{c|}{$\sigma$} & \multicolumn{3}{c||}{$\mu$} \\
                & Resolution &    100 &    200 &    500 &    100 &    200 &    500 \\
\hline
Stability &   &        &        &        &        &        &        \\
\hline
multistable & no adaptation &  0.016 &  0.012 &  0.003 &  0.009 &  0.001 & -0.002 \\
                & adaptation &  0.011 &  0.003 &  0.001 &  0.006 & -0.002 &  0.001 \\
unistable & no adaptation &  0.019 &  0.016 &  0.005 & -0.019 &  0.005 &     0 \\
                & adaptation &  \textbf{0.091} &  \textbf{0.036} &  0.003 &  0.097 &      0 & -0.001 \\
fast metastable & no adaptation &  \textbf{0.061} &  \textbf{0.025} &  \textbf{0.007} & -0.007 &  0.001 &  0.001 \\
                & adaptation &  0.069 &  0.029 &   \textbf{0.01} & -0.009 &  0.002 &  0.001 \\
slow metastable & no adaptation &  0.028 &  0.009 &  0.004 & -0.026 &  0.002 & -0.001 \\
                & adaptation &  0.031 &  0.015 &  0.007 & -0.036 & -0.002 & -0.004 \\
\hline
\end{tabular}
\end{table}

\begin{figure}[H]
    \centering
    \includegraphics[width=0.92\textwidth]{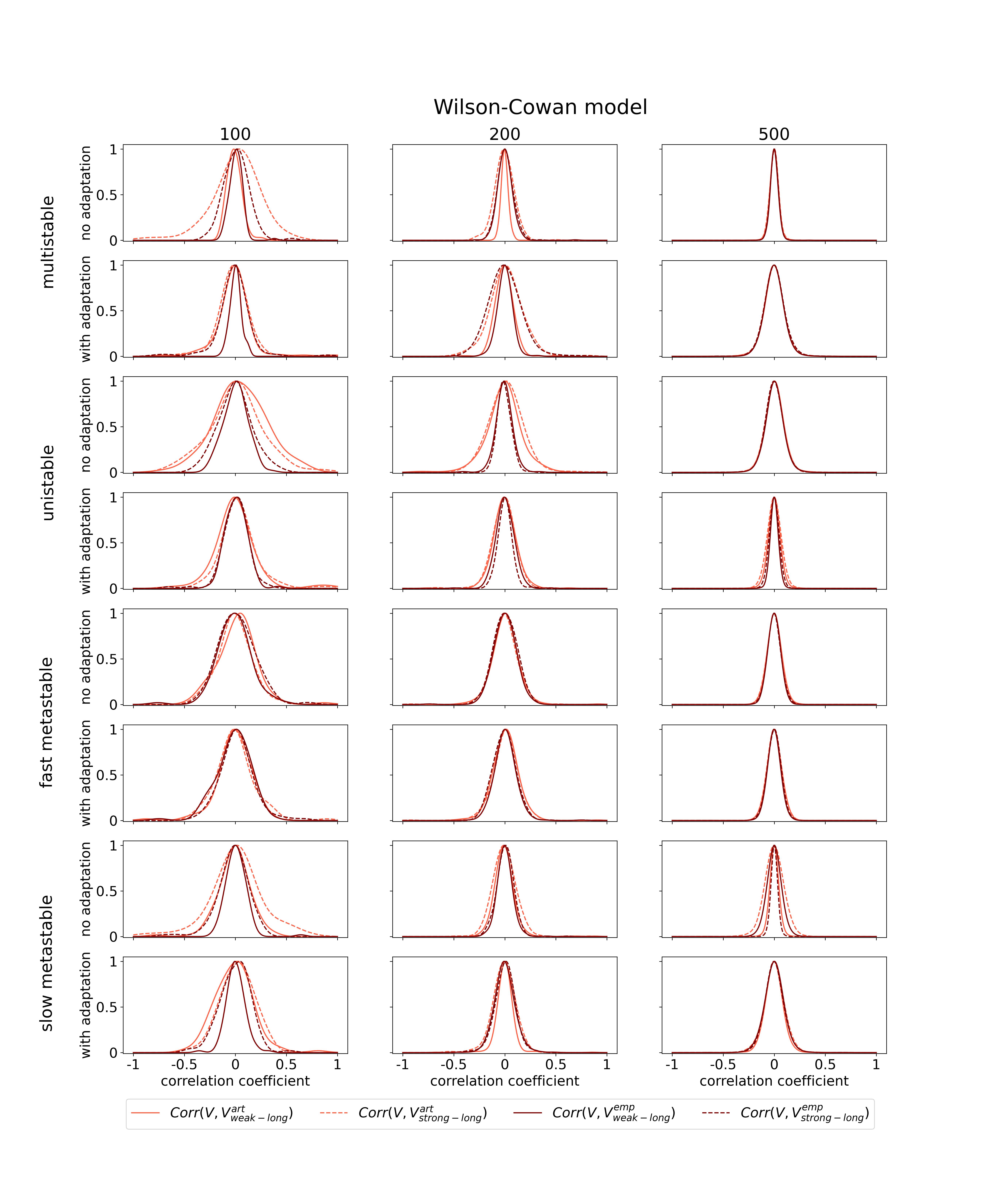}
    \caption{Distribution of the values from the matrices $Corr(V,V^{type}_{strengh})$ of correlation coefficients, each normalized to its maximum value. Correlation coefficients are computed between the spatial modes obtained with the averaged connectivity matrix $C$ and with the spatial modes of the empirically derived (darker colors, $V^{emp}$) and the artificially manipulated (lighter colors, $V^{art}$) matrices, with stronger (dashed, $V_{strong-long}$) and weaker (solid, $V_{weak-long}$) long-range connections. Distributions are estimated using kernel density estimation. Each column corresponds to one parcellation, each pair of rows (upper row without, lower row with adaptation) to the type of stability (multistable, unistable, fast, and slow metastable). Means and standard deviatons are provided in Table \ref{tab:density_params_table_wc}. For parameters, see \ref{fig:stability-state-markers}.}
    \label{fig:long_short_density_wc}
\end{figure}

\begin{table}[H]
    \centering
\caption{Average standard deviation ($\sigma$) and mean ($\mu$) of density estimates of Figure \ref{fig:long_short_density_wc} for the Wilson-Cowan model, per type of stability, with and without adaptation, and per resolution. Density estimates of broadest width per resolution are highlighted with and without adaptation in \textbf{bold}.}
\label{tab:density_params_table_wc}
\begin{tabular}{||ll|ccc|ccc||}
\hline
                &   & \multicolumn{3}{c|}{$\sigma$} & \multicolumn{3}{c||}{$\mu$} \\
                & Resolution &    100 &    200 &    500 &    100 &    200 &    500 \\
\hline
Stability &   &        &        &        &        &        &        \\
\hline
multistable & no adaptation &  0.019 &  0.005 &  0.001 &   0.01 &  0.006 &      0 \\
                & adaptation &  0.021 &  \textbf{0.015} &  \textbf{0.008} & -0.015 &  0.002 &  0.001 \\
unistable & no adaptation &  \textbf{0.041} &  \textbf{0.017} &  \textbf{0.007} & -0.003 & -0.006 & -0.001 \\
                & adaptation &  0.026 &  0.009 &  0.003 &     0 &  0.001 &     0 \\
fast metastable & no adaptation &   0.03 &  0.013 &  0.004 &  0.016 &  0.002 &  0.001 \\
                & adaptation &  \textbf{0.029} &  0.013 &  0.004 &  0.017 & -0.003 & -0.001 \\
slow metastable & no adaptation &   0.03 &  0.008 &  0.004 & -0.026 &  0.005 &      0 \\
                & adaptation &  0.023 &   0.01 &  0.008 & -0.002 &  0.002 &      0 \\
\hline
\end{tabular}
\end{table}

\begin{table}[H]
\centering
\caption{Overview of the parameter values used for the whole-brain aLN sleep model with 100, 200, and 500 nodes. All other parameters are given in Table 1.}
\begin{tabular}{||c|c|c|c|c||}
\hline
\multirow{2}{*}{Parameter} & \multicolumn{3}{c|}{Value} & \multirow{2}{*}{Description} \\
\cline{2-4}
 & 100 & 200 & 500 & \\
\hline
$\mu_E^{ext}$ & 3.3 mV/ms & 3.3 mV/ms & 3.3 mV/ms & \makecell{Mean external \\input to E} \\
$\mu_I^{ext}$ & 3.7 mV/ms & 3.7 mV/ms & 3.7 mV/ms & \makecell{Mean external \\input to I} \\
$\sigma_{ou}$ & 0.37 mV/$ms^{3/2}$ & 0.37 mV/$ms^{3/2}$ & 0.37 mV/$ms^{3/2}$ & \makecell{Noise \\strength} \\
$b$ & 3.2 pA & 3.2 pA & 4.2 pA & \makecell{Adaptation \\strength} \\
$\tau_A$ & 4765 ms & 4765 ms & 4965 ms & \makecell{Adaptation \\time constant} \\
\hline
\end{tabular}
\label{tab:aln_sleep_params}
\end{table}

\begin{table}[H]
\centering
\caption{Overview of the parameter values used for the whole-brain Wilson-Cowan sleep model with 100, 200, and 500 nodes. All other parameters are given in Table 2.}
\begin{tabular}{||c|c|c|c|c||}
\hline
\multirow{2}{*}{Parameter} & \multicolumn{3}{c|}{Value} & \multirow{2}{*}{Description} \\
\cline{2-4}
 & 100 & 200 & 500 & \\
\hline
$\mu_E^{ext}$ & 5.26 & 5.26 & 5.26 & \makecell{Mean external input to E} \\
$\mu_I^{ext}$ & 5.51 & 5.61 & 5.61 & \makecell{Mean external input to I} \\
$\sigma_{ou}$ & 0.49 & 0.49 & 0.49 & \makecell{Noise strength} \\
$K_{gl}$ & 2.18 & 2.18 & 2.18 & Coupling strength \\
$b$ & 21.45 & 27 & 59 & \makecell{Adaptation strength} \\
$\tau_A$ & 1629.46 & 2600 & 2920 & \makecell{Adaptation time constant} \\
$v_{gl}$ & 20 & 20 & 20 & Global signal speed \\
\hline
\end{tabular}
\label{tab:wc_sleep_params}
\end{table}

\begin{figure}[H]
    \centering
    \includegraphics[width=\textwidth]{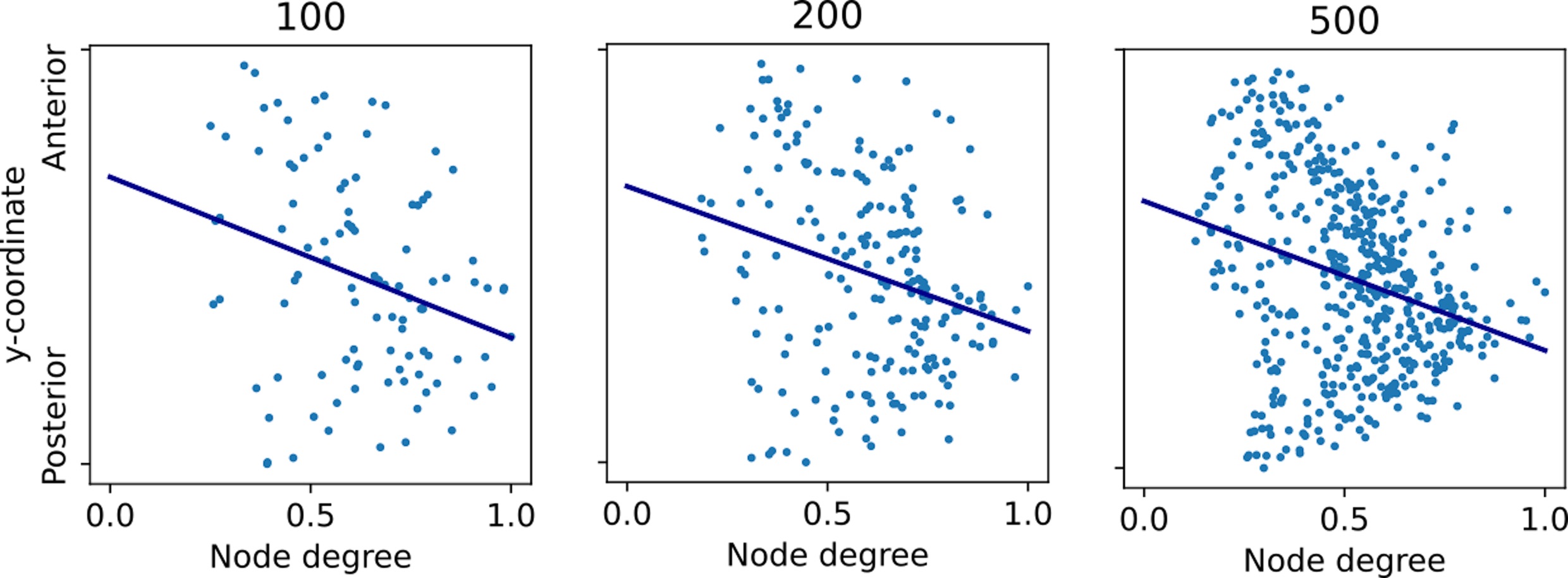}
    \caption{Correlation between node degree and the y-coordinate along the antero-posterior axis for the Schaefer parcellation scheme with 100 (left; \textit{y}-slope = -62.29, r = -0.29, $p =$ 0.003), 200 (middle; \textit{y}-slope = -58.65, r = -0.27, $p <$ 0.001), and 500 nodes (right; \textit{y}-slope = -63.86, r = -0.28, $p <$ 0.001).}
    \label{fig:sc_gradient}
\end{figure}

\begin{figure}[H]
    \centering
    \includegraphics[width=\textwidth]{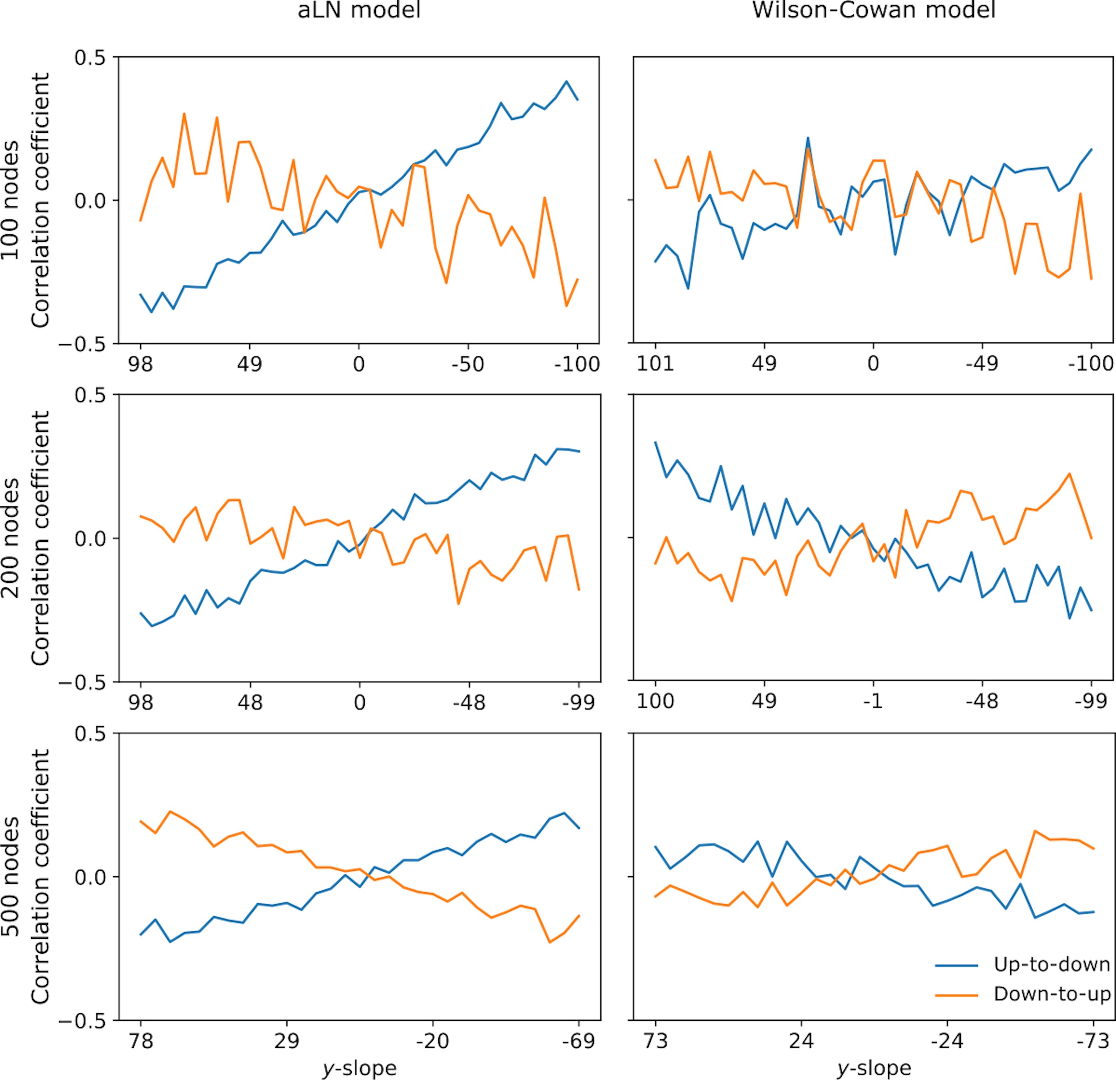}
    \caption{Correlation coefficient between mean transition phases of the nodes from the up to the down state (blue) and vice-versa (orange) and the node coordinates along the antero-posterior $y$-axis as a function of the structural connectivity gradient ($y$-slope) along the antero-posterior axis. These values were set as the targets during the permutation of the structural connectivity matrix for the aLN (left column) and the Wilson-Cowan models (right column) with 100 (top row), 200 (middle row), and 500 nodes (bottom row). The range of slope values that could be achieved through permutation was lower compared to those in Figure 11 and was additionally restricted for the 500 nodes case. Model parameters are given in Tables \ref{tab:aln_sleep_params} and \ref{tab:wc_sleep_params}.}
    \label{fig:sc_gradient_change_2}
\end{figure}

\begin{figure}[H]
    \centering
    \includegraphics[width=\textwidth]{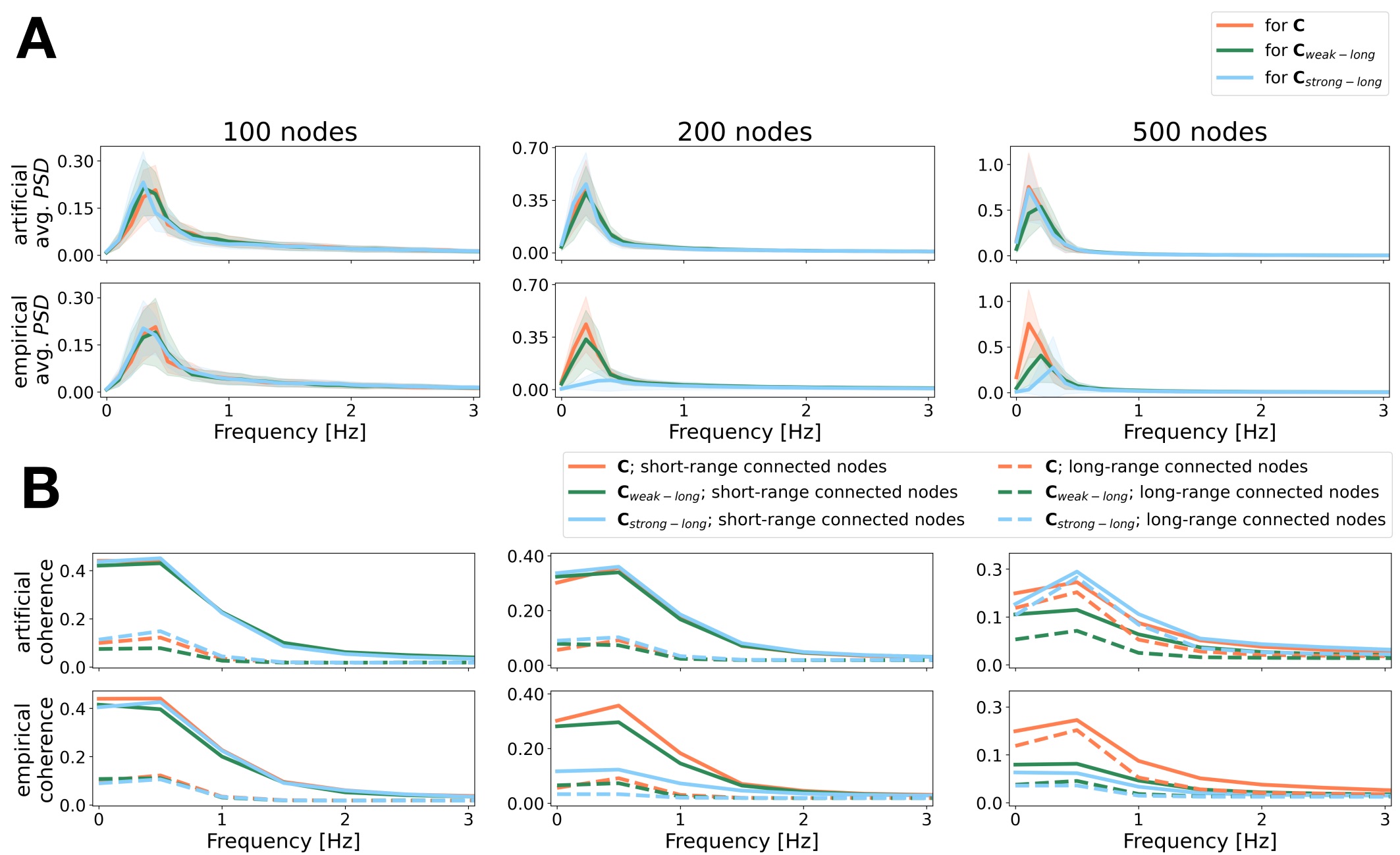}
    \caption{Power and coherence as a function of frequency for SO activity generated by the Wilson-Cowan model. Results are shown for the average connectivity matrix, $C$, (coral), and the connectivity matrices with weaker, $C_{weak-long}$, (green) and stronger, $C_{strong-long}$, (blue) long-range connections. Every column corresponds to one parcellation. (A) Averaged power spectra with standard deviation for each activity induced by the three connectivity matrices. The top (bottom) row shows the results for the artificially changed (empirically selected) connections. (B) Corresponding coherence values plotted separately for nodes that are connected through short- (solid lines) or long-range (dashed lines) connections. Model parameters are given in Table \ref{tab:wc_sleep_params}.}
    \label{fig:coherence_wc}
\end{figure}

\begin{table}[H]
    \centering
    \caption{Values of the dominant temporal frequency $f_{dom}=\text{argmax}_favg\bigl(PSD(f)\bigr)$ of the averaged power spectrum and the corresponding peak power spectrum $P(f_{dom})$ of Figure 12A per parcellation for the aLN model. Values in \textbf{bold} indicate dominant frequencies $f_{dom}$ different from $0.4\ \rm{Hz}$ which appears in most settings. Lowest row displays the standard deviation in feature per column.}
    \begin{tabular}{|| ll|rrr|rrr ||}
\hline
          & {} & \multicolumn{3}{c|}{$f_{dom}$} & \multicolumn{3}{c||}{$PSD(f_{dom})$} \\
          & Resolution &                100 &  200 &  500 &                         100 &     200 &     500 \\
\hline
artificial & $\textbf{C}$ &                0.4 &  0.4 &  0.4 &                      422.57 &  233.90 &  188.83 \\
          & $\textbf{C}_{weak-long}$ &                0.4 &  0.4 &  \textbf{0.5} &                      404.39 &  296.59 &  224.61 \\
          & $\textbf{C}_{strong-long}$ &                0.4 &  0.4 &  0.4 &                      498.04 &  246.58 &  209.31 \\
empirical & $\textbf{C}$ &                0.4 &  0.4 &  0.4 &                      422.57 &  233.90 &  188.83 \\
          & $\textbf{C}_{weak-long}$ &                0.4 &  0.4 &  \textbf{0.5} &                      558.58 &  284.15 &  179.36 \\
          & $\textbf{C}_{strong-long}$ &                0.4 &  0.4 &  \textbf{0.6} &                      482.22 &  124.25 &   97.85 \\
Standard Deviation & & 0.0 & 0.0 & 0.08 & 59 & 61 & 44 \\
\hline
\end{tabular}
    \label{tab:aln_psd_table}
\end{table}

\begin{table}[H]
    \centering
    \caption{Values of the dominant temporal frequency $f_{dom}=\text{argmax}_favg\bigl(PSD(f)\bigr)$ of the averaged power spectrum and the corresponding peak power spectrum $P(f_{dom})$ of Figure \ref{fig:coherence_wc}A per parcellation for the Wilson-Cowan model. Values in \textbf{bold} indicate dominant frequencies $f_{dom}$ different from most of the other dominant frequencies per parcellation. Lowest row displays the standard deviation in feature per column.}
    \begin{tabular}{||ll|rrr|rrr||}
\hline
          & Property & \multicolumn{3}{c|}{$f_{dom}$} & \multicolumn{3}{c||}{$PSD(f_{dom})$} \\
          & Resolution &                100 &  200 &  500 &                         100 &   200 &   500 \\
\hline
artificial & $\textbf{C}$ &                \textbf{0.4} &  0.2 &  0.1 &                        0.21 &  0.44 &  0.75 \\
          & $\textbf{C}_{weak-long}$ &                0.3 &  0.2 &  \textbf{0.2} &                        0.21 &  0.40 &  0.54 \\
          & $\textbf{C}_{strong-long}$ &                0.3 &  0.2 &  0.1 &                        0.23 &  0.46 &  0.73 \\
empirical & $\textbf{C}$ &                \textbf{0.4} &  0.2 &  0.1 &                        0.21 &  0.44 &  0.75 \\
          & $\textbf{C}_{weak-long}$ &                \textbf{0.4} &  0.2 &  \textbf{0.2} &                        0.19 &  0.34 &  0.41 \\
          & $\textbf{C}_{strong-long}$ &                0.3 &  \textbf{0.4} &  \textbf{0.3} &                        0.20 &  0.06 &  0.27 \\
Standard Deviation & & 0.05 & 0.08 & 0.08 & 0.01 & 0.15 & 0.2 \\
\hline
\end{tabular}
    \label{tab:wc_psd_table}
\end{table}

\begin{table}[H]
    \centering 
    \caption{Maximum coherence values for non-zero frequencies for the metastable states of the Wilson-Cowan model for all settings shown in Figure \ref{fig:coherence_wc}B. Both for the artificial and the empirical case, values in \textbf{bold} indicate the highest coherence values per parcellation, per set of nodes connected on a short-range (long-range). The corresponding frequencies were 0.5 Hz for all settings. Parameters are as for Figure \ref{fig:coherence_wc}.}
    \begin{tabular}{||ll|rrrrrr||}
\hline
          & Property & \multicolumn{6}{c||}{$coh(f_{max})$} \\
          & Resolution & \multicolumn{2}{c}{100} & \multicolumn{2}{c}{200} & \multicolumn{2}{c||}{500} \\
          & Distance &            short &  long & short &  long & short &  long \\
\hline
artificial & $\textbf{C}$ &       0.44 &  0.12 & \textbf{0.36} &  0.09 &  0.26 &  0.23 \\
          & $\textbf{C}_{weak-long}$ &       0.43 &  0.08 &  0.34 &  0.07 &  0.17 &  0.11 \\
          & $\textbf{C}_{strong-long}$ &      \textbf{0.45} & \textbf{0.15} & \textbf{0.36} & \textbf{0.10} & \textbf{0.29} & \textbf{0.27} \\
empirical & $\textbf{C}$ &   \textbf{0.44} & \textbf{0.12} & \textbf{0.36} & \textbf{0.09} & \textbf{0.26} & \textbf{0.23} \\
          & $\textbf{C}_{weak-long}$ &       0.40 &  0.11 &  0.30 &  0.07 &  0.12 &  0.07 \\
          & $\textbf{C}_{strong-long}$ &       0.43 &  0.11 &  0.12 &  0.03 &  0.09 &  0.05 \\
\hline
\end{tabular}
    \label{tab:wc_coherence_table}
\end{table}

%\end{document}
\end{document}